%% file: caha.tex
\documentclass[usegraphicx,usedcolumn,usenatbib]{mn2e}
\usepackage{astbibref}
\usepackage{color,epsfig,amssymb,longtable}
\usepackage{array,colortbl,lscape}

\def\HII{H{\sc ii}}

\def\uno{SDSS J145506.06+380816.6}  
\def\dos{SDSS J150909.03+454308.8}  
\def\tres{SDSS J152817.18+395650.4}  
\def\cuatro{SDSS J154054.31+565138.9}  
\def\cinco{SDSS J161623.53+470202.3}  
\def\seis{SDSS J165712.75+321141.4}  
\def\siete{SDSS J172906.56+565319.4}  

\def\unoc{SDSS J1455}  
\def\dosc{SDSS J1509}  
\def\tresc{SDSS J1528}  
\def\cuatroc{SDSS J1540}  
\def\cincoc{SDSS J1616}  
\def\seisc{SDSS J1657}  
\def\sietec{SDSS J1729}  

\begin{document}
\title[Precision abundances of bright H{\sevensize\it II} galaxies]
{Precision abundance analysis of bright \HII\ galaxies} 
\author[G. H\"agele et al.]
{Guillermo~F.~H\"agele$^{1}$\thanks{PhD fellow of Ministerio de Educaci\'on y
    Ciencia, Spain; guille.hagele@uam.es},
\'Angeles~I.~D\'{\i}az$^{1}$\thanks{On sabbatical leave at IoA, Cambridge},  
Elena Terlevich$^{3}$\thanks{Research Affiliate at IoA},
Roberto Terlevich$^{3}$\footnotemark[3], 
\newauthor Enrique~P{\'e}rez-Montero$^{1,3}$\thanks{Post-Doc fellow of
  Ministerio de Educaci\'on y Ciencia, Spain}, 
and M\'onica V. Cardaci$^{1,4}$\footnotemark[1]
\\
$^{1}$ Departamento de F\'{\i}sica Te\'orica, C-XI, Universidad Aut\'onoma de
Madrid, 28049 Madrid, Spain\\ 
$^{2}$ INAOE, Tonantzintla, Apdo. Postal 51, 72000 Puebla, M\'exico\\ 
$^{3}$ Laboratoire d'Astrophysique de Toulouse-Tarbes. Observatoire
  Midi-Pyr\'en\'ees. 14, avenue Edouard Belin. 31400. Toulouse. France\\
$^{4}$ XMM Science Operations Centre, European Space Astronomy Centre of ESA,
P.O. Box 50727, 28080 Madrid, Spain\\ 
}
\date{Accepted 
      Received ;
      in original form }
\pagerange{\pageref{firstpage}--\pageref{lastpage}}
\pubyear{2005}
\maketitle

\begin{abstract}

We present high signal-to-noise spectrophotometric observations of 
seven luminous \HII\ galaxies. The observations have been made with 
the use of a double-arm spectrograph which provides spectra with a 
wide wavelength coverage, from 3400 to 10400\,\AA\ free of second order effects, 
of exactly the same region of a given galaxy. These observations are analysed 
applying a methodology designed to obtain accurate elemental abundances
of oxygen, sulphur, nitrogen, neon, argon and iron in the ionized gas. 
Four electron temperatures and one electron density are derived 
from the observed forbidden line ratios using the five-level atom approximation. 
For our best objects errors of  1\% in t$_e$([O{\sc iii}]),  
3\% in t$_e$([O{\sc ii}]) and 5\% in t$_e$([S{\sc iii}]) are achieved 
with a resulting accuracy of  7\% in total oxygen abundances, O/H.

The ionisation structure of the nebulae can be mapped by the theoretical
oxygen and sulphur ionic ratios, on the one side,  and the corresponding
observed emission line ratios, on the other --  the $\eta$ and $\eta$' plots
--. The combination of both is shown to provide a means to test
photo-ionisation model sequences currently applied to derive elemental
abundances in \HII\ galaxies.

\end{abstract}


\section{Introduction}

When studying evolution two types of ages should be distinguished: the
chronological and the evolutionary ages. In the case of galaxies, estimates of
the chronological age can be obtained analyzing, for example, the age
distribution of their stellar population while the evolutionary age can be
estimated from, for example, the metal content of their interstellar medium.  

\HII\ galaxies, the subclass of Blue Compact Dwarf galaxies (BCDs) which show
spectra with strong emission lines similar to those of giant extragalactic
\HII\ regions \citep[GEHRs;][]{1970ApJ...162L.155S,1980ApJ...240...41F}, have
the lowest metal content of any starforming galaxy suggesting that they are
among the youngest or less evolved galaxies known \citep{2007ApJ...654..226R,
  1972ApJ...173...25S}. After the findings that a considerable number of the
objects observed at intermediate and high redshifts seem to have properties
similar to the \HII\ galaxies we know in the Local  Universe, it has been
suggested that these objects might have been very common in the past and some
of them may have evolved to other kind of objects
\citep{1995ApJ...440L..49K}. In order to detect these evolutionary effects we
need to compare the properties of \HII\ galaxies both in the Local Universe
and at higher redshifts. We therefore need to know the true distribution
functions of their properties among which the chemical abundances are of the
greatest relevance.

Spectrophotometry of bright \HII\ galaxies in the Local Universe allows the
determination of abundances from methods that rely on the measurement of
emission line intensities and atomic physics. This is referred to as the
"direct" method. In the case of more distant or intrinsically fainter
galaxies, the low signal-to-noise obtained with current telescopes precludes
the application of this method and empirical ones based on the strongest
emission lines are required. The fundamental basis of these empirical methods
is reasonably well understood \citep[see e.g.][]{2005MNRAS.361.1063P}. The
accuracy of the results however depends on the goodness of their calibration
which in turn depends on a well sampled set of precisely derived abundances by
the "direct" method so that interpolation procedures are reliable. Enlarging
the calibration range is also important since, at any rate, empirically 
obtained relations should never be used outside their calibration validity
range. 

The precise derivation of elemental abundances however is not a
straightforward matter. Firstly, accurate measurements of the emission lines
are needed. Secondly, a certain knowledge of the ionisation structure of the
region is required in order to derive ionic abundances of the different
elements and in some cases photoionisation models are needed to correct for
unseen ionisation states. An accurate diagnostic requires the measurement of
faint auroral lines covering a wide spectral range and their accurate (better
than 5\%) ratios to Balmer recombination lines. These faint lines are usually
about 1\% of the  H$\beta$ intensity. The spectral range must include from
the UV [O{\sc ii}]\,$\lambda$\,3727\,\AA\ doublet, to the near IR [S{\sc iii}]
$\lambda\lambda$\,9069,9532\,\AA\ lines. This allows the derivation of the
different line temperatures: T$_e$([O{\sc ii}]), T$_e$([S{\sc ii}]),
T$_e$([O{\sc iii}]), T$_e$([S{\sc iii}]), T$_e$([N{\sc ii}]), needed in order
to study the temperature and ionisation structure of each \HII\ galaxy
considered as a multizone ionised region.


Unfortunately most of the available \HII\ galaxy spectra have only a
restricted wavelength  range (usually from about 3600 to 7000\,\AA),
consequence of observations with single arm spectrographs, and do not have the
adequate S/N to accurately measure the intensities of the weak diagnostic
emission lines. Even the Sloan Digital Sky Survey (SDSS; Stoughton et al.\
2002) \nocite{2002AJ....123..485S} spectra do not cover simultaneously the
3727 [O{\sc ii}] and  the 9069 [S{\sc iii}] lines, they only represent an
average inside a 3\,arcsec fibre and reach the required S/N only for the
brightest objects.  

It is important to realise that the combination of accurate spectrophotometry
and wide spectral coverage cannot be achieved  using single arm spectrographs
where, in order to  reach the necessary spectral resolution, the wavelength
range must be split into several independent observations. In those cases, the
quality of the spectrophotometry is at best doubtful mainly because the
different spectral ranges are not observed simultaneously. This problem
applies to both objects and calibrators. Furthermore one can never be sure of
observing exactly the same region of the nebula in each spectral range. To
avoid all these problems the use of double arm spectrographs is required.

In this work we present simultaneous blue and red observations obtained with
the double arm TWIN spectrograph at the 3.5m telescope of the Spanish-German
Observatory of Calar Alto. These data are of a sufficient quality as to allow
the detection and measurement of several temperature sensitive lines and add
to the still scarce base of precisely derived abundances. In the next section
we describe some details regarding the selection of the sample as well as the
observations and data reduction. The results are presented in section
3. Sections 4 and 5 are devoted to the analysis of these results which are
compared with previous data in section 6. Section 7 is devoted to the
discussion of our results and finally, our conclusions are summarized in
section 8.

%
%

\begin{table*}
\centering
\caption[]{Journal of observations.}
\label{jour}
\begin{tabular} {@{}c c c c c c}
\hline
 \multicolumn{1}{c}{Object  ID}  &  spSpec SDSS   & hereafter ID &  Date &  Exposure (s)  & Seeing (\arcsec)\\
\hline
\uno    & spSpec-52790-1351-474  & \unoc    & 2006 June 25  &  5 $\times$ 1800  & 0.9-1.2 \\
\dos    & spSpec-52721-1050-274  & \dosc    & 2006 June 23  &  4 $\times$ 1800  & 0.8-1.1 \\
\tres   & spSpec-52765-1293-580  & \tresc   & 2006 June 22  &  4 $\times$ 1800  & 0.8-1.2 \\
\cuatro & spSpec-52072-0617-464  & \cuatroc & 2006 June 24  &  6 $\times$ 1800  & 1.0-1.4 \\
\cinco  & spSpec-52377-0624-361  & \cincoc  & 2006 June 23  &  5 $\times$ 1800  & 0.8-1.1 \\
\seis   & spSpec-52791-1176-591  & \seisc   & 2006 June 25  &  5 $\times$ 1800  & 0.9-1.2 \\
\siete  & spSpec-51818-0358-472  & \sietec  & 2006 June 22  &  5 $\times$ 1800  & 0.8-1.1 \\
\hline
\end{tabular}
\end{table*}

%
%

\begin{table*}
\centering
\caption[]{Right ascension, declination, redshift and SDSS photometric
  magnitudes obtained using the SDSS explore tools$^a$.}
\label{obj}
\begin{tabular} {l c c c c r r c c}
\hline
 \multicolumn{1}{c}{Object  ID}   & \multicolumn{1}{c}{RA} & \multicolumn{1}{c}{Dec} &  redshift  & \multicolumn{1}{c}{u} & \multicolumn{1}{c}{g} & \multicolumn{1}{c}{r} & \multicolumn{1}{c}{i}   &    \multicolumn{1}{c}{z} \\
\hline
\unoc    &  14$^h$\,55$^m$\,06\fs06 & 38$^{\circ}$\,08\arcmin\,16\farcs67 &  0.028 & 18.25 & 17.57 & 17.98 & 18.23 & 18.18 \\
\dosc    &  15$^h$\,09$^m$\,09\fs03 & 45$^{\circ}$\,43\arcmin\,08\farcs88 &  0.048 & 18.57 & 17.72 & 18.19 & 17.87 & 17.94 \\
\tresc   &  15$^h$\,28$^m$\,17\fs18 & 39$^{\circ}$\,56\arcmin\,50\farcs43 &  0.064 & 18.54 & 17.88 & 18.17 & 17.52 & 17.99 \\
\cuatroc &  15$^h$\,40$^m$\,54\fs31 & 56$^{\circ}$\,51\arcmin\,38\farcs98 &  0.011 & 19.11 & 18.91 & 18.97 & 19.53 & 19.46 \\
\cincoc  &  16$^h$\,16$^m$\,23\fs53 & 47$^{\circ}$\,02\arcmin\,02\farcs36 &  0.002 & 16.84 & 16.45 & 16.77 & 17.35 & 17.43 \\
\seisc   &  16$^h$\,57$^m$\,12\fs75 & 32$^{\circ}$\,11\arcmin\,41\farcs42 &  0.038 & 17.63 & 17.03 & 17.27 & 17.15 & 17.15 \\
\sietec  &  17$^h$\,29$^m$\,06\fs56 & 56$^{\circ}$\,53\arcmin\,19\farcs40 &  0.016 & 18.05 & 17.26 & 17.21 & 17.38 & 17.24 \\
\hline
\multicolumn{9}{l}{$^a$http://cas.sdss.org/astro/en/tools/explore/obj.asp}
\end{tabular}
\end{table*}


\section{Observations and data reduction}
\label{Obs}

\subsection{Object selection}

SDSS constitutes a very valuable base for statistical studies of the
properties of galaxies. At this moment, the Fifth Data
Release\footnote{http://www.sdss.org/dr6/} (DR6), the last one up to now,
represents the completion of the SDSS-I project (Adelman-McCarthy et al.\
2007).\nocite{2007arXiv0707.3413A} The DR6 contains five-band photometric data
for  about 2.87$\times$10$^8$  objects selected over 9583\,$deg^2$ and more
than 1.27 million spectra of galaxies, quasars and stars selected from
7425\,$deg^2$.



Using the implementation of the SDSS database in the INAOE Virtual Observatory
superserver\footnote{http://ov.inaoep.mx/}, we selected from the SDSS DR3 the
brightest nearby narrow emission line galaxies with very strong lines and
large equivalent widths of the H$\alpha$ line.  
Specifically our selection criteria were:
\begin{itemize}
\item[-] H$\alpha$ flux, F(H$\alpha$) $>$ 4 $\times$ 10$^{-14}$ erg\,cm$^{-2}$\,s$^{-1}$
\item[-] H$\alpha$ equivalent width, EW(H$\alpha$) $>$ 50 \AA
\item[-] H$\alpha$ width, 2.8 $<$ FWHM(H$\alpha$) $<$ 16 \AA
\item[-] redshift, {\it z}, 10$^{-3}$ $< z <$ 0.2
\end{itemize}
AGN-like objects were removed from this list by using diagnostic diagrams of
the kind presented in \citet*{1981PASP...93....5B}. The obtained list
contains about 10500 \HII\ like objects  \citep{jesustesis}. They show
spectral properties indicating a wide range of gaseous abundances and ages of
the underlying stellar populations.  

The objects with the highest (H$\alpha$) fluxes and equivalent widths
observable from the Calar Alto Observatory at the epoch of observation were
selected and for seven of them the corresponding data were secured. 

The journal of observations is given in table \ref{jour}  and some general
characteristics of the objects from the SDSS web page are listed in table
\ref{obj}. Column 3 of table \ref{jour} gives the short name by which we will
refer to the observed  \HII\ galaxies in what follows.

%
%

\begin{table}
\centering
\caption[]{CAHA TWIN configuration for the observations.}
\label{config}
\begin{tabular} {l c c c c}
\hline
 & Spectral range  &       Disp.             & R$_{\rm{FWHM}}$$^a$ & Spatial res.         \\
 &      (\AA)             & (\AA\,px$^{-1}$) &        & (\arcsec\,px$^{-1}$) \\
\hline
blue & 3400-5700   &       1.09             &  1970   &   0.56                \\
red  & 5800-10400  &       2.42             &  1560   &   0.56                \\
\hline
\multicolumn{5}{l}{$^a$R$_{\rm{FWHM}}$\,=\,$\lambda$/$\Delta\lambda_{\rm{FWHM}}$}
\end{tabular}
\end{table}

\subsection{Observations}

Blue and red spectra were obtained simultaneously using the  double beam
Cassegrain Twin Spectrograph (TWIN) mounted on the 3.5m telescope of the Calar
Alto Observatory at the Complejo Astron\'omico Hispano Alem\'an (CAHA),
Spain. They were acquired in June 2006, during a four night observing run and
under excellent seeing and photometric conditions. Site\#22b and Site\#20b,
2000\,$\times$\,800\,px 15\,$\mu$m, detectors were attached to the blue and
red arms of the spectrograph, respectively. The T12 grating was used in the
blue covering the wavelength range 3400-5700\,\AA\ (centered at
$\lambda_c$\,=\,4550\,\AA), giving a spectral dispersion of
1.09\,\AA\,pixel$^{-1}$ (R\,$\simeq$\,4170). On the red arm, the T11 grating
was mounted  providing a spectral range from 5800 to 10400\,\AA\
($\lambda_c$\,=\,8100\,\AA) and a spectral dispersion of
2.42\,\AA\,pixel$^{-1}$ (R\,$\simeq$\,3350).  
The pixel size for this set-up configuration is 0.56\,arcsec for both spectral
ranges.  The slit width was $\sim$1.2\,arcsec, which, combined with the
spectral dispersions, yielded spectral resolutions of about 3.2 and 7.0\,\AA\
FWHM in the blue and the red respectively. All observations were made at
paralactic angle to avoid effects of differential refraction in the UV. The
instrumental configuration, summarized in table \ref{config}, covers the whole
spectrum from 3400 to 10400\,\AA\ (with a gap between 5700 and 5800\,\AA)
providing at the same time a moderate spectral resolution. This guarantees the
simultaneous measurement  of the nebular lines from [O{\sc
    ii}]\,$\lambda\lambda$\,3727,29 to [S{\sc
    iii}]\,$\lambda\lambda$\,9069,9532\,\AA\  at both ends of the spectrum, in
the very same region of the galaxy. A good signal-to-noise ratio was also
required to allow the detection and measurement of weak lines such as  [O{\sc
    iii}]\,$\lambda$\,4363, [S{\sc ii}]\,$\lambda\lambda$\,4068, 6717 and
6731, and [S{\sc iii}]\,$\lambda$\,6312. The signal-to-noise ratios attained
for each final spectrum are given in Table \ref{snr}. 

\begin{figure*}
\includegraphics[width=.48\textwidth,height=.31\textwidth,angle=0]{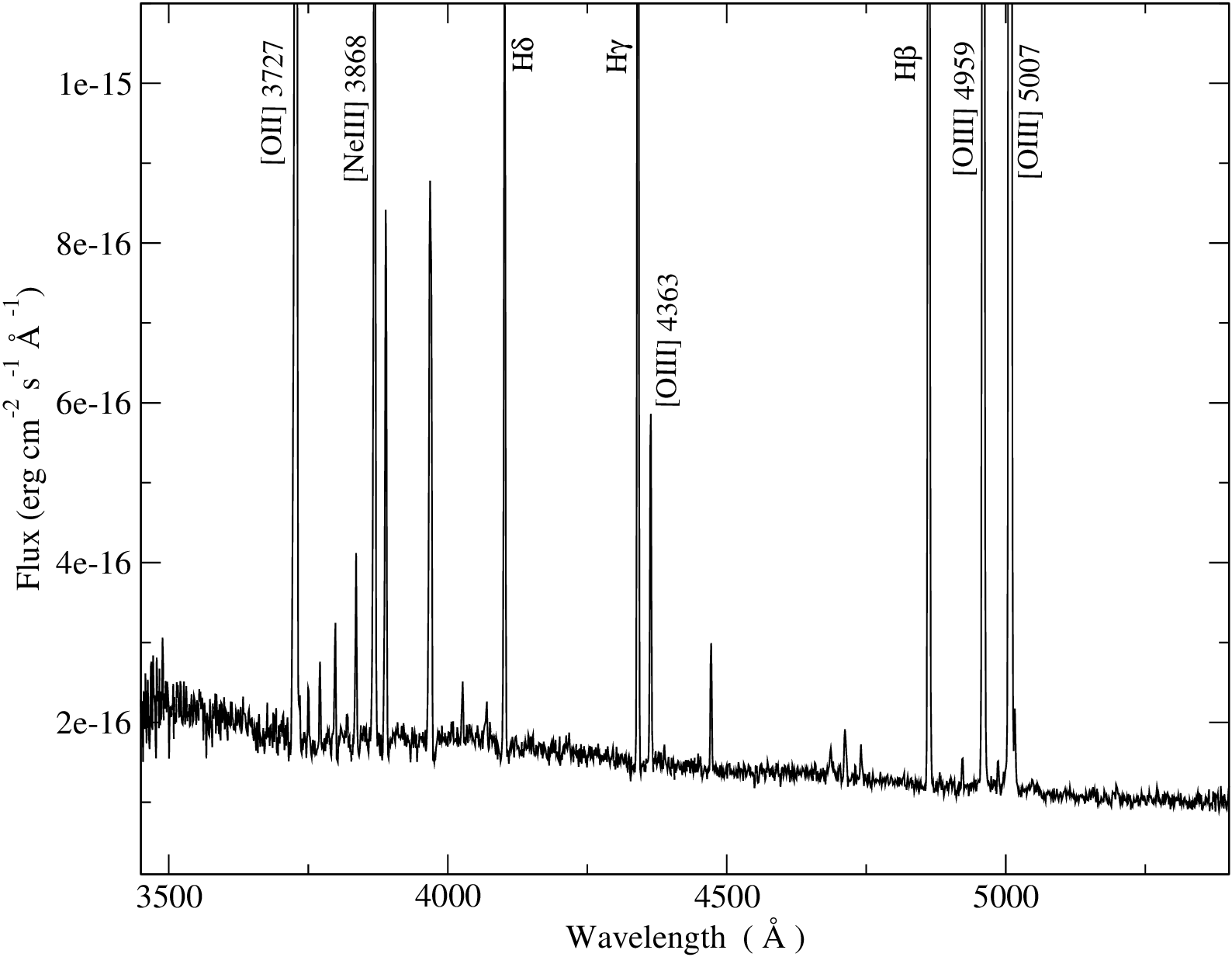}\hspace{0.5cm}
\includegraphics[width=.48\textwidth,height=.31\textwidth,angle=0]{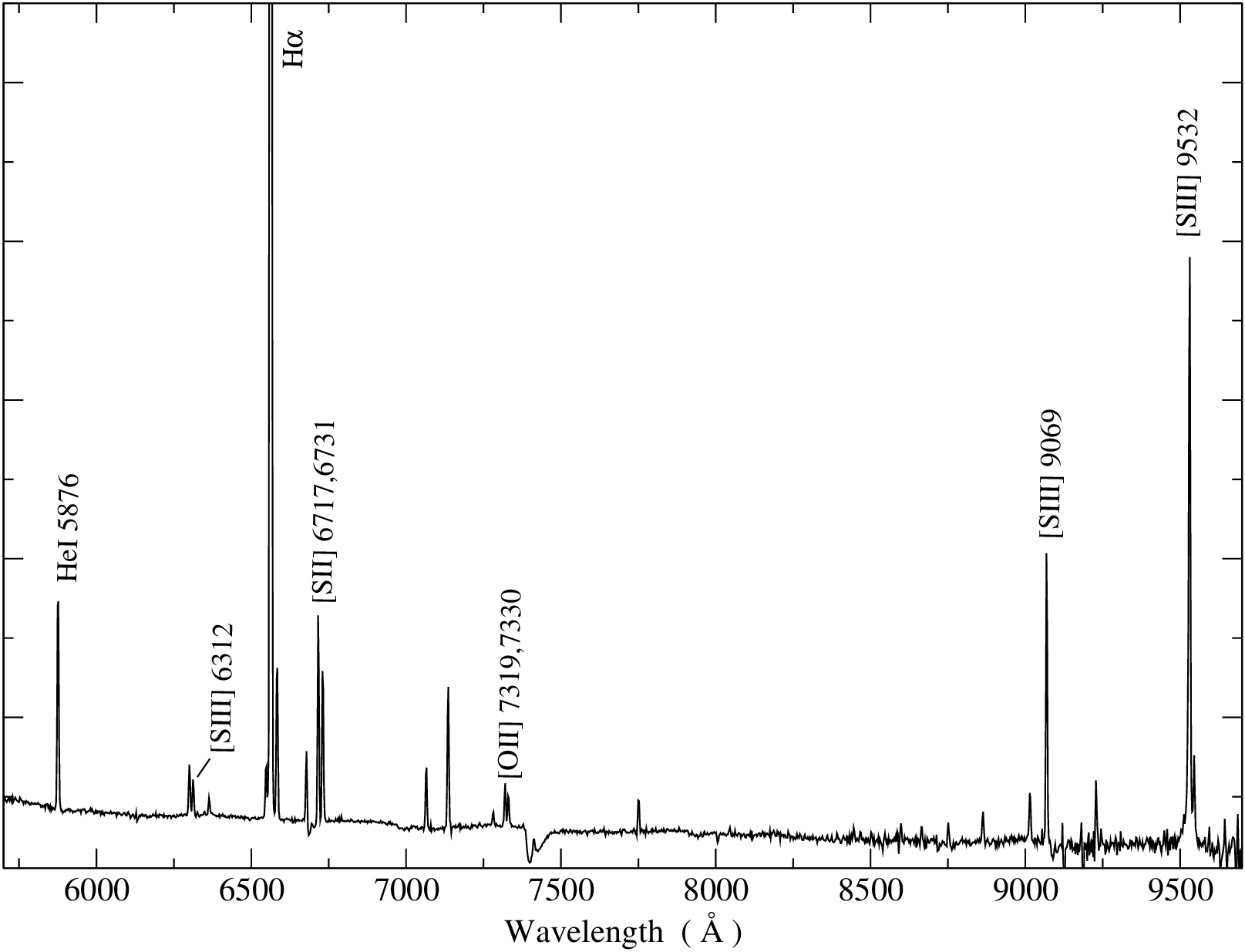}\\
\vspace{0.2cm}
\includegraphics[width=.48\textwidth,height=.31\textwidth,angle=0]{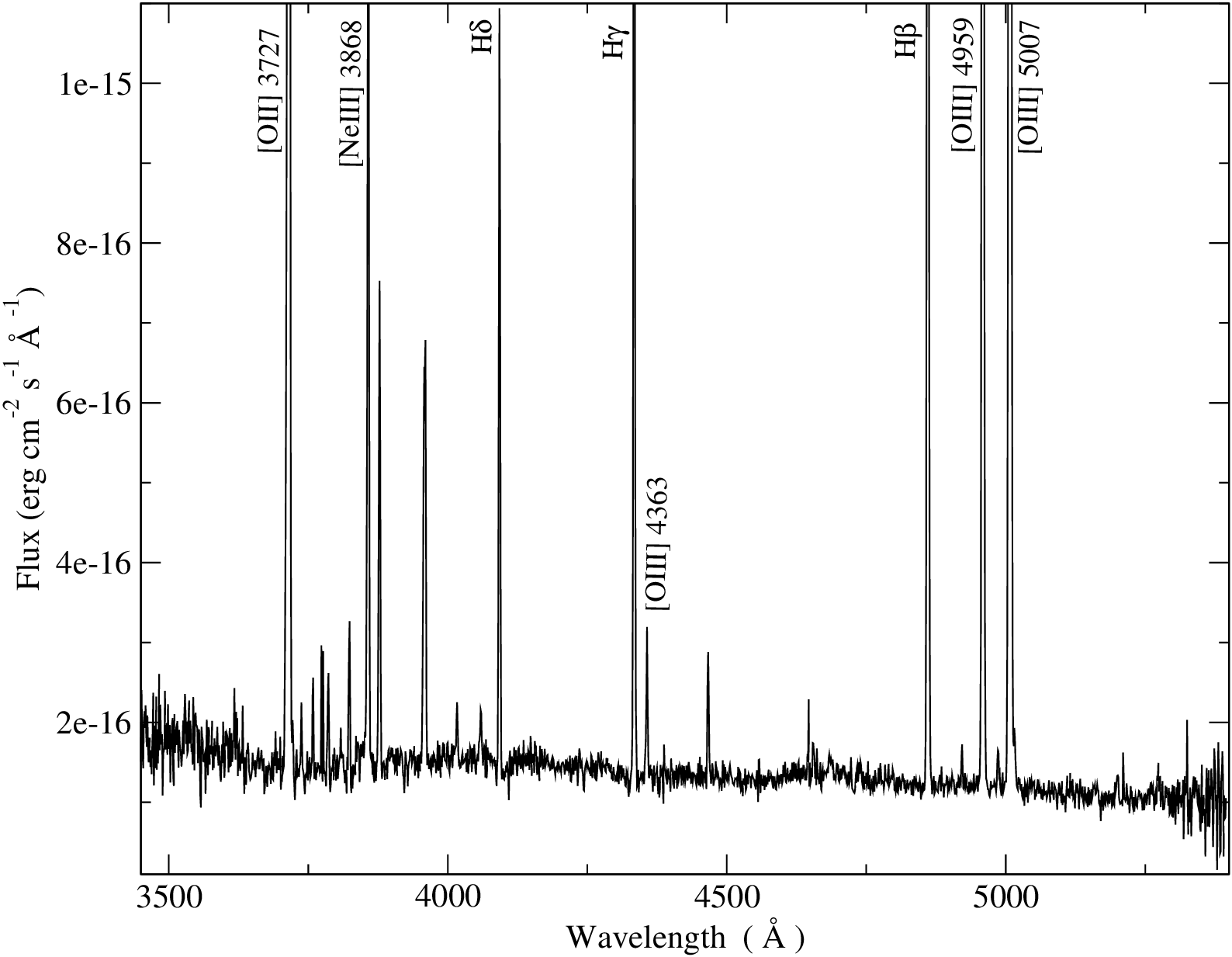}\hspace{0.5cm}
\includegraphics[width=.48\textwidth,height=.31\textwidth,angle=0]{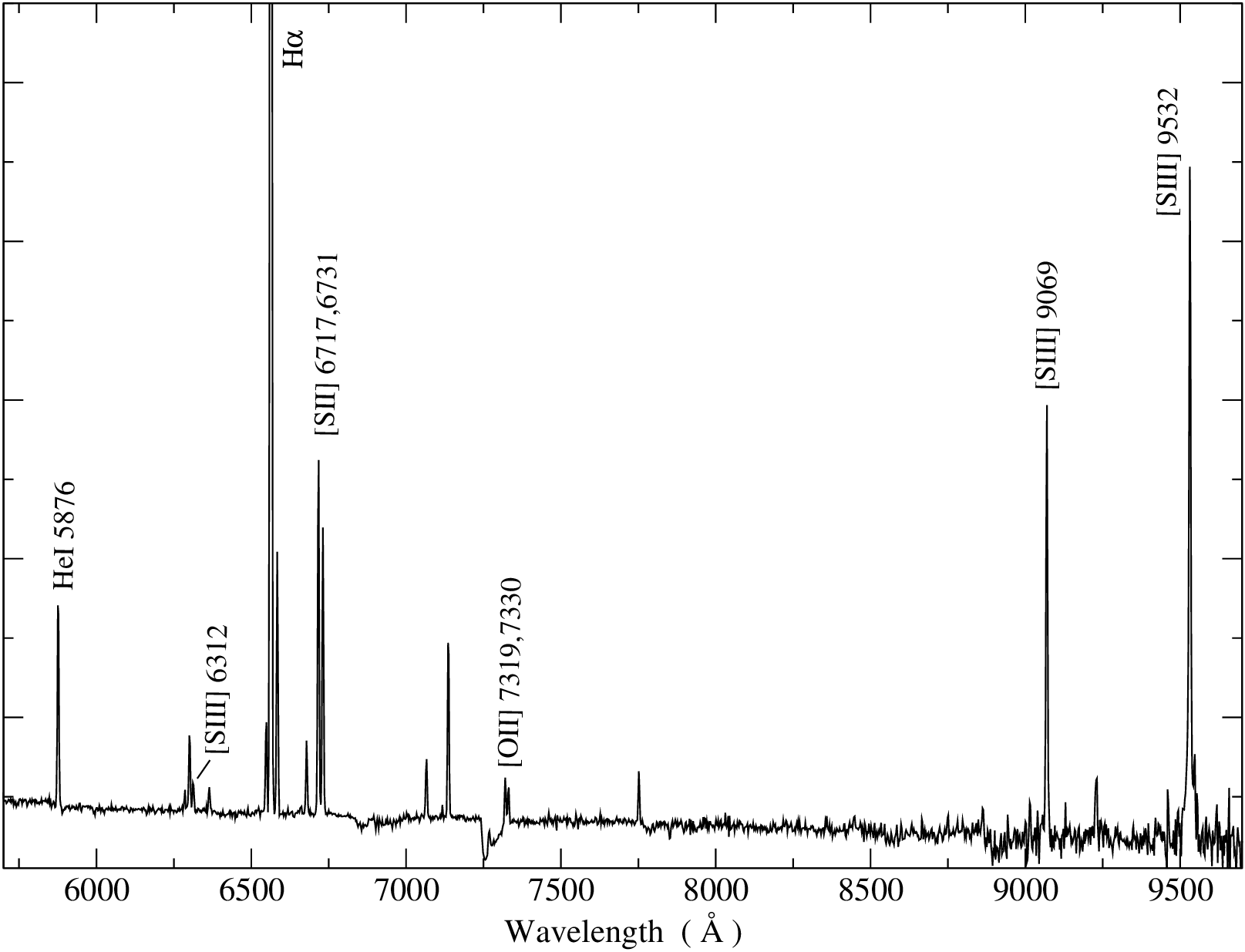}\\
\vspace{0.2cm}
\includegraphics[width=.48\textwidth,height=.31\textwidth,angle=0]{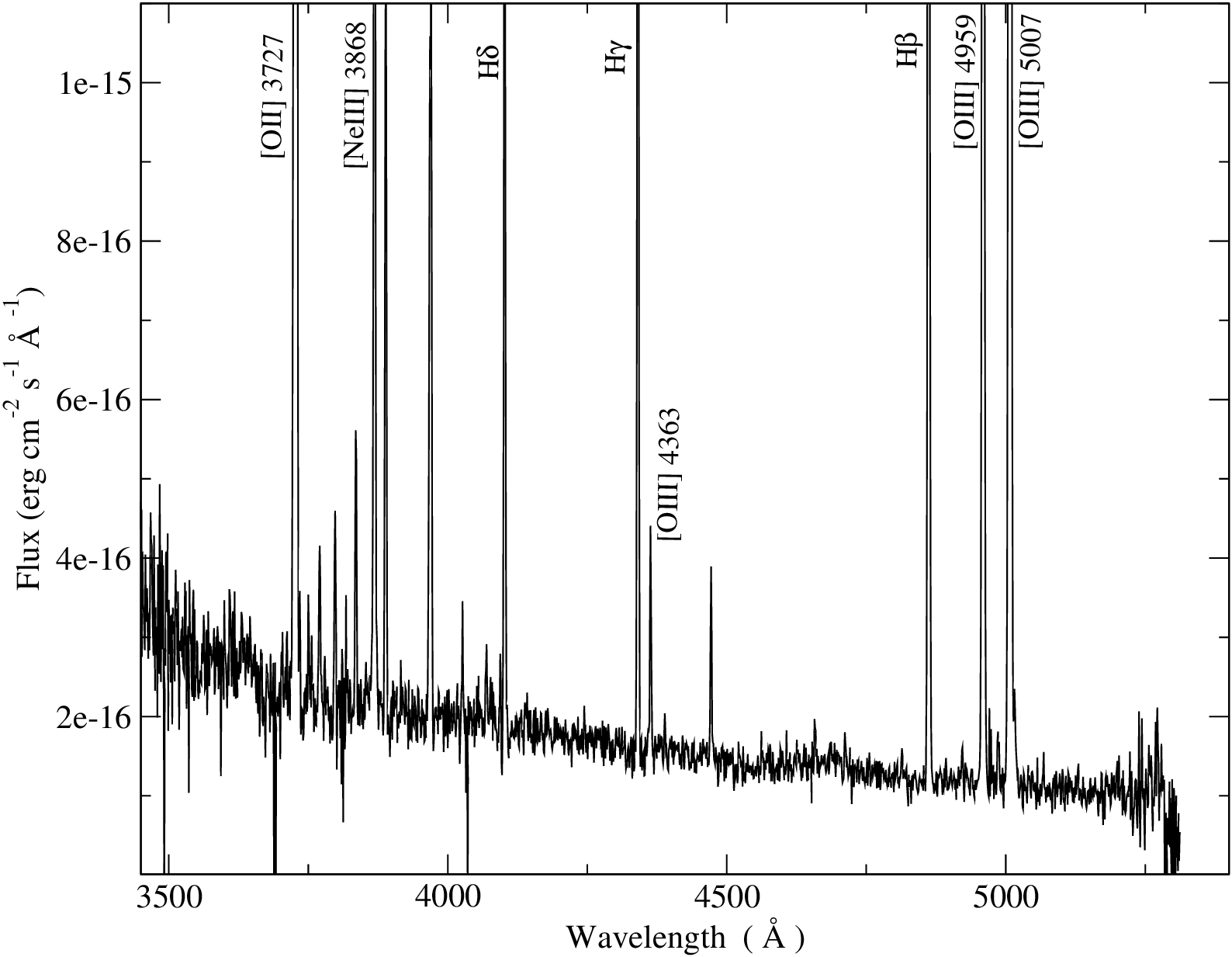}\hspace{0.5cm}
\includegraphics[width=.48\textwidth,height=.31\textwidth,angle=0]{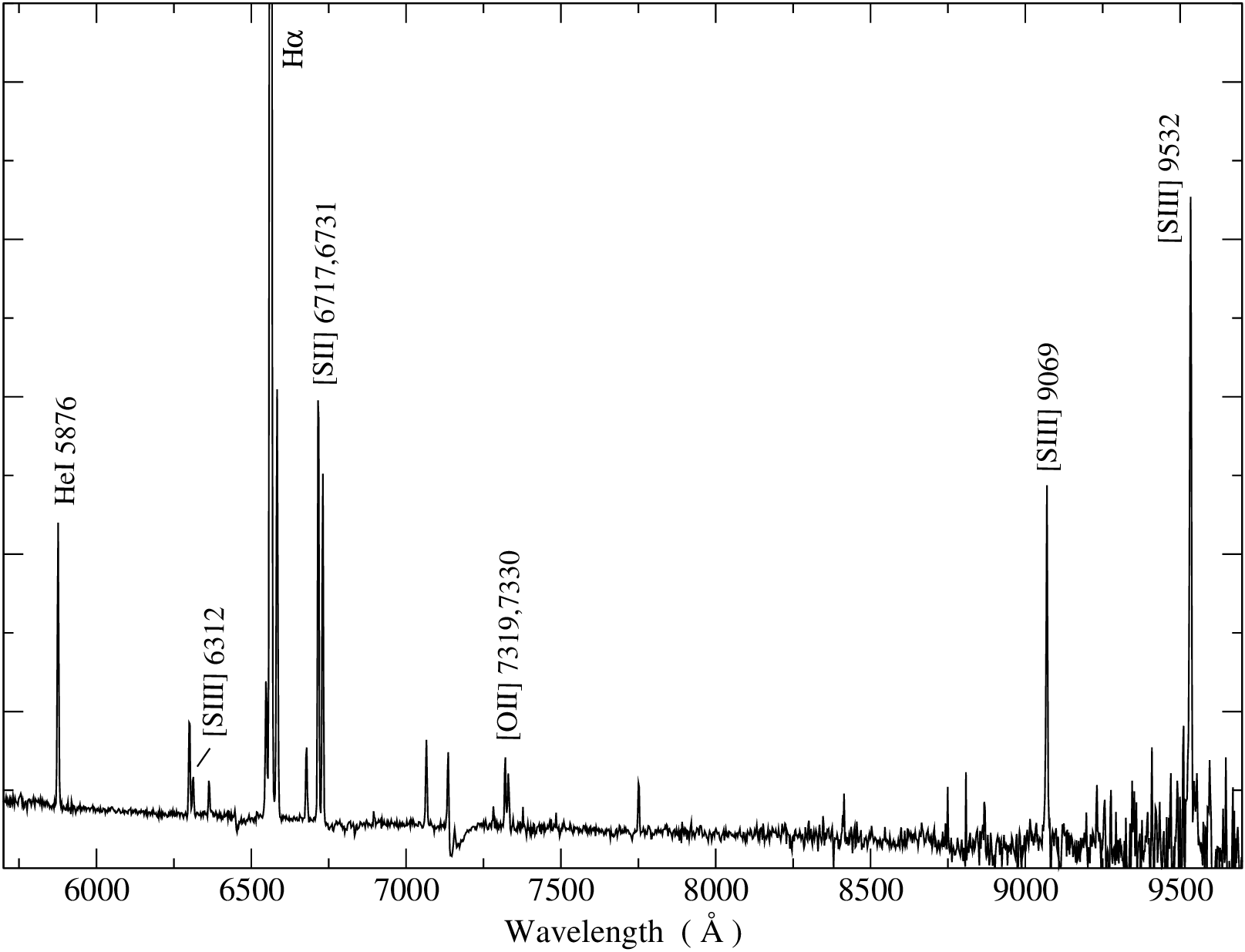}\\
\vspace{0.2cm}
\includegraphics[width=.48\textwidth,height=.31\textwidth,angle=0]{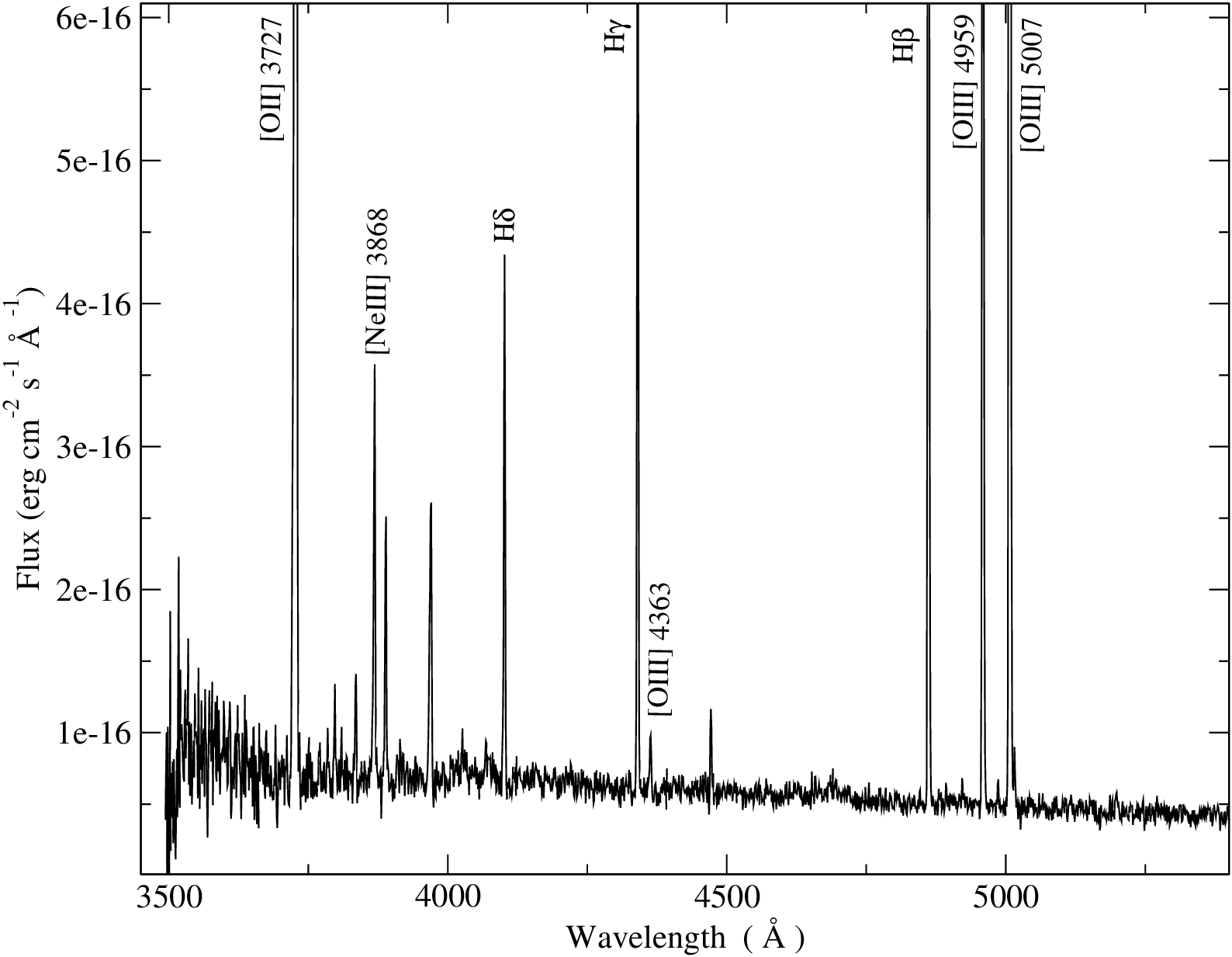}\hspace{0.5cm}
\includegraphics[width=.48\textwidth,height=.31\textwidth,angle=0]{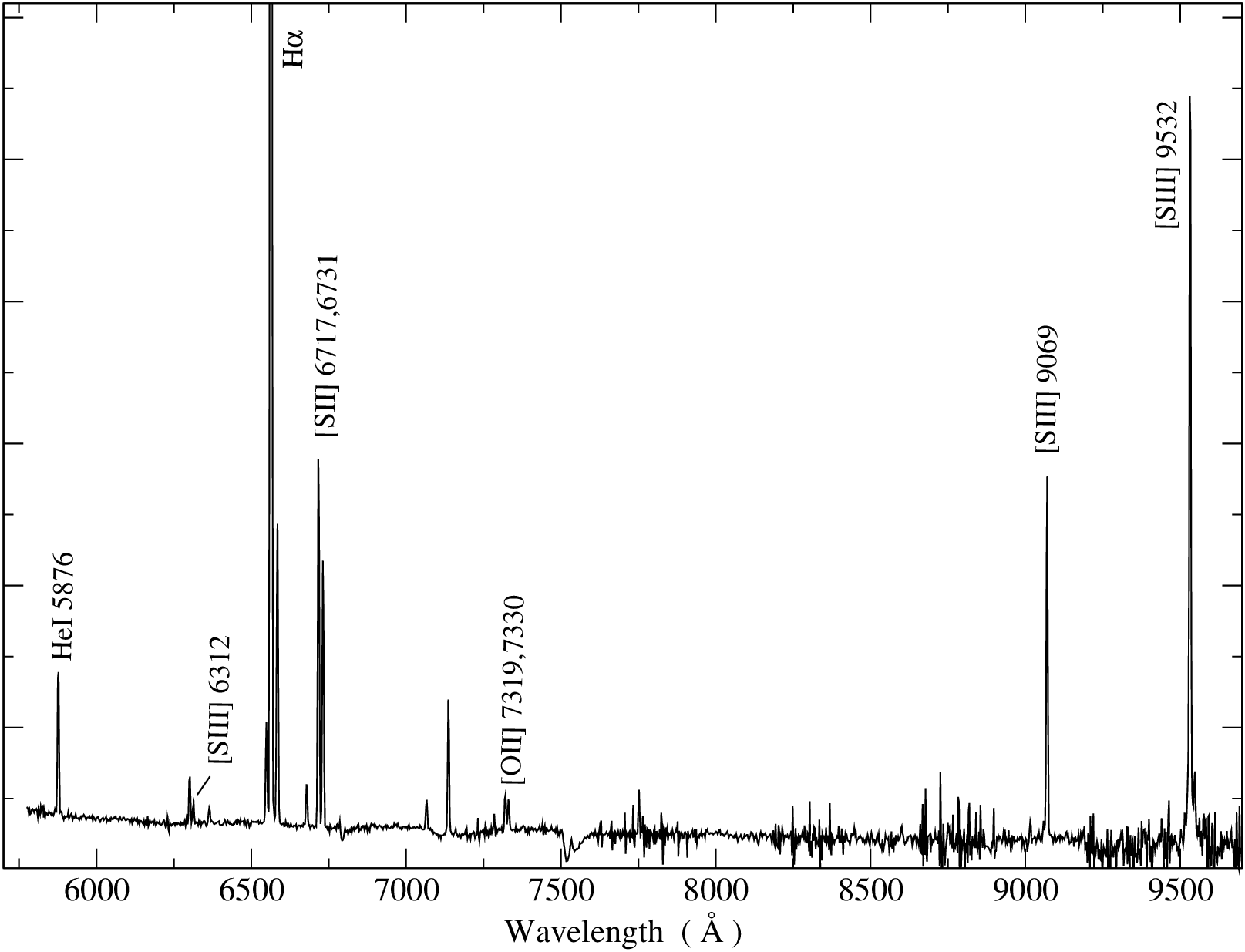}
\caption{ Blue and red CAHA spectra of \unoc, \dosc, \tresc\ and \cuatroc\
  in the rest frame. The flux scales are the same in both spectral ranges.}
\label{uno}
\end{figure*}

\setcounter{figure}{0}
\begin{figure*}
\includegraphics[width=.48\textwidth,height=.31\textwidth,angle=0]{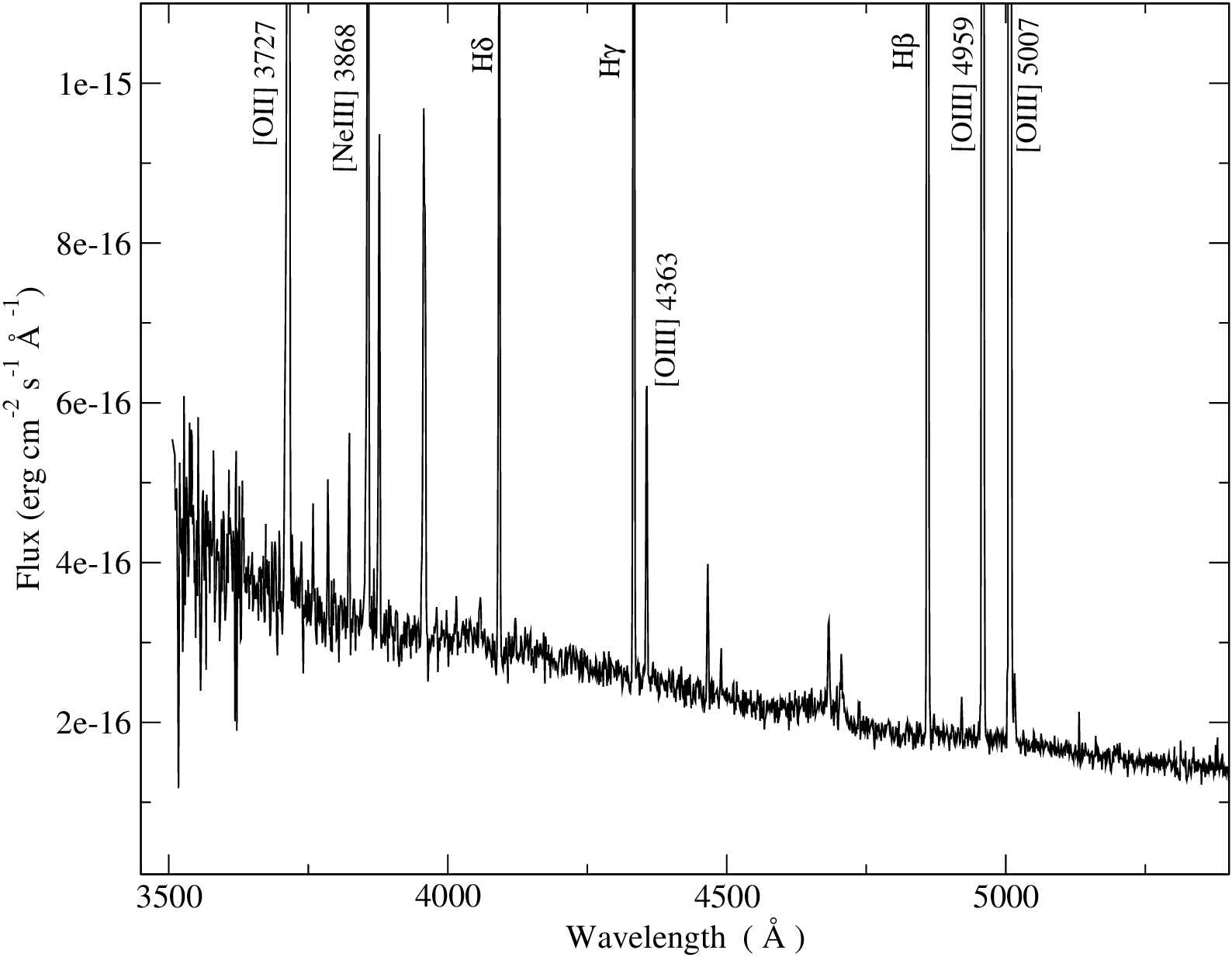}\hspace{0.5cm}
\includegraphics[width=.48\textwidth,height=.31\textwidth,angle=0]{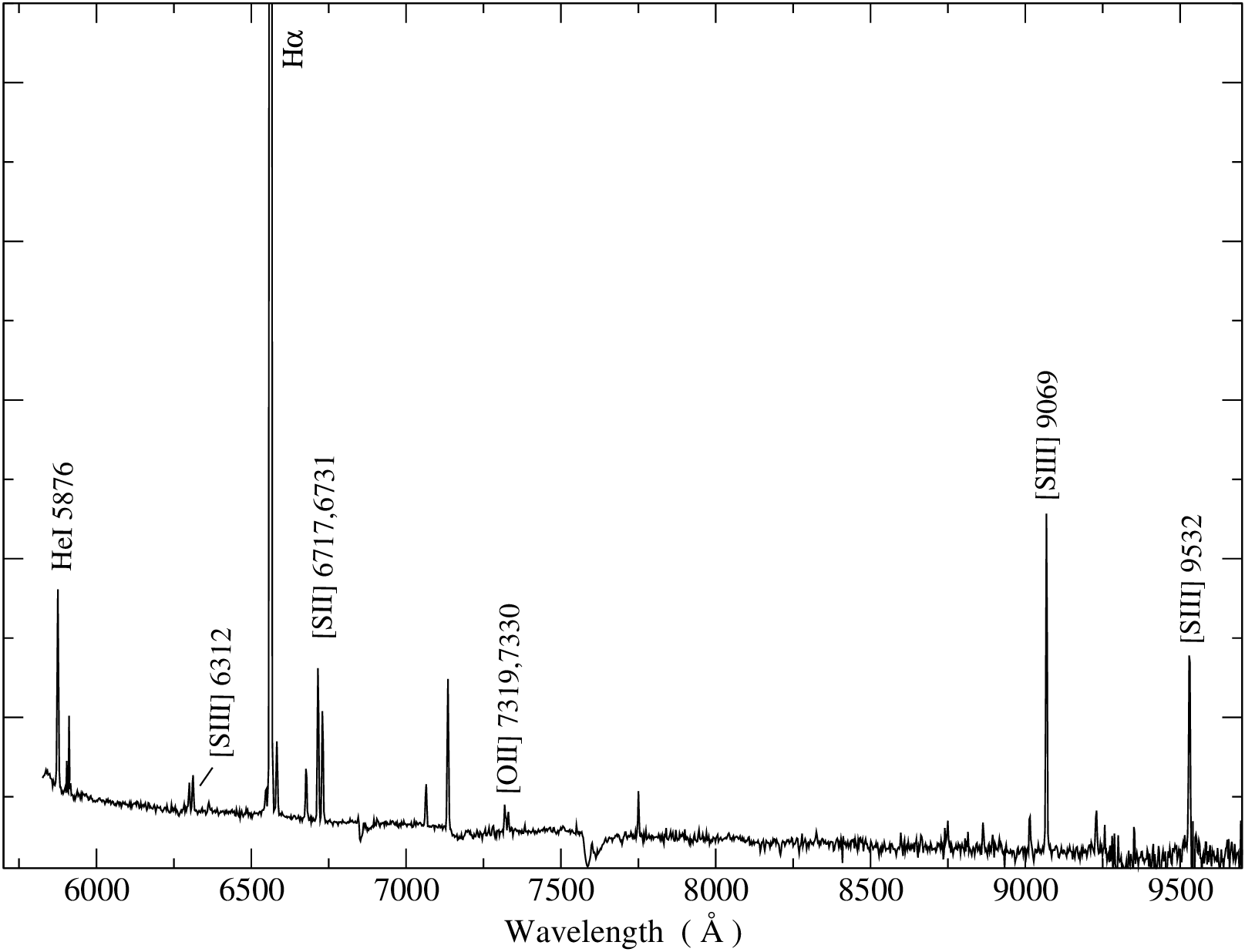}\\
\vspace{0.2cm}
\includegraphics[width=.48\textwidth,height=.31\textwidth,angle=0]{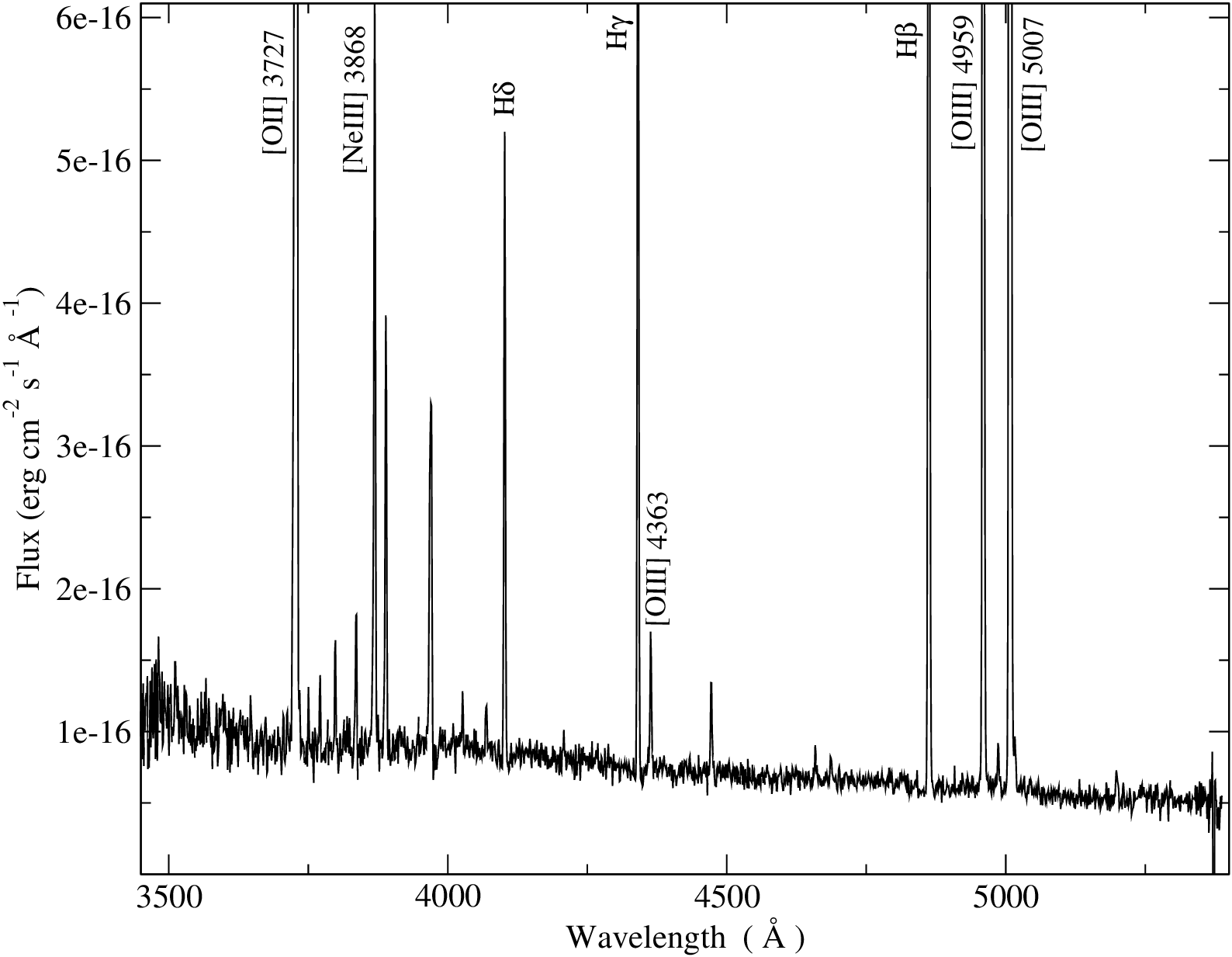}\hspace{0.5cm}
\includegraphics[width=.48\textwidth,height=.31\textwidth,angle=0]{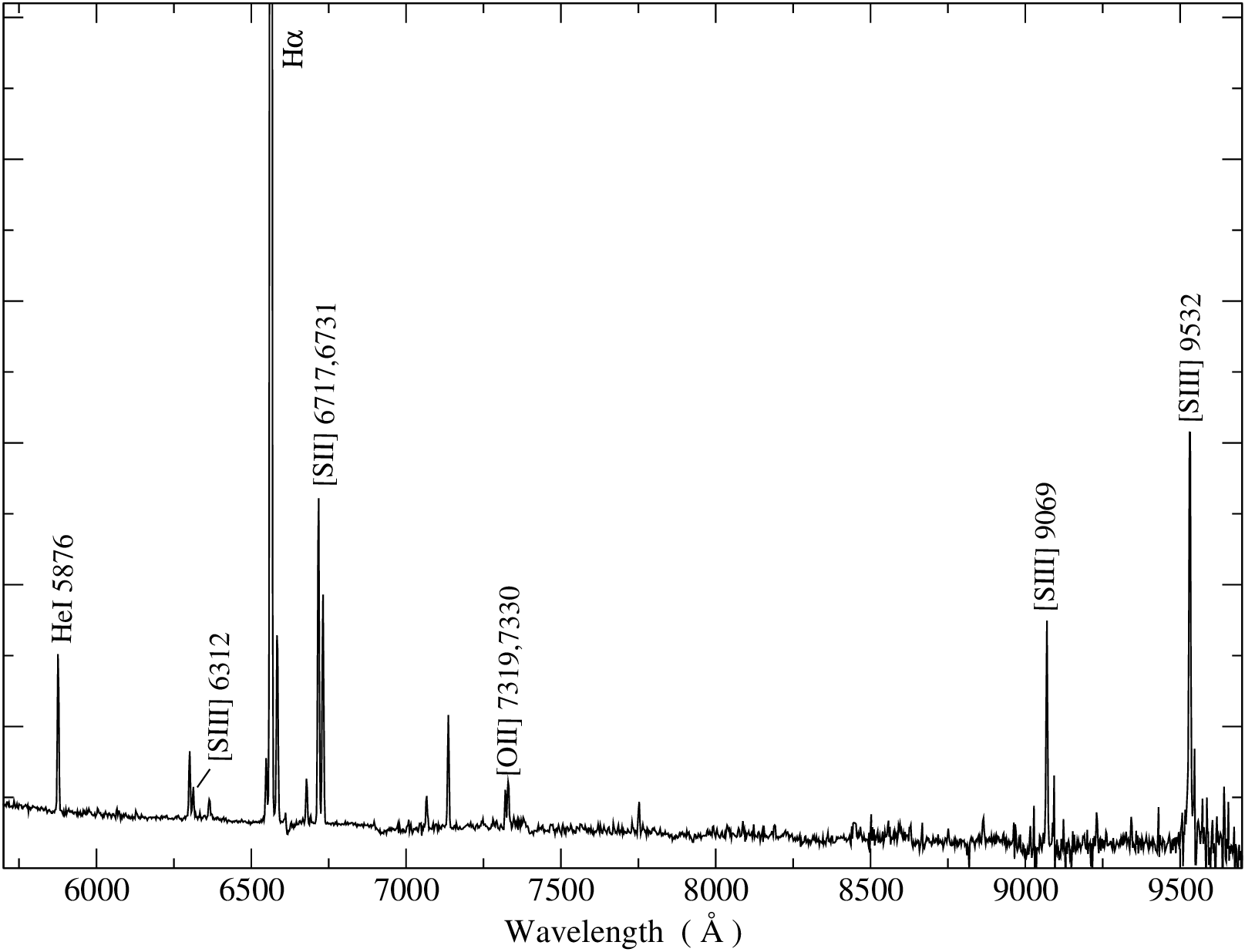}\\
\vspace{0.2cm}
\includegraphics[width=.48\textwidth,height=.31\textwidth,angle=0]{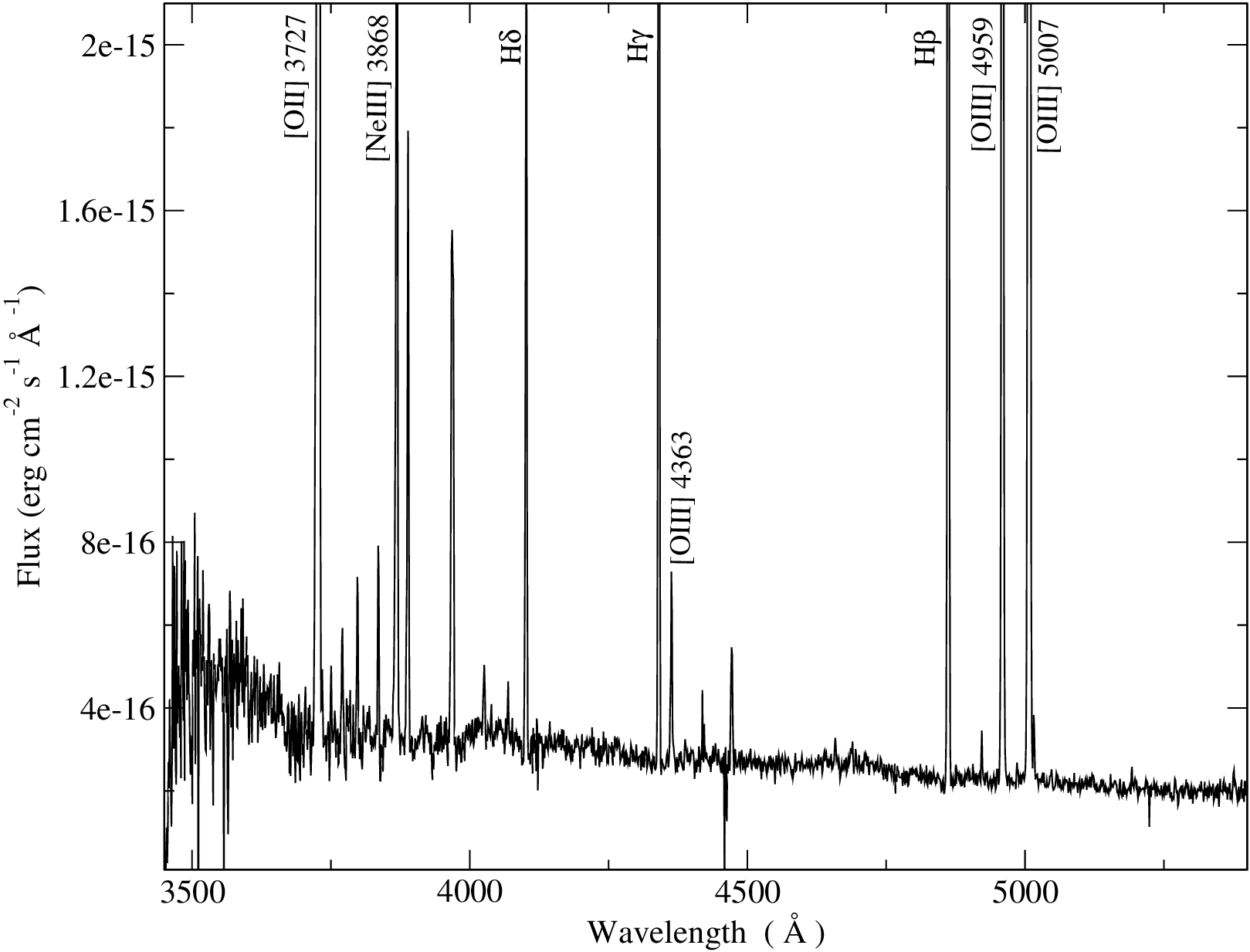}\hspace{0.5cm}
\includegraphics[width=.48\textwidth,height=.31\textwidth,angle=0]{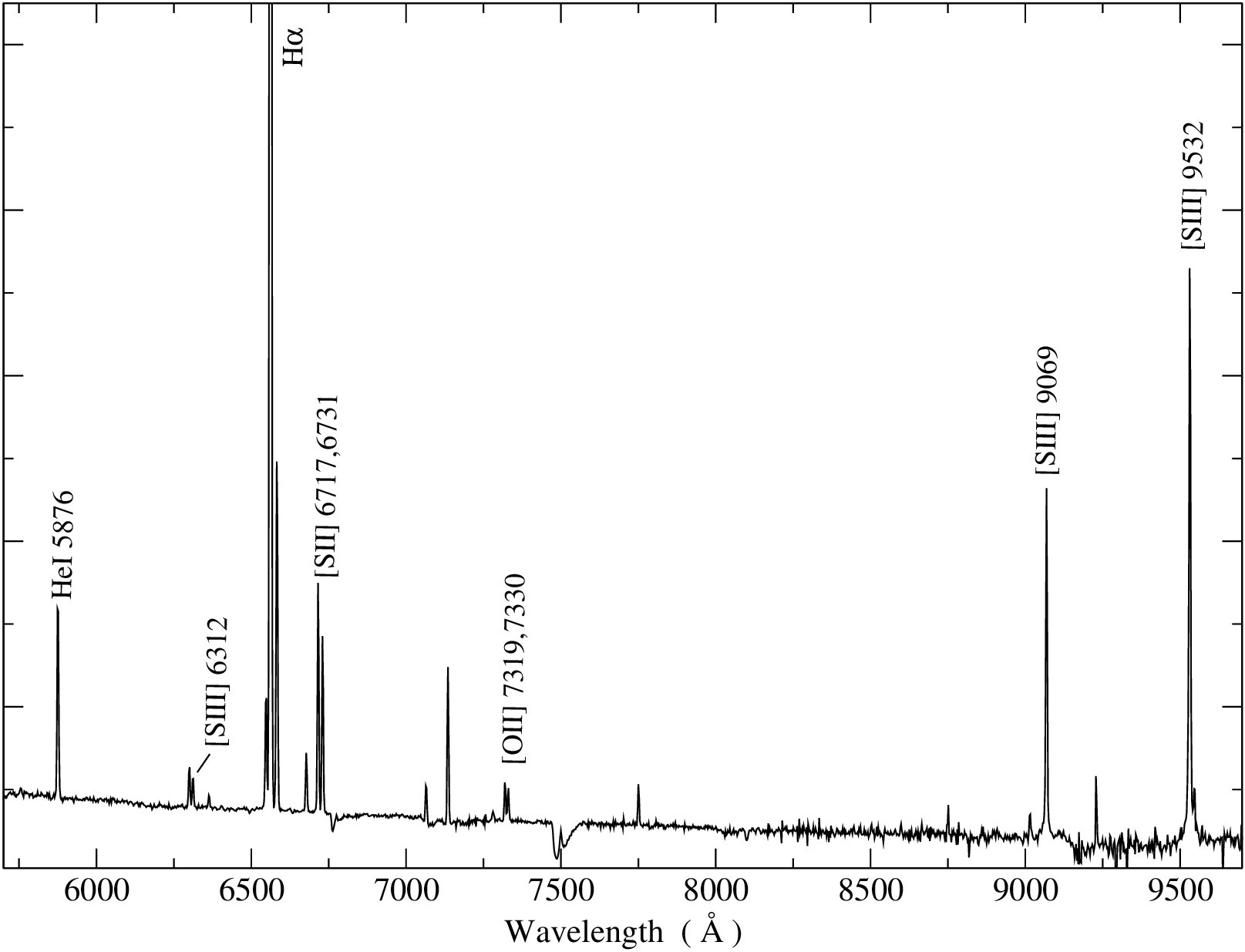}
\caption{({\it cont}) Blue and red CAHA spectra of \cincoc, \seisc\ and \sietec.
  in the rest frame. The flux scales are the same in both spectral ranges.}
\end{figure*}

%
%

\begin{table*}
\centering
\caption[]{Signal-to-noise ratio attained for each final spectrum.}
\label{snr}
\begin{tabular} {l cccccccc}
\hline
Object  ID & 5100-5200 & 6000-6100 & [S{\sc ii}]\,$\lambda$\,4068 & [O{\sc iii}]\,$\lambda$\,4363 & [N{\sc ii}]\,$\lambda$\,5755 & [S{\sc iii}]\,$\lambda$\,6312 & [O{\sc ii}]\,$\lambda$\,7319 & [O{\sc ii}]\,$\lambda$\,7330 \\
\hline
 \unoc    &  20 & 50 & 27 & 180 & --- & 157  & 160 & 126 \\
 \dosc    &  15 & 40 & 25 & 40  & --- & 95   & 99  & 82  \\
 \tresc   &  10 & 33 & 18 & 60  & --- & 125  & 91  & 73  \\
 \cuatroc &  10 & 35 & 15 & 24  & --- & 92   & 78  & 62  \\
 \cincoc  &  15 & 40 & 12 & 87  & --- & 61   & 82  & 53  \\
 \seisc   &  15 & 35 & 21 & 64  & --- & 107  & 67  & 47  \\
 \sietec  &  15 & 35 & 12 & 77  & 29  & 105  & 147 & 122 \\
\hline
\end{tabular}
\end{table*}

\subsection{Data reduction}

Several bias and sky flat field frames were taken at the beginning and end of
each night. In addition, two lamp flat fields and one calibration lamp
exposure were performed at each telescope position. The calibration lamp used
was HeAr. The images were processed and analyzed with IRAF\footnote{IRAF: the
Image Reduction and Analysis Facility is distributed by the National Optical
Astronomy Observatories, which is operated by the Association of Universities
for Research in Astronomy, Inc. (AURA) under cooperative agreement with the
National Science Foundation (NSF).} routines in  the usual manner. The
procedure includes the removal of cosmic rays, bias substraction, division by
a normalized flat field and wavelength calibration.  

Finally, the spectra were corrected for atmospheric extinction and flux
calibrated. Four standard star observations were performed each night,
allowing a good spectrophotometric calibration with an estimated accuracy of
about 3\%, estimated from the differences between the different standard star
flux calibration curves.

\section{Results}
\label{measure}

The spectra of the observed \HII\ galaxies with some of the relevant
identified emission lines are shown in Fig. \ref{uno}. The spectrum of each
observed galaxy is split into two panels, with the blue part on the left and
the red part on the right.   

The  emission line fluxes were measured using the SPLOT task in IRAF following
the procedure described in H\"agele et al.\ (2006; hereafter Paper
I).\nocite{2006MNRAS.372..293H} Following \cite{2003MNRAS.346..105P}, the
statistical errors associated with the observed emission fluxes have been
calculated using the expression  \[ \sigma_{l}\,=\,\sigma_{c}N^{1/2}[1 +
  EW/(N\Delta)]^{1/2} \] \noindent where $\sigma_{l}$ is the error in the
observed line flux, $\sigma_{c}$ represents the standard deviation in a box
near the measured emission line and stands for the error in the continuum
placement, N is the number of pixels used in the measurement of the line flux,
EW is the line equivalent width, and $\Delta$ is the wavelength dispersion in
\AA\ per pixel \citep{1994ApJ...437..239G}. There are several emission lines
affected by cosmetic faults or charge transfer in the CCD, internal
reflections in the spectrograph, telluric emission lines or atmospheric
absorption lines. These lines were excluded from any subsequent analysis. 
In the case of \tresc, the [Ar{\sc iii}]\,$\lambda$\,7136 is affected by a
sky absorption band. Its observed flux has been scaled to that of the [Ar{\sc
    iii}]\,7751\AA\ line according to the theoretical relation, [Ar{\sc
    iii}]\,7136/[Ar{\sc iii}]\,7751\,=\,4.17, derived from the IONIC
task in the STSDAS package of IRAF for a wide temperature range, from 5000 to
50000\,K. For \cincoc\ the [S{\sc iii}]\,$\lambda$\,9532 line is affected by
strong narrow water-vapour lines and therefore its value has been set to its
theoretical ratio to the weaker [S{\sc iii}]\,$\lambda$\,9069\,\AA\ line,
taken to be 2.44.


An underlying stellar population is easily appreciable by the presence of
absorption features that depress the Balmer and Paschen emission lines. A
pseudo-continuum has been defined at the base of the hydrogen emission lines
to measure the line intensities and minimise the errors introduced by the
underlying population (see Paper I). We can clearly see the wings of the
absorption lines implying that, even though we have used a pseudo-continuum,
there is still an absorbed fraction of the emitted flux that we are not able
to measure with an acceptable accuracy \citep[see discussion
in][]{1988MNRAS.231...57D}. This fraction is not the same for all lines, nor
are the ratios between the absorbed fractions and the emission. In Paper I we
estimated that the differences between the measurements obtained using the
defined the pseudo-continuum and those made using a multi-Gaussian fit to the
absorption and emission components, when this fitting is possible, are, for
all the Balmer lines, within the observational errors. 
This is also expected to be the case for the objects presented here given our
selection criterion of large H$\alpha$ equivalent width. At any rate,  for the
Balmer and Paschen emission lines we have doubled the derived error,
$\sigma_{l}$, as a conservative approach to include the uncertainties
introduced by the presence of the underlying stellar population.  

The absorption features of the underlying stellar population will also affect
the helium emission lines to some extent. However, the wings of these
absorption lines are narrower than those of hydrogen \cite[see, for example,
][]{2005MNRAS.357..945G}. Therefore it is difficult to set adequate
pseudo-continua at both sides of the lines to measure their fluxes.

The reddening coefficient [$c$(H$\beta$)] has been calculated assuming the
galactic extinction law of \cite{1972ApJ...172..593M} with $R_v$=3.2 and
obtained  by performing a least square fit to the difference between the
theoretical and observed Balmer and Paschen decrements vs.\ the reddening law
whose slope is the logarithmic reddening at the H$\beta$ wavelength: 
\[
log\Big(\frac{I(\lambda)}{I(H\beta)}\Big)\,=\,log\Big(\frac{F(\lambda)}{F(H\beta)}\Big)+c(H\beta)\,f(\lambda)
\]
The theoretical Balmer line intensities have been computed using
\cite{1995MNRAS.272...41S} with an iterative method to estimate $t_e$ and
$N_e$ in each case. As $N_e$ introduces only a second order effect, for
simplicity we assume $N_e$ equal to N([S{\sc ii}]). Due to the large error
introduced by the presence of the underlying stellar population, only the four
strongest Balmer emission lines (H$\alpha$, H$\beta$, H$\gamma$ and H$\delta$)
have been used. 

Table \ref{ratiostot 1}  gives the equivalent widths and the reddening
corrected emission lines for each observed galaxy together with the reddening
constant and its error taken as the uncertainties of the least square fit and
the reddening corrected H$\beta$ intensity. The adopted reddening curve,
$f(\lambda)$, normalized to H$\beta$, is given in column 2 of the table. The
errors in the emission lines were obtained by propagating in quadrature the
observational errors in the emission line fluxes and the reddening constant
uncertainties.


The relative errors in the emission lines vary from a few percent for the more
intense nebular emission lines (e.g. [O{\sc
    iii}]\,$\lambda\lambda$\,4959,5007, [S{\sc
    ii}]\,$\lambda\lambda$\,6717,6731 or the strongest Balmer emission lines)
to 10-35\,\% for the weakest lines that have less contrast with the continuum
noise (e.g. He{\sc i}\,$\lambda\lambda$\,3820,7281, [Ar{\sc
    iv}]\,$\lambda$\,4740 or O{\sc i}\,$\lambda$\,8446). For the auroral
lines, the fractional errors are between $\sim$3 and $\sim$10\,\%.

 \onecolumn
 \landscape

\input{tables/ratios-pap.tex}    

 \endlandscape
 \twocolumn

\section{Electron densities and temperatures from forbidden lines}
\label{physi}

The physical conditions of the ionised gas, including electron temperatures
and electron density, have been derived using the five-level statistical
equilibrium atom approximation in the task TEMDEN of the STSDAS package of the
software IRAF \citep*{1987JRASC..81..195D, 1995PASP..107..896S}. The
atomic coefficients used with their corresponding references are given in
table \ref{atomic}.

The electron density, $N_e$, has been derived from the [S{\sc
    ii}]\,$\lambda\lambda$\,6717\,/\,6731\,\AA\ line ratio. In all the
observed galaxies  the electron densities have been found to be lower than 200
cm$^{-3}$, well below the critical density for collisional deexcitation. We
were not able to estimate the density from line ratios such as [Ar{\sc
    iv}]\,$\lambda\lambda$\,4713,4740\,\AA, representative of the higher
ionisation zones, hence we are not able to determine any existing distribution
in density.


%
%

\begin{table}
\begin{minipage}{85mm}
\vspace{-0.3cm}
\normalsize
\caption{Sources of the effective collision strengths of each ion.}
\begin{center}
\begin{tabular}{@{}ll}
\hline
\hline
  Ion & references  \\ 
\hline
 O{\sc ii}   &  \citet{1976MNRAS.177...31P} \\ 

 O{\sc iii}, N{\sc ii}   &  \citet{1994AAS..103..273L}  \\ 

 S{\sc ii}   &  \citet*{1996ADNDT..63...57R}  \\ 

 S{\sc iii}  &  \citet{1999ApJ...526..544T}  \\ 

 Ne{\sc iii} & \citet{1994AAS..108....1B}  \\ 

 Ar{\sc iii} & \citet*{1995A+AS..111..347G}  \\ 

 Ar{\sc iv}  & \citet*{1987A+A...188..251Z}  \\ 
\hline
\end{tabular}
\end{center}
\label{atomic}
\end{minipage}
\end{table}

%
%

\begin{table}
\begin{minipage}{85mm}
\vspace{-0.3cm}
\normalsize
\caption{Emission-line ratios used to derive electron densities and temperatures.}
\begin{center}
\begin{tabular}{@{}ll}
\hline
\hline
  & ratios  \\
\hline
N$_e$([S{\sc ii}])     &    R$_{S2}$\,=\,I(6717)\,/\,I(6731) \\
t$_e$([O{\sc iii}])    &    R$_{O3}$\,=\,(I(4959)+I(5007))\,/\,I(4363) \\
t$_e$([O{\sc ii}])     &    R$_{O2}$\,=\,I(3727)\,/\,(I(7319)+I(7330)) \\
t$_e$([S{\sc iii}])    &    R$_{S3}$\,=\,(I(9069)+I(9532))\,/\,I(6312) \\
t$_e$([S{\sc ii}])     &    R$_{S2}'$\,=\,(I(6717)+I(6731))\,/\,(I(4068)+I(4074)) \\
t$_e$([N{\sc ii}])     &    R$_{N2}$\,=\,(I(6548)+I(6584))\,/\,I(5755) \\
\hline
\end{tabular}
\end{center}
\label{ratios}
\end{minipage}
\end{table}

For all the objects we have derived the electron temperatures of [O{\sc ii}],
[O{\sc iii}], [S{\sc ii}] and [S{\sc iii}]. Only for one object, \sietec, it
was possible to derive T$_e$([N{\sc ii}]).  The emission-line ratios used to
calculate each temperature are summarized in table \ref{ratios}. Adequate
fitting functions have been derived from the TEMDEM task and are given below: 
\begin{eqnarray*}
t_e([O{\textrm{\sc iii}}])\, =\, 0.8254\, - \,0.0002415\,R_{O3} \,+\, \frac{47.77}{R_{O3}}\\
t_e([S{\textrm{\sc iii}}])\,=\,\frac{R_{S3}\,+\,36.4}{1.8\,R_{S3}\,-3.01}\\
t_e([O{\textrm{\sc ii}}]) \,=\,0.23 +0.0017\,R_{O2}\,+\,\frac{38.3}{R_{O2}}\,+\,f_1(n_e)\\
t_e([S{\textrm{\sc ii}}] \,=\,1.92\,-\,0.0375\,R_{S2}'\,-\,\frac{14.5}{R_{S2}'}\,+\,\frac{105.64}{R_{S2}'^2}\,+\,f_2(n_e)\\
t_e([N{\textrm{\sc ii}}])\,=\,0.537\,+0.000253\,R_{N2}\,+\,\frac{42.13}{R_{N2}}
\end{eqnarray*}
\noindent where $n_e$\,=\,10$^{-4}\,N_e$ and $f_1(n_e)$,$f_2(n_e)$\,$<<$\,1
for $N_e$\,$<$\,1000\,cm$^{-\rm 3}$. The above expressions are valid in the
temperature range between 7000 and 23000\,K and the errors involved in the
fittings are always lower than observational errors by factors between 5 and
10.

In order to calculate the errors associated with the derived electron
temperatures and densities, we have propagated the emission line intensity
errors listed in Table \ref{ratiostot 1} through our calculations.  

Both the [O{\sc ii}]\,$\lambda\lambda$\,7319,7330\,\AA\ and the [N{\sc
    ii}]\,$\lambda$\,5755\,\AA\  lines have a contribution by direct
recombination which increases with temperature. Such emission, however, can be
quantified and corrected for as: 
\begin{eqnarray*}
\frac{I_R(7319+7330)}{I(H\beta)}\,=\,9.36\,t^{0.44}\,\frac{O^{2+}}{H^+} \\
\frac{I_R(5755)}{I(H\beta)}\,=\,3.19\,t^{0.30}\,\frac{N^{2+}}{H^+} 
\end{eqnarray*}
\noindent  where $t$ denotes the electron temperature in units of 10$^4$\,K
\citep{2000MNRAS.312..585L}. Using the calculated [O{\sc iii}] electron
temperatures, we have estimated these contributions to be less than 4\% in all
cases and therefore we have not corrected for this effect, but we have
including it as an additional source of error. In the worst cases this amounts
to about 10 \% of the total error. The expressions above, however, are only
valid in the range of temperatures between 5000 and 10000\,K in the case of
[O{\sc ii}] and between 5000 and 20000 K in the case of [N{\sc ii}]. While the
[O{\sc iii}] temperatures found in our objects are inside the range of
validity for [N{\sc ii}], they are slightly over that range for [O{\sc
    ii}]. At any rate, the relative contribution of recombination to
collisional intensities decreases rapidly with increasing temperature.

%
%


\input{tables/temden-pap.tex}    



The derived electron densities and temperatures for the seven observed objects
are given in Table \ref{temden} along with their corresponding errors.

\section{Chemical abundances}
\label{abund}

We have derived the ionic chemical abundances of the different species using
the strongest available emission lines detected in the analysed spectra and
the task IONIC of the STSDAS package in IRAF. This package is also based on
the five-level statistical equilibrium atom approximation (De Robertis, Dufour
\& Hunt, 1987; Shaw \& Dufour, 1995).  

The total abundances have been derived by taking into account, when required,
the unseen ionization stages of each element, using the appropriate ICF for
each species: 
\[\frac{X}{H} = ICF(X^{+i}) \cdot \frac{X^{+i}}{H^+}\]

\subsection{Ionic abundances}

\subsubsection{Helium}

We have used the well detected and measured He{\sc i}\,$\lambda\lambda$\,4471,
5876, 6678 and 7065\,\AA\ lines,  to calculate the abundances of once ionized
helium. For three of the objects also the He{\sc ii}\,$\lambda$\,4686\,\AA\
line was measured allowing the calculation of twice ionized He. The He lines
arise mainly from pure recombination, although they could have some
contribution from collisional excitation and be affected by self-absorption
\citep[see][for a complete treatment of these effects]{2001NewA....6..119O,
2004ApJ...617...29O}. We have taken the electron temperature of [O{\sc iii}]
as representative of the zone where the He emission arises since at any rate
ratios of recombination lines are weakly sensitive to electron temperature. We
have used the equations given by Olive \& Skillman  to derive the
He$^{+}$/H$^{+}$ value, using the theoretical emissivities scaled to H$\beta$
from \cite{1999ApJ...514..307B} and the expressions for the collisional
correction factors from \cite{1995ApJ...442..714K}. We have not made, however,
any corrections for fluorescence (three of the used helium lines have a small
dependence with optical depth effects but the observed  objects have low
densities) nor for the presence of an underlying stellar population. To
calculate the abundance of twice ionized helium we have used equation (9) from
\cite{1983ApJ...273...81K}. The results obtained for each line and their
corresponding errors are presented in table \ref{absHe}, along with the
adopted value for He$^{+}$/H$^{+}$ that is the average, weighted by the
errors, of the different ionic abundances derived from each He{\sc i} emission
line. This value is dubbed ``adopted B99" in the Table.

We have also calculated the average values of He$^+$/H$^+$ from the HeI lines
$\lambda\lambda$4471, 5876, 6678, 7065\,\AA\footnote{Although measured, the
HeI line at $\lambda$3889\,\AA\ is a blend with H8 so we decided not to
include it for this work.} and the corresponding errors, obtained using Olive
and Skillman (2004) minimization technique with our derived values of
n$_e$([SII]) and T$_e$([OIII]) and \cite{2005ApJ...622L..73P} He
emissivities. We solved simultaneously for underlying stellar absorptions and
optical depth. The values are shown in Table \ref{absHe} under P05. We found no
significant differences for the objects in this paper in the Helium abundances
obtained using the two different sets of He emissivities. On the other hand
following this method is crucial when trying to determine He abundances to
better than 2 percent (like e.g.~for determining a value of the primordial
He). We chose to wait for a complete error budget determination from the
atomic physics parameters (Porter in preparation, private communication)
before adopting the latter values for the He abundances.

{\small
\begin{table*}
\vspace{-0.3cm}
\caption{Ionic and total chemical abundances for helium.} 
\label{absHe}

\begin{center}
\begin{tabular}{@{}cc ccccccc}
\hline
  & $\lambda$\,(\AA)   & \unoc & \dosc & \tresc & \cuatroc & \cincoc & \seisc & \sietec  \\
\hline
\input{tables/abundHe-pap.tex}
\hline
\end{tabular}
\end{center}
\end{table*}
}


\begin{table*}
{\small
\caption{Ionic chemical abundances derived from forbidden emission lines.}
\label{ion-abs}

\begin{center}
\begin{tabular}{@{}cccccccc}
\hline
  & \unoc & \dosc & \tresc & \cuatroc & \cincoc & \seisc & \sietec  \\
\hline
\input{tables/ionic-abund-pap.tex}

\hline
\end{tabular}
\end{center}}
\end{table*}


\subsubsection{Forbidden lines}

We have derived appropriate fittings to the IONIC task results following the
functional form given by \cite{1992MNRAS.255..325P}. These expressions are
listed below: 

\begin{eqnarray*}
12\,+\,log(O^+/H^+)\,=\,log\frac{I(3727+3729)}{I(H\beta)}\,+\,5.992\,+\\
+\,\frac{1.583}{t_e}\,
-\,0.681\,log\,t_e\,+\,log(1+2.3\,n_e)\\
12\,+\,log(O^{2+}/H^+)\,=\,log\frac{I(4959+5007)}{I(H\beta)}\,+\,6.144\,+\\
+\,\frac{1.251}{t_e}\,
-\,0.550\,log\,t_e\\
12\,+\,log(S^+/H^+)\,=\,log\frac{I(6717+6731)}{I(H\beta)}\,+\,5.423\,+\\
+\,\frac{0.929}{t_e}\,
-\,0.280\,log\,t_e\,+\,log(1+1.0\,n_e)\\
12\,+\,log(S^{2+}/H^+)\,=\,log\frac{I(9069+9532)}{I(H\beta)}\,+\,5.80\,+\\
+\,\frac{0.77}{t_e}\,
-\,0.22\,log\,t_e\\
12\,+\,log(N^+/H^+)\,=\,log\frac{I(6548+6584)}{I(H\beta)}\,+\,6.273\,+\\
+\,\frac{0.894}{t_e}\,
-\,0.592\,log\,t_e\\
log\,(N^+/O^+)\,=\,log\frac{I(6548+6584)}{I(3727+3729)}\,+\,0.281\,-\\
-\,\frac{0.689}{t_e}\,
+\,0.089\,log\,t_e\\
12\,+\,log(Ne^{2+}/H^+)\,=\,log\frac{I(3868)}{I(H\beta)}\,+\,6.486\,+\\
+\,\frac{1.558}{t_e}\,
-\,0.504\,log\,t_e\\
12\,+\,log(Ar^{2+}/H^+)\,=\,log\frac{I(7137)}{I(H\beta)}\,+\,6.157\,+\\
+\,\frac{0.808}{t_e}\,
-\,0.508\,log\,t_e\\
12\,+\,log(Ar^{3+}/H^+)\,=\,log\frac{I(4740)}{I(H\beta)}\,+\,5.705\,+\\
+\,\frac{1.246}{t_e}\,
-\,0.156\,log\,t_e\\
12\,+\,log(Fe^{2+}/H^+)\,=\,log\frac{I(4658)}{I(H\beta)}\,+\,3.504\,+\\
+\,\frac{1.298}{t_e}\,
-\,0.483\,log\,t_e)
\end{eqnarray*}
\noindent where $t_e$ denotes the appropriate line electron temperature, in
units of 10$^{\rm 4}$ K, corresponding to the assumed ionisation structure as
explained below. 

\noindent {\bf -Oxygen} The oxygen ionic abundance ratios, O$^{+}$/H$^{+}$ and
O$^{2+}$/H$^{+}$, have been derived from the [O{\sc
    ii}]\,$\lambda\lambda$\,3727,29\,\AA\ and  [O{\sc
    iii}]\,$\lambda\lambda$\,4959, 5007\,\AA\ lines respectively using for
each ion its corresponding temperature. 

\noindent {\bf -Sulphur.} In the same way,  we have derived S$^+$/H$^{+}$ and
S$^{2+}$/H$^{+}$, abundances using T$_e$([S{\sc ii}]) and T$_e$([S{\sc iii}])
values and the fluxes of the [S{\sc ii}] emission lines at
$\lambda\lambda$\,6717,6731\,{\AA} and  the near-IR [S{\sc
    iii}]\,$\lambda\lambda$\,9069, 9532\,\AA\ lines respectively.    

\noindent {\bf -Nitrogen.} The ionic abundance of nitrogen, N$^{+}$/H$^{+}$
has been derived from the intensities of the $\lambda\lambda$\,6548,
6584\,\AA\ lines and  the derived electron temperature of [N{\sc ii}] in the
case of \sietec\ . For the rest of the objects the assumption  T$_e$([N{\sc
    ii}])\,=\,T$_e$([O{\sc ii}]) has been made.  

\noindent {\bf -Neon.} Ne$^{2+}$ has been derived from the [Ne{\sc iii}] 
emission line at $\lambda$3868\,{\AA} assuming T$_e$([Ne{\sc
    iii}])\,=\,T$_e$([O{\sc iii}]) \citep{1969BOTT....5....3P}.

\noindent {\bf -Argon.} The main ionization states of Ar in ionized regions
are Ar$^{2+}$ and Ar$^{3+}$. The abundance of Ar$^{2+}$ has been calculated
from the measured [Ar{\sc iii}]\,$\lambda$\,7136\,\AA\ line emission assuming
that T$_e$([Ar{\sc iii}])\,$\approx$\,T$_e$([S{\sc iii}])
\citep{1992AJ....103.1330G}, while the  ionic abundance of Ar$^{3+}$ has been
calculated from the emission line of [Ar{\sc iv}]\,$\lambda$\,4740\,\AA\ under
the assumption that T$_e$([Ar{\sc iv}])\,$\approx$\,T$_e$([O{\sc iii}]). 

\noindent {\bf -Iron.} Finally, for iron we have used the emission line of
[Fe{\sc iii}]\,$\lambda$\,4658\,\AA\ to calculate Fe$^{2+}$ assuming
T$_e$([Fe{\sc iii}])\,=\,T$_e$([O{\sc iii}]).  

The ionic abundances of the different elements with respect to ionised
hydrogen along with their corresponding errors are given in Table
\ref{ion-abs}.


\begin{table*}
{\small
\caption{ICFs and total chemical abundances for elements heavier than
  Helium.}
 
\label{total-abs}

\begin{center}
\begin{tabular}{@{}cccccccc}
\hline
  & \unoc & \dosc & \tresc & \cuatroc & \cincoc & \seisc & \sietec  \\
\hline
\input{tables/total-abund-pap.tex}
\hline
\end{tabular}
\end{center}}
\end{table*}

\subsection{Ionization correction factors and total abundances}

For the three objects for which the He{\sc ii} line has been measured (\unoc,
\cincoc\ and \seisc), the total abundance of He has been found by adding
directly the two ionic abundances:  
\[
\frac{He}{H}\,=\,\frac{He^++He^{2+}}{H^+}
\]

As was pointed out by \cite{1994ApJ...431..172S}, the potential fraction of
unobservable neutral helium is a long lasting problem and represents a source
of uncertainty in the derivation of the helium total abundance. The correction
factor for He$^0$ can be approximated by 1.0 for \HII\ regions ionised by very
hot stars (T$_{eff}$\,$\geq$\,40000\,K). An estimate of the ionising stellar
temperature for our objects can be obtained from the $\eta$ parameter
\citep{1988MNRAS.231..257V}\footnote{The $\eta$ parameter is defined as the
ratio of the O$^+$/O$^{2+}$ and S$^+$/S$^{2+}$ ionic ratios.}. The values of
log$\eta$ for the three objects are -0.10, -0.07 and -0.57, much smaller than
the upper limit of log($\eta$)\, 0.9 for which \cite{1992MNRAS.255..325P}
claim that the correction factor for neutral helium is equal to 1.0.  

In table \ref{absHe} we present the total helium abundance values for
these three objects together with their corresponding errors.

\label{oxygen}
At the temperatures derived for our observed galaxies,
most of the oxygen is in the form of O$^+$ and O$^{2+}$, therefore the
approximation: 
\[ 
\frac{O}{H}\,=\,\frac{O^++O^{2+}}{H^+}
\]
\noindent has been used. 
\label{sulphur}
This is not however the case for sulphur for which a relatively important
contribution from S$^{3+}$ may be expected depending of the nebular
excitation. The total sulphur abundance has been calculated using an ICF for
S$^+$+S$^{2+}$ according to Barker's (1980) \nocite{1980ApJ...240...99B} fit
to the photoionization models by \cite{1978A&A....66..257S}:  
\[
ICF(S^++S^{2+}) = \left[ 1-\left(\frac{O^{2+}}{O^{+}+O^{2+}}
  \right)^\alpha\right]^{-1/\alpha} 
\]
\noindent  Although it is customary to write Barker's expression as a function of the 
O$^{+}$/(O$^{+}$+O$^{2+}$) ionic fraction, we have reformulated it in terms of 
O$^{2+}$/(O$^{+}$+O$^{2+}$)  since the errors associated to O$^{2+}$ are
considerably smaller than for O$^{+}$.  A value of 2.5 for $\alpha$ gives the best fit to the
scarce observational data on S$^{3+}$ abundances \citep{2006A&A...449..193P}.

\label{N/O}
We have derived the N/O abundance ratio under the assumption
that   
\[
\frac{N}{O}\,=\,\frac{N^+}{O^+}
\]
\noindent and the N/H ratio as:
\[ log \frac{N}{H} = log\frac{N}{O} + log\frac{O}{H} \]

\label{Neon}
The ionisation correction factor for neon has been calculated according to 
the expression given by \cite{2007MNRAS.submitted}: 
\[
ICF(Ne^{2+})\,=\,0.142\,x+0.753+\frac{0.171}{x}
\]
where $x$\,=\,O$^{2+}$/(O$^{+}$+O$^{2+}$). This expression has been derived from 
photoionisation models \citep{1998PASP..110..761F},
taking as ionizing sources the spectral energy distribution of O and B stars
\citep{2001A&A...375..161P}. 

Given the high excitation of the observed objects there are no significant
differences between the total neon abundance derived using this ICF and those
estimated using the classical approximation: Ne/O$\approx$Ne$^{2+}$/O$^{2+}$.

\label{Argon}
As in the case of neon, the total abundance of argon has been calculated using
the ionization correction factors (ICF(Ar$^{2+}$) and the
ICF(Ar$^{2+}$+Ar$^{3+}$)) given by \cite{2007MNRAS.submitted}. We have used
the first one only when we cannot derive a value for Ar$^{3+}$. The
expressions for these ICFs are:   
\[
ICF(Ar^{2+})\,=\,0.507\,(1-x)+0.749+\frac{0.064}{(1-x)}
\]
\[
ICF(Ar^{2+}+Ar^{3+})\,=\,0.364\,(1-x)+0.928+\frac{0.006}{(1-x)}
\]
where $x$\,=\,O$^{2+}$/(O$^{+}$+O$^{2+}$). 

\label{Iron}

The ICF for iron twice ionized has been taken from \cite{2004IAUS..217..188R}: 
\[
ICF(Fe^{2+})\,=\,\Big(\frac{O^+}{O^{2+}}\Big)^{0.09} \cdot
\Big[1+\frac{O^{2+}}{O^{+}}\Big]
\]

In table \ref{total-abs} we list all the total chemical abundances and the
ICFs derived for elements heavier than helium in the present work. 


\label{data comparison}
\section{Comparison with previous data}

Five of the seven \HII\ galaxies presented here (\unoc, \dosc,\cuatroc,
\cincoc\ and \sietec) have been previously studied by
\cite{2006A&A...448..955I} from SDSS/DR3 spectra. \dosc\ was also analysed by
\cite{1992A&A...253..349P} together with \tresc\ using spectra in the
$\lambda\lambda$ 3400-7000\,\AA\ range obtained with the 2.1\,m telescope at
KPNO through a 3.2\,arcsec slit. \cincoc\ and \sietec\ have also been studied
by \cite{2004ApJS..153..429K} from SDSS/DR1 spectra. For each observed object
the reddening corrected emission line intensities reported in these studies
are given for comparison in Table \ref{ratiostot 1}. Only the line intensities
with errors less than 40 per cent are listed.

A good general agreement between our measurements and the ones in the
literature for the strong emission lines and most of the weak ones is
found. There are however some noticeable differences. 

For \dosc , published values for the intensities of the lines bluer than
H$\beta$ are larger than measured in this work. For the [O{\sc
    ii}]\,$\lambda$\,3727\,\AA\ line the discrepancy amounts to 37 \% for the
data by \cite{2006A&A...448..955I} and 22 \%  for those by
\cite{1992A&A...253..349P}. In all cases the derived values of the reddening
constant are similar. 
It is worth noting that in the SDSS image the object seems to have a 3\,arcsec
size, thus the SDSS fiber aperture and the KPNO observations might contain a
substantial part of the galaxy, while our data comes from the brightest
central knot. This could explain part of the found differences.  

For \tresc\ the values given by \cite{1992A&A...253..349P} for the  [O{\sc
i}]\,$\lambda$\,6300\,\AA\ and [S{\sc ii}]\,$\lambda\lambda$\,6717,6731\,\AA\
lines are larger and smaller respectively than measured in this work. This is
most noticeable for the [O{\sc i}] line with their value being larger than
ours by almost a factor of three.

For \cincoc\ the [Ne{\sc iii}]\,$\lambda$\,3868\,\AA, line intensity given in
\cite{2006A&A...448..955I} is larger than measured here by 33\%.  On the
contrary, a smaller intensity than given here by about 30\% is measured for
this line in \sietec\ by \cite{2006A&A...448..955I}.  




\section{Discussion}

\subsection{Gaseous physical conditions and element abundances}

\subsubsection{Densities and temperatures}

Four electron temperatures -- T$_e$([O{\sc iii}]), T$_e$([O{\sc ii}]),
T$_e$([S{\sc iii}]) and  T$_e$([S{\sc ii}])-- have been estimated in the seven
observed objects. In addition, T$_e$([N{\sc ii}]) has been estimated in
\sietec; [N{\sc ii}]\,$\lambda$\,5755 is detected, but has poor signal, in
\unoc, \dosc, \tresc\ and \seisc\  and falls in the gap between the blue and
red spectra for the other two objects (\cuatroc\ and \cincoc). The good
quality of the data allows us to reach accuracies of the order of 2\%, 4\%,
5\% and 7\%  for T$_e$([O{\sc iii}]), T$_e$([O{\sc ii}]), T$_e$([S{\sc iii}]),
and  T$_e$([S{\sc ii}]) and T$_e$([N{\sc ii}]) respectively. The worse
measurement of a line temperature is T$_e$([S{\sc ii}]) for \cincoc, with a
$\sim$\,10\% error.


\begin{table*}
\centering
\caption[]{Published [O{\sc iii}] temperature and abundances for the observed objects.}
\label{comp}
\begin{tabular} {@{}l c c c c c c c}
\hline
          &   t$_e$([O{\sc iii}]) &  12+log(O/H)      &  log(S/O) & log(N/O)& log(Ne/O) & log(Ar/O)   ref.\\
\hline	   							        	     
 \unoc    & 1.36\,$\pm$\,0.04 & 8.03\,$\pm$\,0.03 &     -1.65 &  -1.37  &  -0.73    &  -2.45    & $^{1}$ \\
 \dosc    & 1.16\,$\pm$\,0.05 & 8.22\,$\pm$\,0.04 &     -1.78 &  -1.40  &  -0.69    &  -2.50    & $^{1}$ \\
          & 1.20\,$\pm$\,0.08 & 8.17\,$\pm$\,0.10 &     ---   &  -1.24  &  -0.69    &    ---    & $^{2}$ \\
          &      1.10$^a$     & 8.26$^a$          &     ---   &  -1.24$^a$ &  ---   &    ---    & $^{2}$ \\
 \tresc   &      1.10$^a$     & 8.26$^a$          &     ---   &  -1.46$^a$ &  ---   &    ---    & $^{2}$ \\
 \cuatroc & 1.04\,$\pm$\,0.07 & 8.26\,$\pm$\,0.07 &     -1.78 &  -1.34  &  -0.80    &  -2.36    & $^{1}$ \\
          & 1.06\,$\pm$\,0.08 & 8.13\,$\pm$\,0.08 &     ---   &  ---    &    ---    &   ---     & $^{3}$ \\
 \cincoc  & 1.39\,$\pm$\,0.05 & 7.98\,$\pm$\,0.04 &     -1.52 &  -1.56  &  -0.69    &  -2.33    & $^{1}$ \\
 \sietec  & 1.15\,$\pm$\,0.04 & 8.21\,$\pm$\,0.03 &     -1.76 &  -1.18  &  -0.74    &  -2.40    & $^{1}$ \\
          & 1.15\,$\pm$\,0.03 & 8.17\,$\pm$\,0.03 &     ---   &  ---    &    ---    &   ---     & $^{3}$ \\
\hline
\multicolumn{6}{l}{$^{1}$Izotov et al.\ (2006); $^{2}$Peimbert \& Torres-Peimbert (1992); $^{3}$Kniazev et al.\ (2004).}\\
\multicolumn{6}{l}{$^a$based on an empirical method from Pagel et al.\ (1979).}
\end{tabular}
\end{table*}

%
%

\begin{table*}
\centering
\caption[]{Electron temperatures for HII Galaxies in units of 10$^{-4}$ K.}
\label{temp}
\begin{tabular}{l c c c c l}
\hline
Object  & t$_e$([O{\sc ii}])  & t$_e$([O{\sc iii}]) & t$_e$([S{\sc ii}]) & t$_e$([S{\sc iii}]) & Ref.  \\
\hline
SBS0335-052    & 1.34$\pm$0.03 & 2.03$\pm$0.03 &          --           &          --            & ICS01 \\
SBS0832+699    & 0.95$\pm$0.09 & 1.66$\pm$0.03 &          --           &          --            & ITL94 \\
SBS1135+581   & 1.36$\pm$0.07 & 1.31$\pm$0.02 & 1.01$\pm$0.03 &          --            & ITL94 \\
SBS1152+579   & 1.42$\pm$0.09 & 1.63$\pm$0.02 & 1.70$\pm$0.10 &          --            & ITL94 \\
SBS1415+437   & 1.41$\pm$0.07 & 1.71$\pm$0.01 & 1.39$\pm$0.06 &          --            & IT98 \\
SBS0723+692A & 1.45$\pm$0.13 & 1.58$\pm$0.01 & 1.74$\pm$0.14 &          --            & ITL97 \\
SBS0749+568   & 1.08$\pm$0.06 & 1.52$\pm$0.08 &          --            & 1.78$\pm$0.23 & ITL97, PMD03 \\ 
SBS0907+543   & 1.52$\pm$0.09 & 1.44$\pm$0.04 &          --            &          --            & ITL97 \\
SBS0917+527   & 1.18$\pm$0.09 & 1.51$\pm$0.03 & 1.19$\pm$0.08 &          --            & ITL97 \\
SBS0926+606   & 1.31$\pm$0.05 & 1.43$\pm$0.02 & 1.07$\pm$0.06 & 1.49$\pm$0.13 & ITL97, PMD03 \\
SBS0940+544N & 1.33$\pm$0.13 & 2.03$\pm$0.04 &          --           &          --            & ITL97 \\
SBS1222+614   & 1.26$\pm$0.08 & 1.45$\pm$0.01 & 1.79$\pm$0.16 &          --            & ITL97 \\
SBS1256+351   & 1.32$\pm$0.07 & 1.35$\pm$0.01 & 0.97$\pm$0.04 &          --            & ITL97 \\
SBS1319+579A & 1.40$\pm$0.08 & 1.29$\pm$0.01 & 0.65$\pm$0.04 &          --            & ITL97 \\
SBS1319+579C & 1.27$\pm$0.07 & 1.23$\pm$0.03 & 0.77$\pm$0.06 &          --            & ITL97 \\
SBS1358+576   & 1.30$\pm$0.13 & 1.47$\pm$0.02 & 1.18$\pm$0.08 &          --            & ITL97 \\
SBS1533+574B & 1.39$\pm$0.09 & 1.24$\pm$0.03 & 0.78$\pm$0.07 &          --            & ITL97 \\
Pox36               & 0.92$\pm$0.05 & 1.25$\pm$0.06 & 0.79$\pm$0.05 &          --            & IT04 \\
CGC007-025      & 1.04$\pm$0.04 & 1.66$\pm$0.02 & 1.72$\pm$0.10 &          --            & IT04 \\ 
Mrk 450-1          & 1.23$\pm$0.05 & 1.16$\pm$0.01 & 1.04$\pm$0.04 &          --            & IT04 \\
Mrk 450-2          & 1.35$\pm$0.13 & 1.24$\pm$0.03 & 1.07$\pm$0.10 &          --            & IT04 \\
HS0029+1748    & 1.23$\pm$0.11 & 1.28$\pm$0.06 &          --            &          --            & IT04 \\
HS0122+0743    & 1.27$\pm$0.11 & 1.79$\pm$0.03 & 1.30$\pm$0.13 &          --            & IT04 \\
HS0128+2832    & 1.50$\pm$0.06 & 1.25$\pm$0.01 & 1.22$\pm$0.07 &          --            & IT04 \\
HS1203+3636A  & 0.95$\pm$0.06 & 1.07$\pm$0.02 & 1.51$\pm$0.12 &          --            & IT04 \\
HS1214+3801    & 1.25$\pm$0.05 & 1.33$\pm$0.01 & 1.18$\pm$0.06 &          --            & IT04 \\
HS1312+3508    & 1.01$\pm$0.08 & 1.31$\pm$0.03 &          --            &          --            & IT04 \\
HS2359+1659    & 1.15$\pm$0.11 & 1.18$\pm$0.02 &          --            &          --            & IT04 \\
Mrk 35               & 0.97$\pm$0.03 & 1.02$\pm$0.01 & 1.06$\pm$0.02 &          --            & IT04 \\
UM 238              & 0.83$\pm$0.08 & 1.24$\pm$0.02 & 0.83$\pm$0.08 &          --            & IT04 \\
UM 439              & 1.23$\pm$0.10 & 1.28$\pm$0.06 &       --              &          --             & IT04 \\
IIZw 40              & 1.27$\pm$0.06 & 1.34$\pm$0.03 &       --               & 1.30$\pm$0.04 & GIT00,PMD03 \\
Mrk 22               & 1.16$\pm$0.09 & 1.35$\pm$0.03 & 0.96$\pm$0.08 &  1.94$\pm$0.21& ITL94, PMD03 \\
Mrk 36               & 1.37$\pm$0.12 & 1.53$\pm$0.05 &        --              & 1.55$\pm$0.17 & IT98, PMD03 \\
UM 461              & 1.64$\pm$0.16 & 1.62$\pm$0.05 &        --              & 1.93$\pm$0.10 & IT98, PMD03 \\
UM 462              & 1.19$\pm$0.03 & 1.38$\pm$0.02 & 1.00$\pm$0.07 & 1.61$\pm$0.17 & IT98, PMD03 \\
Mrk5                  & 1.32$\pm$0.08 & 1.22$\pm$0.06 &        --              & 1.30$\pm$0.11 & IT98, PMD03 \\
VIIZw 403          &  1.42$\pm$0.12 & 1.52$\pm$0.03 &        --              & 1.28$\pm$0.10 & ITL97, PMD03 \\
Mrk 209             &  1.28$\pm$0.08 & 1.62$\pm$0.01 & 1.25$\pm$0.13 & 1.59$\pm$0.13 & ITL97, PMD03 \\ 
Mrk 1434           &  1.24$\pm$0.08 & 1.55$\pm$0.02 &        --              & 1.72$\pm$0.14 & ITL97, PMD03 \\
Mrk 709             &  1.50$\pm$0.15 & 1.67$\pm$0.06 &        --              & 1.62$\pm$0.16 & T91, PMD03 \\
UGC 4483          &         --             & 1.68$\pm$0.06 &        --              & 1.57$\pm$0.17 & S94 \\  
IZw18NW           & 1.28$\pm$0.40  & 1.96$\pm$0.09 &        --              & 2.49$\pm$0.51 & SK93 \\
IZw18SE            & 1.18$\pm$0.50  & 1.72$\pm$0.12 &        --              & 1.97$\pm$0.34 & SK93 \\
SDSS0364          & 1.31$\pm$0.03 & 1.24$\pm$0.01 & 1.04$\pm$0.07 & 1.26$\pm$0.04 & Paper I \\ 
SDSS0390          & 1.03$\pm$0.02 & 1.25$\pm$0.02 & 0.86$\pm$0.06 & 1.31$\pm$0.05 & Paper I \\
SDSS0417          & 1.35$\pm$0.04 & 1.28$\pm$0.02 & 1.03$\pm$0.05 & 1.36$\pm$0.05 & Paper I \\
KISSR 1845       & 1.08$\pm$0.09 & 1.32$\pm$0.04 &       --               &          --            & M04 \\  
KISSR 396         & 1.66$\pm$0.10 & 1.40$\pm$0.05 &       --               &          --            & M04 \\
KISSB 171         & 1.16$\pm$0.06 & 1.18$\pm$0.04 &       --               &          --            & L04 \\
KISSB 175         & 1.12$\pm$0.10 & 1.34$\pm$0.02 &       --               &          --            & L04 \\
KISSR 286         & 1.05$\pm$0.06 & 1.10$\pm$0.02 &       --               &          --            & L04 \\
\hline

\noalign {\noindent \small T91: \cite{1991A&AS...91..285T}; SK93: \cite{1993ApJ...411..655S}; S94: \cite{1994ApJ...431..172S} ;ITL94: \cite{1994ApJ...435..647I}; ITL97: \cite{1997ApJS..108....1I}; IT98: \cite{1998ApJ...500..188I}; ICS: \cite{2001A&A...378L..45I}; GIT00: \cite{2000ApJ...531..776G}; PMD03: \cite{2003MNRAS.346..105P}; IT04: \cite{2004A&A...415...87I}; M04: \cite{2004AJ....127..686M}; L04: \cite{2004ApJ...616..752L}; Paper I: \cite{2006MNRAS.372..293H}}
\end{tabular}
\end{table*}

\begin{figure}
\includegraphics[width=8.5cm,angle=0,clip=]{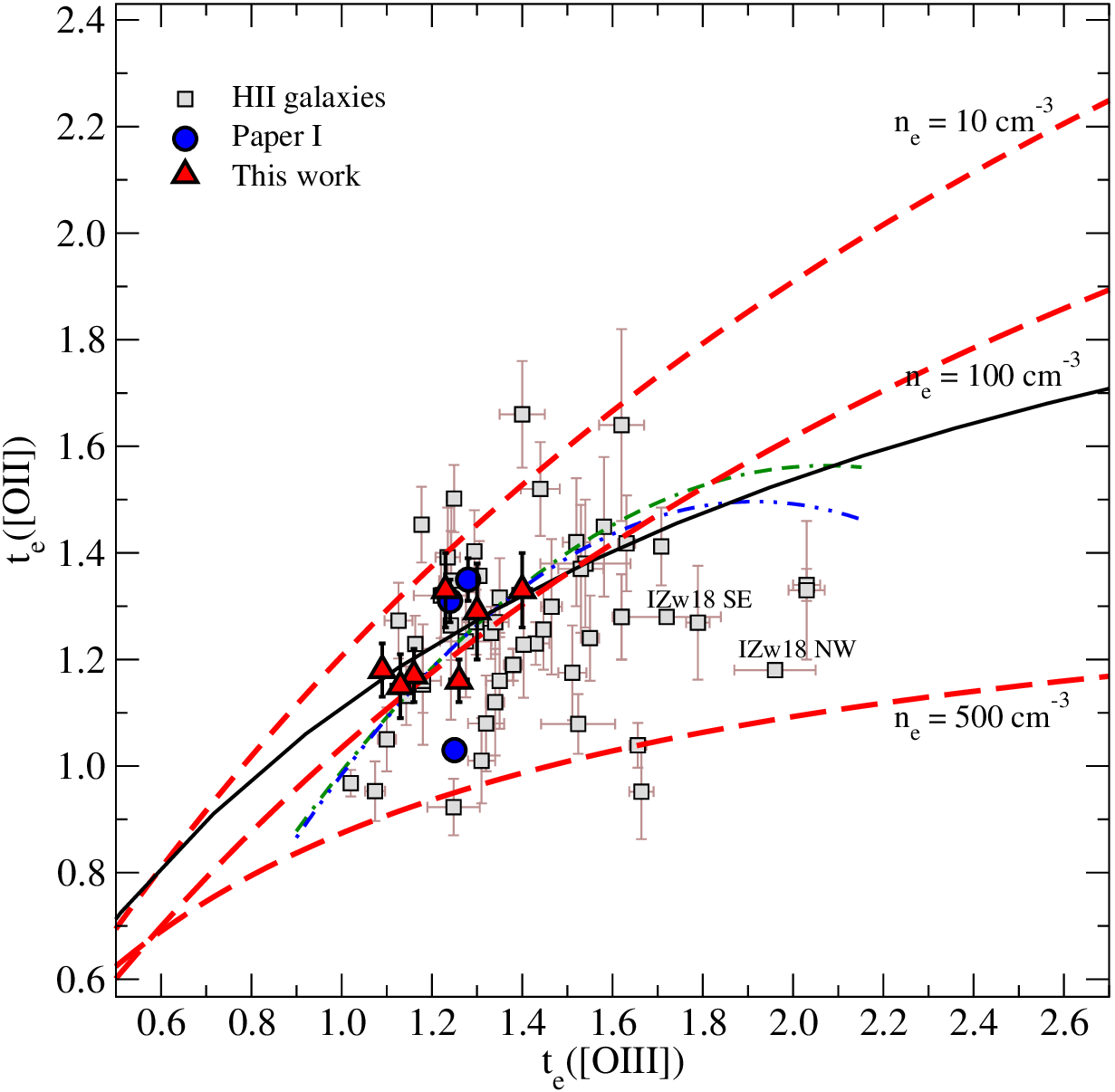}\\
\caption{Relation between t$_e$([O{\sc ii}]) and t$_e$([O{\sc iii}]) for the 
  objects in this paper and the data in Table \ref{temp}. 
  The dashed lines (red in the electronic version) correspond to
  photoionisation models from P\'erez-Montero \& D\'iaz (2003) for electron
  densities N$ _{e} $\,=\,10, 100 and 500\,cm$^{-3}$. The model sequences from
  \citet[][solid line]{1990A&AS...83..501S} and  Izotov et al.\ (2006) for low and intermediate
  metallicity \HII\ regions (double dashed-dotted line and the dashed-double dotted
  line, green and blue respectively in the electronic version) 
  are also shown. Temperatures are in units of 10$^4$\,K.} 
\label{T([OII]-T([OIII])}
\end{figure}

The seven observed objects show temperatures within a relatively narrow range,
between 10900 and 14000\,K for T$_e$([O{\sc iii}]). This was also the case for
the objects analysed in Paper I. This could be due to the adopted selection
criteria, high H$\beta$ flux and large equivalent width of H$\alpha$, which
tends to select objects with abundances and electron temperatures close to the
median values shown by \HII\ galaxies. In table \ref{comp} we have listed the
previously reported t$_e$([O{\sc iii}]) values for our observed objects. We
find a general good agreement between those values and our measurements. Only
for two objects, \dosc\ and \sietec, we find differences of 900\,K --in
average-- and 1100\,K, respectively. Figure \ref{T([OII]-T([OIII])} shows the
relation between the [O{\sc ii}] and [O{\sc iii}] temperatures measured for
these objects. Also shown in the figure are the corresponding values for \HII\
galaxies as derived from the emission line intensities compiled from the
literature. This derivation has been done following the same prescriptions
given in the present work. These values are given in Table \ref{temp} together
with the references that have been used. We have restricted our compilation to
objects for which the temperatures could be derived with an accuracy better
than 10\%. An exception has been made for IZw18. In this case, the large
errors in the derived [O{\sc ii}] temperatures (30\% and 40\% for T$_e$([O{\sc
    ii}]) in the NW and SE knots respectively), are probably related to the
low oxygen abundance  in this galaxy and hence the weakness of the involved
emission lines. These points are labelled in the figure and, due to their
large values, no error bars in t$_e$([O{\sc ii}]) are shown. 

The relation between the [O{\sc ii}] and [O{\sc iii}] temperatures does not
show a clear trend, showing a scatter which is larger than observational
errors. Given that the [O{\sc ii}] temperature is somewhat dependent on
density one could be tempted to adscribe this scatter to a density effect. The
density effect can be seen by looking at the model sequences which are
overplotted and which correspond to photo-ionisation models from
P\'erez-Montero \& D\'\i az (2003) for electron densities N$_{e}$\,=\,10, 100
and 500\,cm$^{-3}$. Higher density models show lower values of t$_e$([O{\sc
    ii}]) for a given  t$_e$([O{\sc iii}]). The effect is more noticeable at
high electron temperatures. In fact, the data points populate the region of
the diagram spanned  by model sequences with most objects located between the
model sequences corresponding to N$_e$ 100 and 500 cm$^{\rm -3}$. Our observed
objects, however,  lie between the model sequences for N$_{e}$ 10 and 100
cm$^{\rm -3}$. This is actually consistent with the derived values of N([S{\sc
    ii}]).


All in all, the data show that there is not a unique relation between the
[O{\sc ii}] and [O{\sc iii}] temperatures which allows a reliable derivation
of one of this temperatures when the other one cannot be secured. This is
actually a standard procedure in principle adopted for the analysis of low
resolution and poor signal-to-noise data and now extended to data of much
higher quality. The solid line in Figure \ref{T([OII]-T([OIII])} shows the
relation based on the photoionisation models by Stasi\'nska (1990), adopted in
many abundance studies of ionised nebulae. A substantial part of the sample
objects show [O{\sc ii}] temperatures which are lower or higher than predicted
by this relation for as much as 3000 K. At a value of T$_e$([O{\sc iii}] of
10000 K, this differences translate into higher and lower O$^+$/H$^+$ ionic
ratios, respectively, by a factor of 2.5. However, when using model sequences
to predict [O{\sc ii}] temperatures no uncertainties are attached to the
t$_e$([O{\sc ii}]) vs.\ t$_e$([O{\sc iii}]) relation and the outcome is a
reported T$_e$([O{\sc ii}]) which carries only the usually small observational
error of T$_e$([O{\sc iii}]) which translates into very small errors in the
oxygen ionic and total abundances. Thus it is possible to find in the
literature values of T$_e$([O{\sc ii}]) with quoted fractional errors lower
than 1\% and absolute errors actually less than that quoted for T$_e$([O{\sc
    iii}]) \citep{1998ApJ...500..188I}, which translate into ionic
O$^+$/H$^+$ ratios with errors of only 0.02\,dex. 

\begin{figure}
\includegraphics[width=8.6cm,angle=0,clip=]{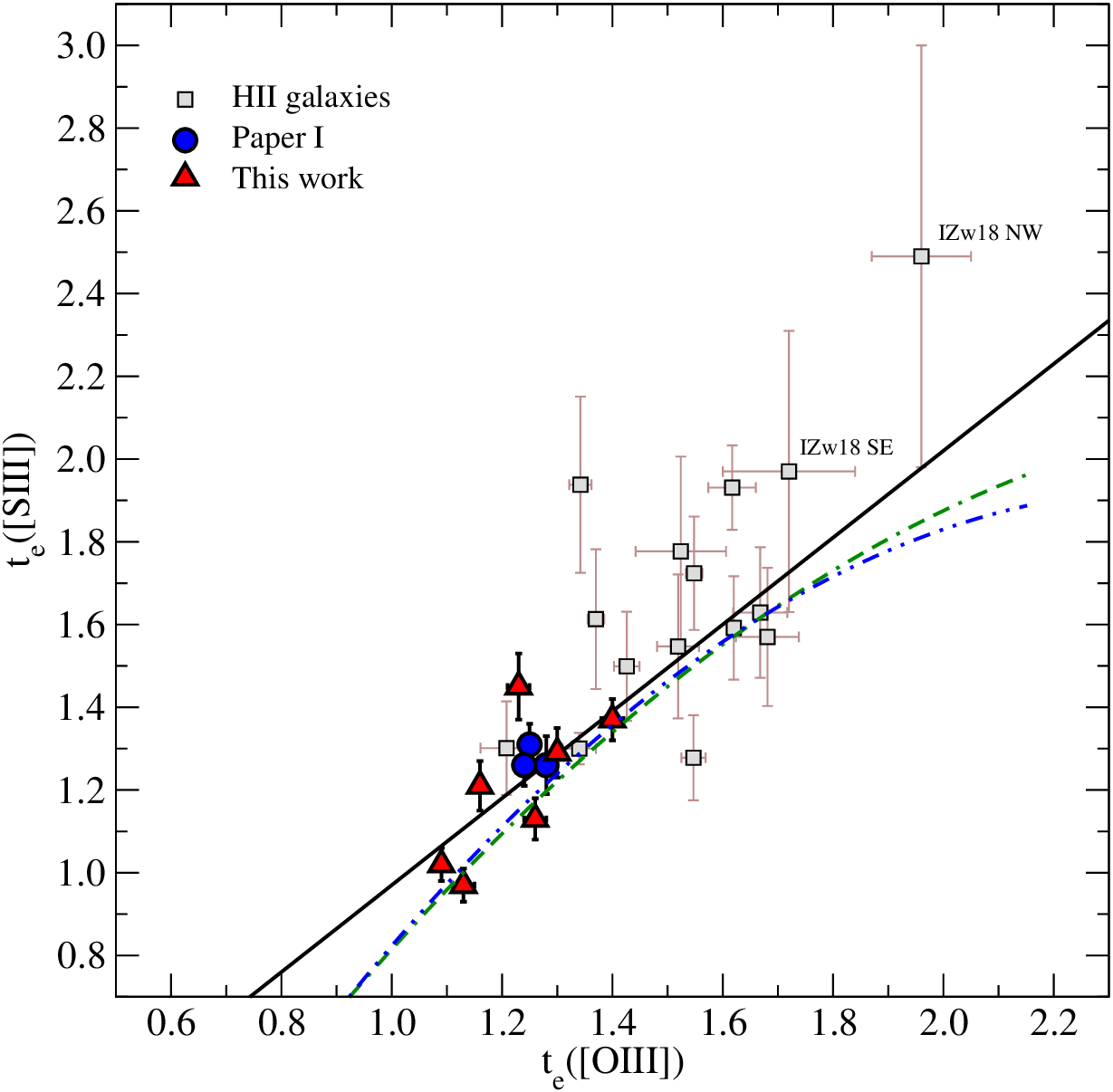}\\
\caption{Relation between t$_e$([S{\sc iii}]) and t$_e$([O{\sc iii}]) for the
  objects in this paper and those from Table \ref{temp}. The solid line
  corresponds to the photoionisation model sequence of P\'erez-Montero \&
  D\'\i az (2005). The double dashed-dotted line and the dashed-double dotted
  line (green and blue respectively in the electronic version) represent the
  models presented by Izotov et al.\ (2006) for low and intermediate
  metallicity \HII\ regions. The temperatures are in units of 10$^4$\,K.} 
\label{T([SIII]-T([OIII])}
\end{figure}

Recently, this procedure has been justified by \cite{2006A&A...448..955I}
based on the comparison of a selected SDSS  data sample of "\HII-region like"
objects with phtoionisation models computed by the authors. Different
expressions of T$_e$([O{\sc ii}]) as a function of T$_e$([O{\sc iii}]) are
given for different metallicity regimes and it is argued that, despite a large
scatter, the relation between T$_e$([O{\sc ii}]) and T$_e$([O{\sc iii}])
derived from observations follows generally the one obtained by
models. However, no clear trend is shown by the data and the large errors
attached to the electron temperature determinations, in many cases around
$\pm$\,2000\,K for T$_e$([O{\sc ii}]), actually preclude the test of such a
statement. In fact, most of the data with the smallest error bars lie below
and above the theoretical relation. While it may well be that these objects
belong to a different family from typical \HII\ galaxies, this has not been
actually shown to be the case. The model sequences of
\cite{2006A&A...448..955I} for the cases of low and intermediate metallicities
are shown in Figure \ref{T([SIII]-T([OIII])} as double dashed-dotted  and the
dashed-double dotted lines (green and blue in the electronic version)
respectively. These models diverge from previous sequences at temperatures
below 10000 K and above 18000 K. Of great concern is the model degenerate
behaviour at high temperatures. Unfortunately, only one object in our sample
has an [O{\sc iii}] temperature larger than 20000 K (SBS0940+544N,
T$_e$([O{\sc iii}]\,=\,20300\,$\pm$\,400\,K) and its [S{\sc iii}] temperature
has a 10\% error. Therefore it is not possible with the present data to
address this important issue.  

\begin{figure}
\includegraphics[width=8.5cm,angle=0,clip=]{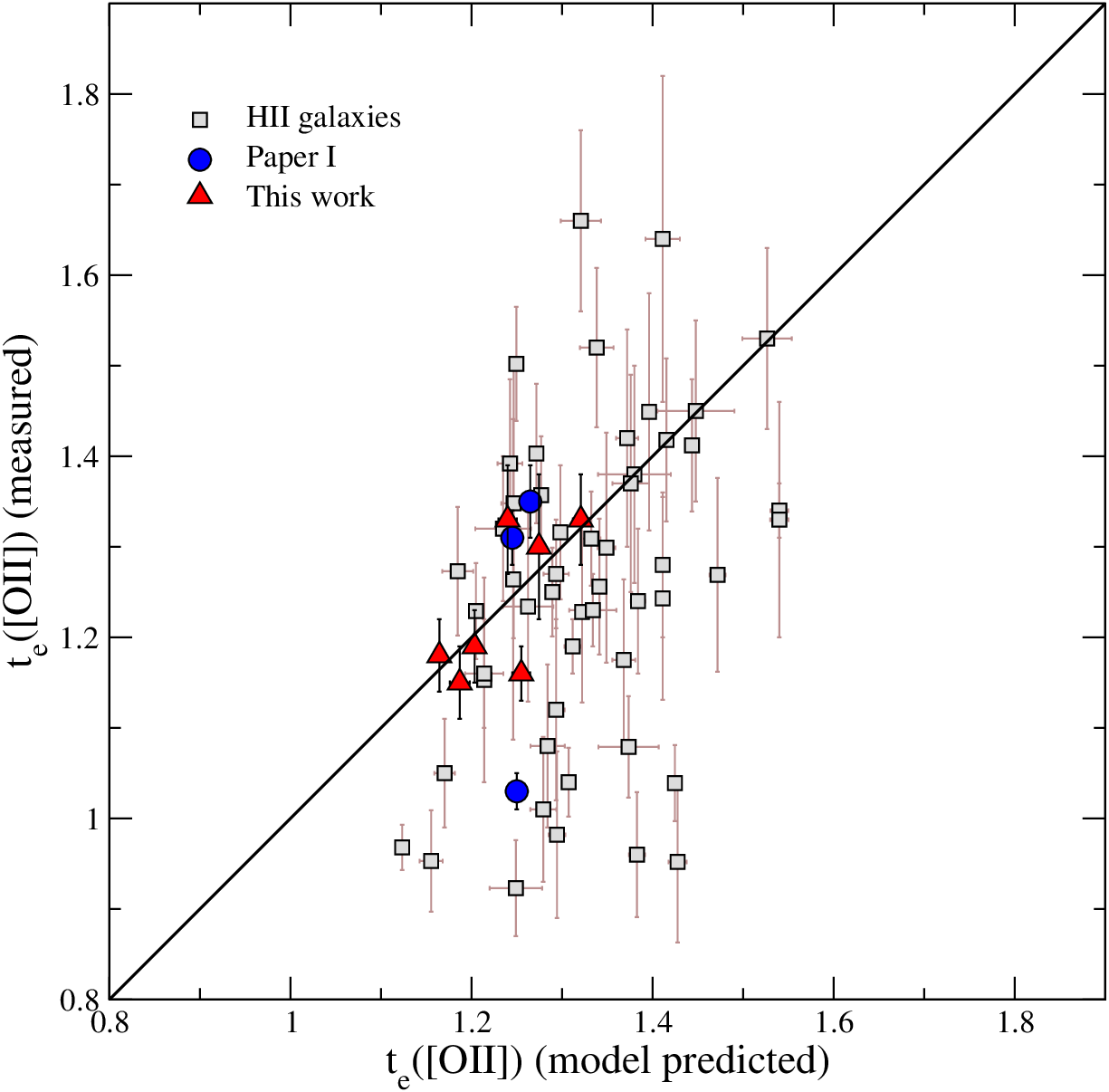}\\
\caption{Comparison between the t$_e$([O{\sc ii}]) values derived from direct
  measurements with those derived from t$_e$([O{\sc iii}]) using Stasi\'nska
  (1990) photoionization models.}  
\label{tmodel}
\end{figure}

 
Figure \ref{tmodel} shows the comparison between the t$_e$([O{\sc ii}]) values
derived from direct measurements with those derived from t$_e$([O{\sc iii}])
using \cite{1990A&AS...83..501S} photoionization models. These values agree
for five of our observed objects and are lower and higher by 900\,K
respectively for \seisc\ and \sietec. These differences translate into higher
and lower O$^+$/H$^+$ ionic ratios by factors of $\sim$1.32 and $\sim$1.26
respectively, and higher and lower total oxygen abundances by $\sim$0.05 and
$\sim$0.06\,dex, respectively. 

In general, model predictions overestimate t$_e$([O{\sc ii}]) and hence
underestimate the O$^+$/H$^+$ ratio. This is of relatively low concern in
objects of high excitation for which O$^+$/O is less than $\sim$ 10\%, but
caution should be taken when dealing with lower excitation objects where total
oxygen abundances could be underestimated by up to 0.2 dex.

%
%

\begin{figure}
\includegraphics[width=8.5cm,angle=0,clip=]{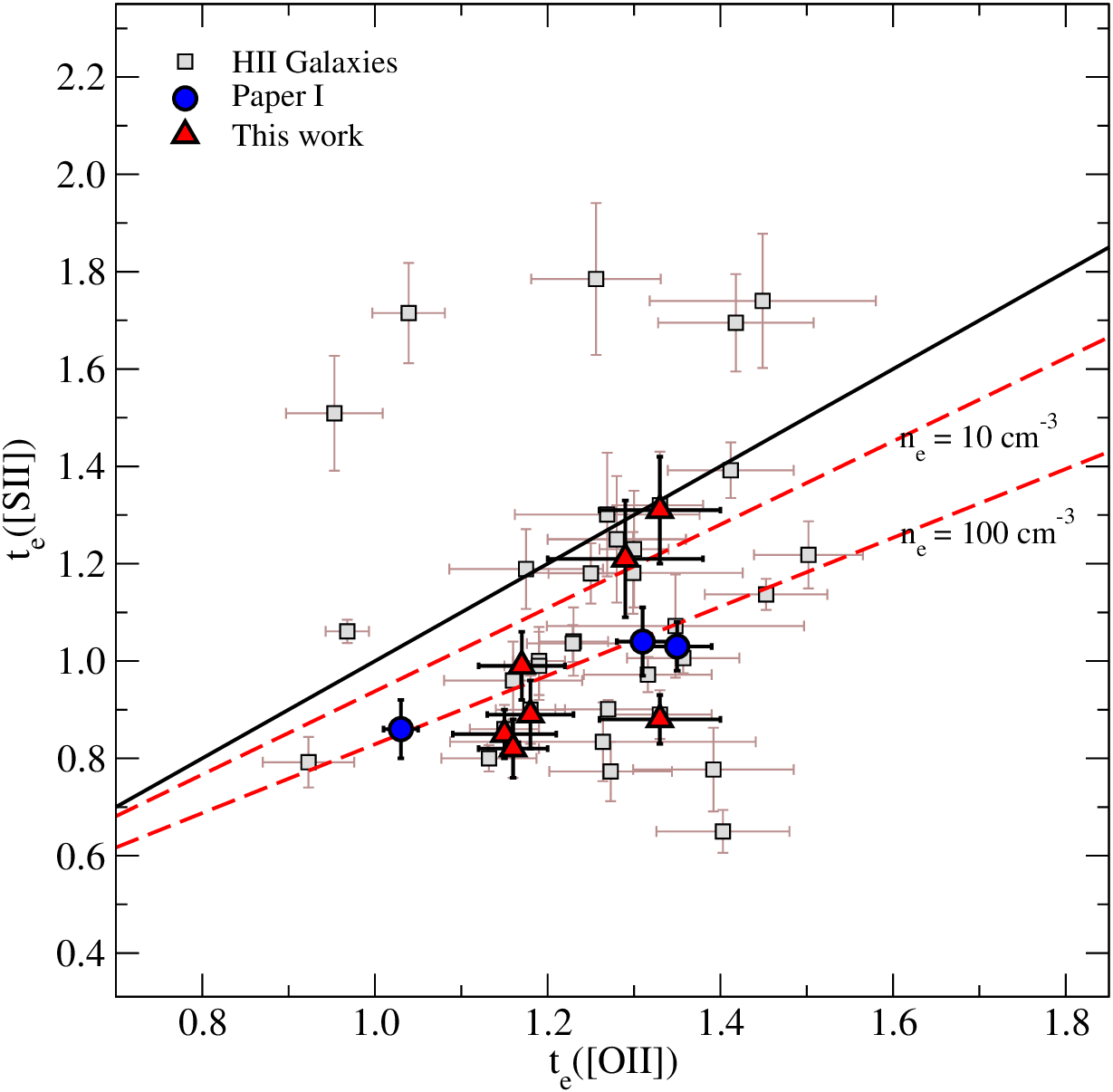}\\
\caption{Relation between t$_e$([S{\sc ii}]) and t$_e$([O{\sc ii}]) for the
  objects in this paper and those from Table \ref{temp}. The solid line
  represents the   one to one relation. The dashed lines (red in the
  electronic version) correspond to the photoionization models from
  P\'erez-Montero \& D\'\i az (2003) for electron densities N$ _{e}$\,=\,10
  and 100\,cm$^{-3}$. Temperatures are in units of  10$^4$\,K.} 
\label{T([SII]-T([OII])}
\end{figure}


In the usually assumed structure of ionised nebulae, low ionisation lines
arise from the same region and therefore the temperatures of [O{\sc ii}],
[S{\sc ii}] and [N{\sc ii}] are expected to show similar values, once
allowance is made for a possible density effect for the first two. In
Fig. \ref{T([SII]-T([OII])} we show the relation between t$_e$([S{\sc ii}])
and t$_e$([O{\sc ii}])  for the objects in this paper and those from Table
\ref{temp}. The relations derived by P\'erez-Montero \& D\'\i az (2003) using
photoionization models for electron densities N$ _{e} $\,=\,10 and
100\,cm$^{-3}$ are also shown. Our measurement for the [S{\sc ii}] and [O{\sc
    ii}] temperatures are located in the region predicted by the models,
although any dependence on density is difficult to appreciate. Some objects
however are seen to lie well above the one-to-one relation. These objects are:
SBS1222+614,  HS1203+3636A and CGC007-025. The latter one shows a value of the
[O{\sc iii}] temperature close to the [S{\sc ii}] one, with the [O{\sc ii}]
temperature being lower by about 6000 K. On the other hand, the other two
objects show [O{\sc iii}] and [O{\sc ii}] temperatures fitting nicely on the
model sequence for N$_e$\,=\,100\,cm$^{-3}$.




Regarding T$_e$([N{\sc ii}]), the $\lambda$ 5575 \AA\ line is usually very
weak and difficult to measure in \HII\ galaxies due to their low
metallicities, and therefore only a few measurements exist and with very large
errors, larger than 50 \% in some cases. If we restrict ourselves to the data
with errors of less than 10 \%, the [O{\sc ii}] and [N{\sc ii}] temperatures
differ by at most 1500 K. More high quality data would be needed in order to
confirm this usually assumed relation. 
 
The situation seems to be better for the [S{\sc iii}] temperature. Figure
\ref{T([SIII]-T([OIII])} shows the relation between t$_e$([S{\sc iii}]) and
t$_e$([O{\sc iii}]) for our objects, the \HII\ galaxies from Paper I and the
compilation of published data on \HII\ galaxies for which measurements of the
nebular and auroral lines of [O{\sc iii}] and [S{\sc iii}] exist, thus
allowing the simultaneous determination of T$_e$([O{\sc iii}]) and
T$_e$([S{\sc iii}]) (see Table \ref{temp}). We have omitted from the
compilation the data from \cite{2006A&A...457..477K} that were taken with a
1.52 m telescope and hence do not have the required signal-to-noise to provide
electron temperature measurements with large errors. The solid line in the
figure shows the model sequence from P\'erez-Montero \& D\'\i az (2005), which
differs slightly from the semi-empirical relation by
\cite{1992AJ....103.1330G}, while the other two lines correspond to the
relations given by \cite{2006A&A...448..955I} for low and intermediate
metallicity \HII\ regions. 
The three model sequences are coincident for temperatures in the range from
12000 and 17000 K, and very little, if any, metallicity dependence is
predicted. Although the number of objects is small (only 25, including the
galaxies in the present work) and the errors for the data found in the
literature are large, most objects seem to follow the trend shown by model
sequences. The most discrepant object is Mrk~22 that shows a [S{\sc iii}]
temperature larger than predicted by about 6000 K. Obviously, more high
quality data are needed in order to confirm the relation between the [O{\sc
    iii}] and [S{\sc iii}] and obtain an empirical fit with well determined
{\em rms} errors.


\subsubsection*{Abundances}

The abundances derived for the observed objects show the characteristic low
values found in strong line \HII\ galaxies
\citep{1991A&AS...91..285T,2006MNRAS.365..454H}: 12+log(O/H) between 7.94 and
8.19. The mean error values for the oxygen and neon abundances are 0.04 dex
and slightly larger, 0.07, for sulphur and argon. 

Six of the objects have published abundance determinations which are listed in
Table \ref{comp}. Our results are in general good agreement with those in the
literature. In the case of \cuatroc\ the difference between the value derived
by \cite{2006A&A...448..955I} and ours is 0.19\,dex, but our derived value is
in agreement --within the errors-- with that obtained by
\cite{2004ApJS..153..429K}. We have also found a small difference of 0.13\,dex
--in average-- in the oxygen abundance of \sietec. These differences are
similar to those found by us in our previous work, Paper I, between our
derived abundances using WHT spectra and the values estimated by
\cite{2004ApJS..153..429K} as part of the first edition of the SDSS \HII\
galaxies with oxygen abundance catalog. 



\begin{figure*}
\includegraphics[width=8.5cm,angle=0,clip=]{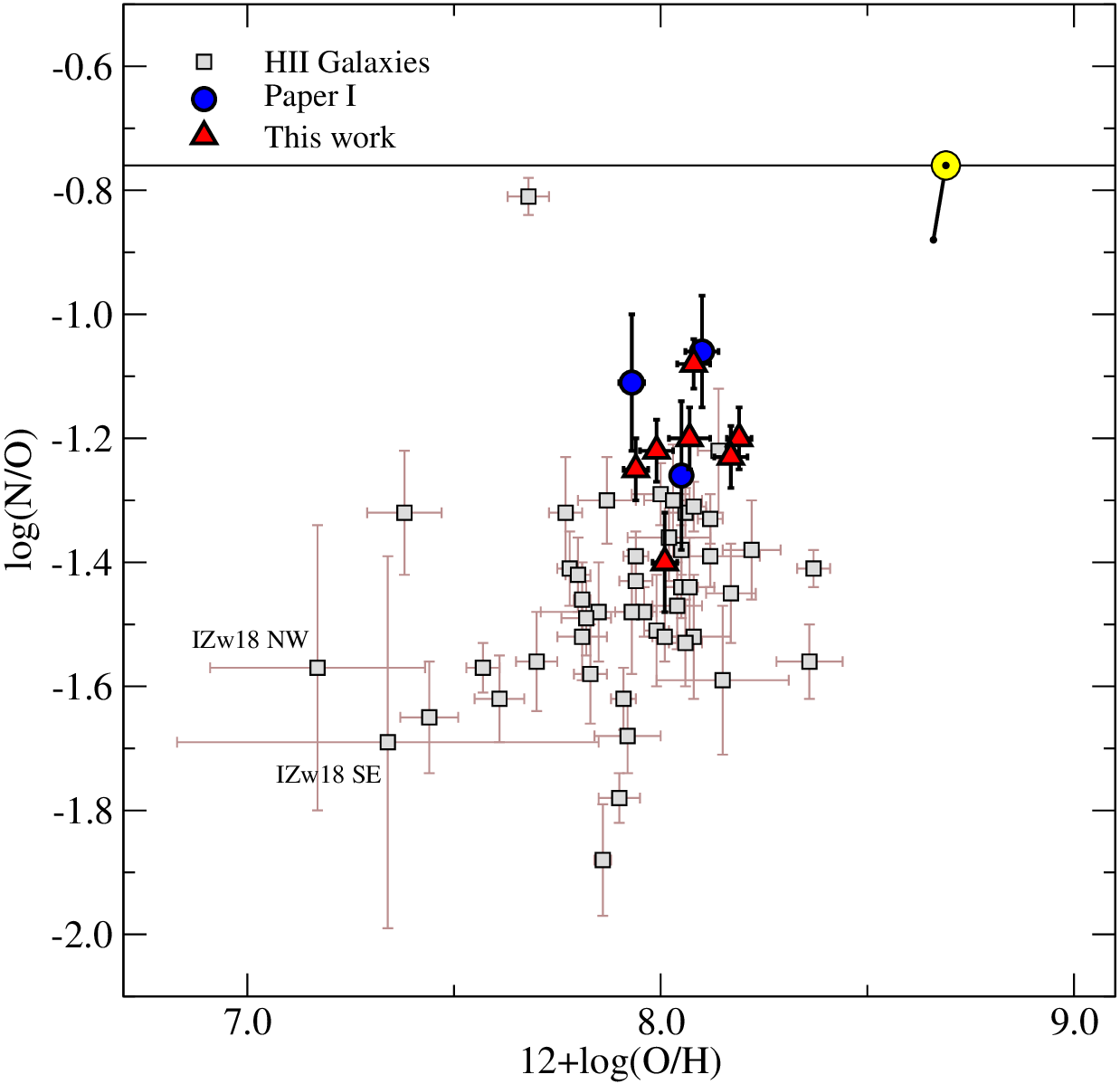}\hspace{0.3cm}
\includegraphics[width=8.5cm,angle=0,clip=]{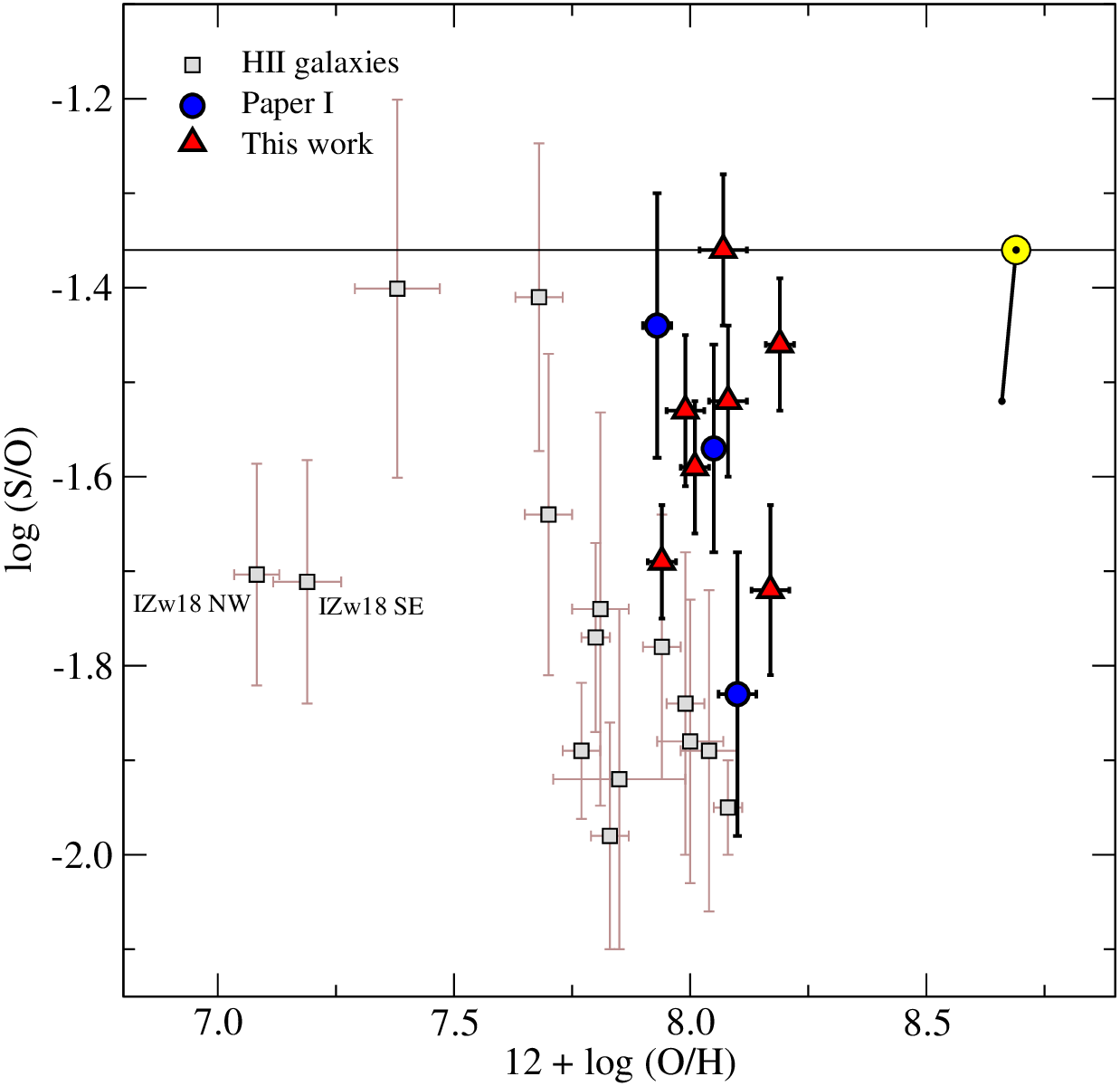}
\caption{Left panel: N/O ratio as a function of 12+log(O/H) for the observed
  objects and the objects from Paper I (filled triangles and circles, red and
  blue respectively in the electronic version of the paper) and the \HII\
  galaxies (open squares) from Table \ref{temp}. Right panel: same as in the
  left panel, but for the S/O ratio. The solar values are shown with the usual
  sun symbol  oxygen from \citet{2001ApJ...556L..63A}, nitrogen from
  \citet{2001AIPC..598...23H} and sulphur from
  \citet{1998SSRv...85..161G}. These values are linked by a solid line with
  the solar ratios from \citet{2005ASPC..336...25A}.} 
\label{abundratios}
\end{figure*}



The logarithmic N/O ratios found for the galaxy for which there are T$_e$([O{\sc
    ii}]) and T$_e$([N{\sc ii}]) determinations is -1.27$\pm$0.05. If the
assumption that t$_e$([O{\sc ii}])\,=\,t$_e$([N{\sc ii}]) is made, an N/O
ratio larger by a factor of 1.5 is obtained. For the rest of the galaxies, for
which this assumption has been made, log(N/O) ratios are between -1.40 and
-1.20 with an average error of 0.05. For all the objects, the derived values
are on the highest side of the distribution for this kind of objects (see
Figure \ref{abundratios}). In general, the common procedure of obtaining
t$_e$([O{\sc ii}]) from t$_e$([O{\sc iii}]) using Stasi\'nska's (1990)
relation and assuming t$_e$([O{\sc ii}]) = t$_e$([N{\sc ii}]), yields N/O
ratios larger than using the measured t$_e$([O{\sc ii}]) values, since
according to Figure \ref{tmodel}, in most cases, the model sequence
overpredicts t$_e$([O{\sc ii}]). An overprediction of this temperature by a 30
\% at T$_e$([O{\sc ii}])=13000\,K would increase the N/O ratio by a factor of
2. Therefore, the effect of our observed objects showing relatively high N/O
ratios seems to be real.



The log(S/O) ratios found for the objects are also listed in table
\ref{total-abs}. These values vary between  -1.69 and -1.36  with an average
error of 0.07, consistent with solar
(log(S/O)$_{\odot}$\,=\,-1.36\footnote{Oxygen from Allende-Prieto et al.\
  (2001) and sulphur from \citet{1998SSRv...85..161G}.}) within the
observational errors, except for \unoc\ and \tresc\  for  which S/O is lower
by a factor of about 1.8. 
Comparing with the \cite{2006A&A...448..955I} derived S/O logarithmic ratios
(see Table \ref{comp}) we find that for three of the observed objects they
found S/O ratios lower than ours by us much as 0.4 dex or a factor of about
2.5. 



\begin{figure*}
\includegraphics[width=8.5cm,angle=0,clip=]{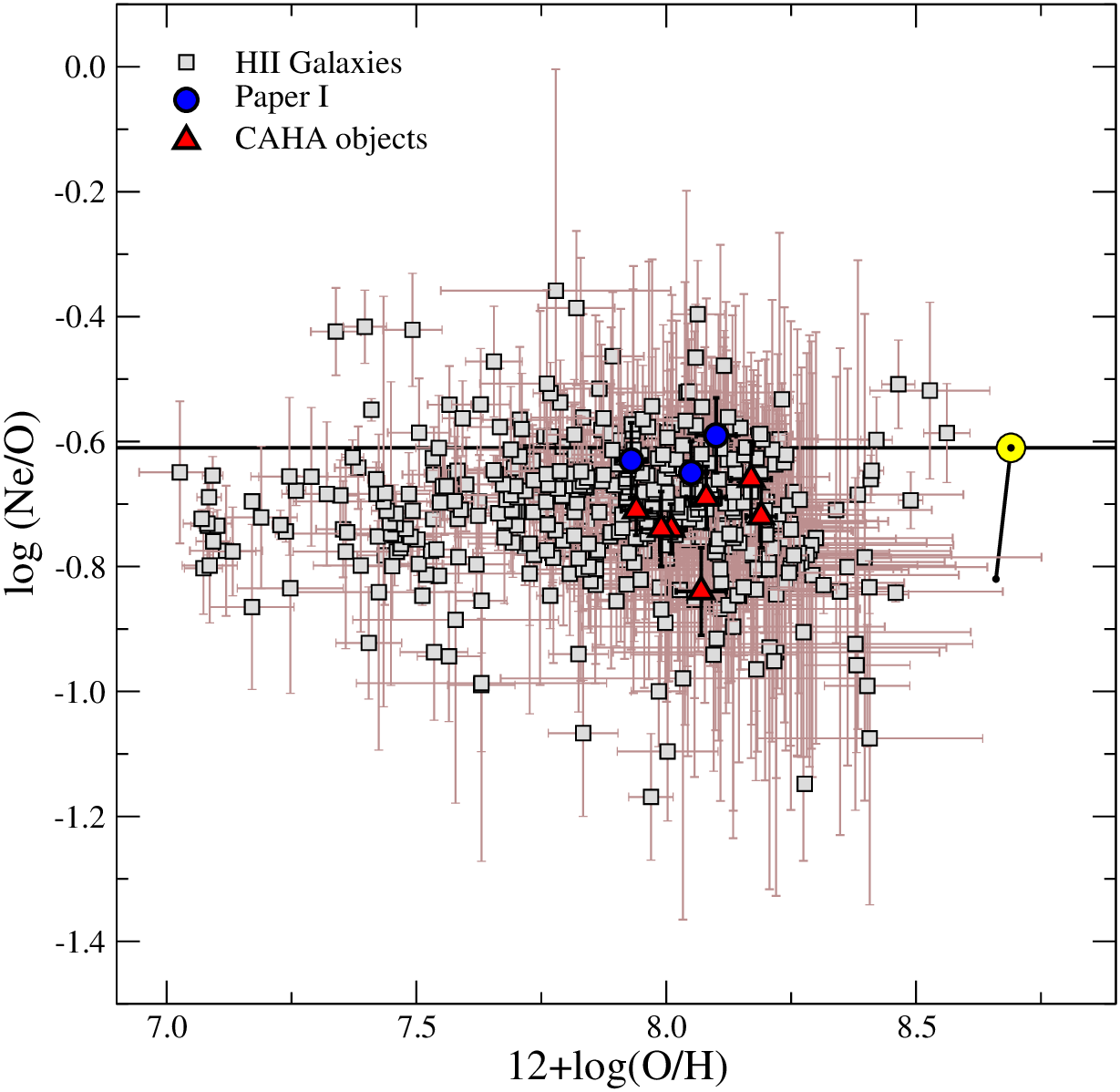}\hspace{0.3cm}
\includegraphics[width=8.5cm,angle=0,clip=]{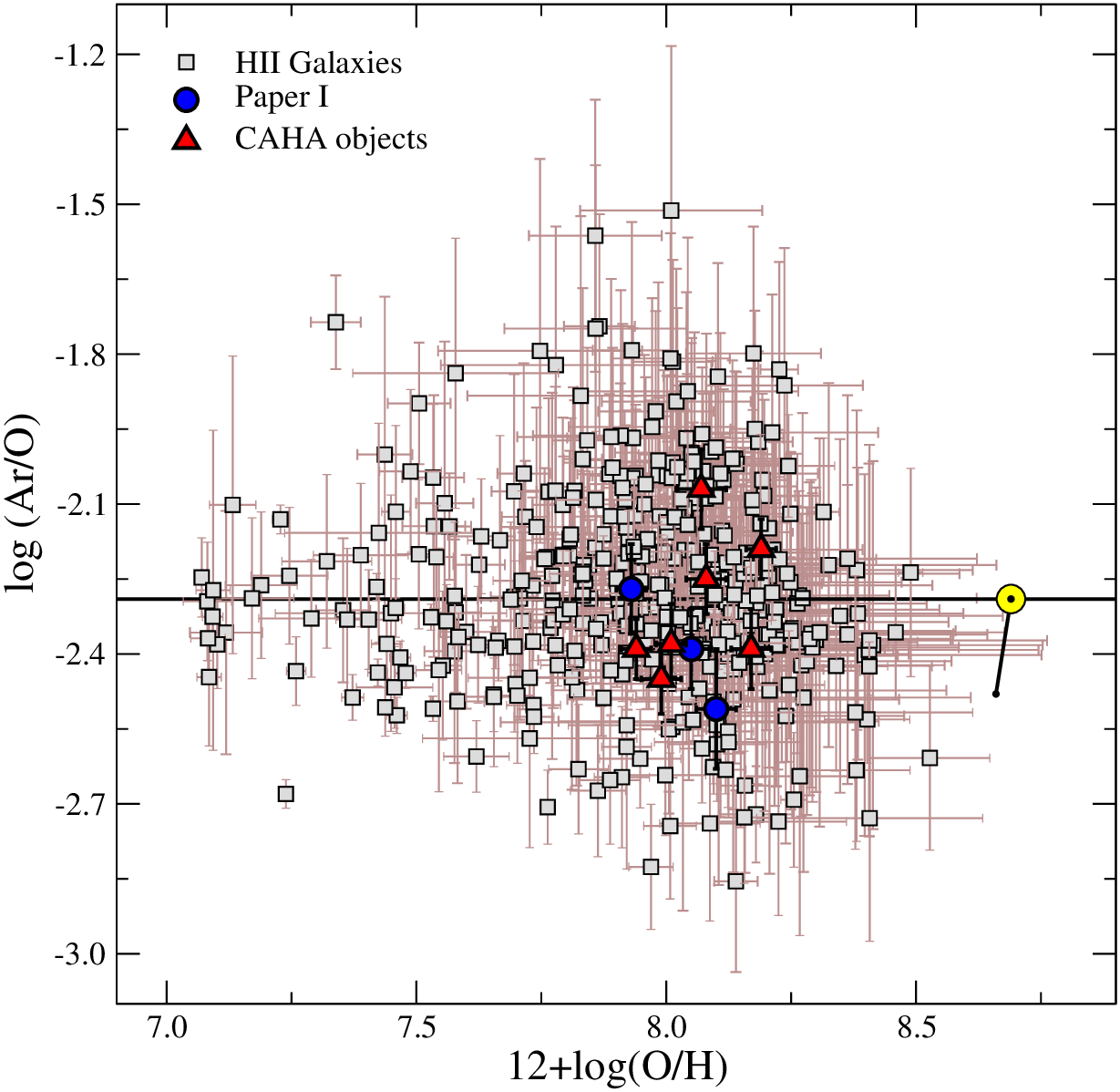}
\caption{Left panel: Ne/O ratio as a function of 12+log(O/H) for the observed
  objects and the objects from Paper I (filled triangles and circles, red and
  blue respectively in the electronic version of the paper) and \HII\ galaxies
  (open squares) from P\'erez-Montero et al.\ (2007). Right panel: same as in
  the left panel, for the Ar/O ratio. The solar values are shown by the usual
  sun symbol, with oxygen as before and neon and argon from
  \citet{1998SSRv...85..161G}.} 
\label{abundratios2}
\end{figure*}



The logarithmic Ne/O  ratio varies between -0.84 and -0.66,  with a constant
value (see Fig. \ref{abundratios2}) within the errors (table \ref{total-abs})
consistent with solar (log(Ne/O)\,=\,0.61\,dex$^5$) , if  the object with the
lowest ratio is excluded. An excellent agreement with the literature values is
found.

The values calculated using the classical approximation for the ICF
(Ne/O\,=\,Ne$^{2+}$/O$^{2+}$), although systematically larger, are within
errors very close to those derived using the ICF for neon from
\cite{2007MNRAS.submitted}. This is to be expected, given the high degree of
ionization of the objects in the sample. 



Finally, the Ar/O ratios found for the observed objects show a larger
dispersion than in the case of Ne/O (see Fig. \ref{abundratios2}), with a mean
value consistent with solar$^5$. Comparing our estimations for the logarithmic
Ar/O ratios with those derived by \cite{2006A&A...448..955I}, we find a good
agreement for three objects, and larger values for \dosc\ and \cuatroc, 0.31
and 0.29 respectively. We must note that for these two objects we have not
been able to measure the ionic abundances of Ar$^{3+}$. 


\begin{figure*}
\includegraphics[width=.47\textwidth,angle=0]{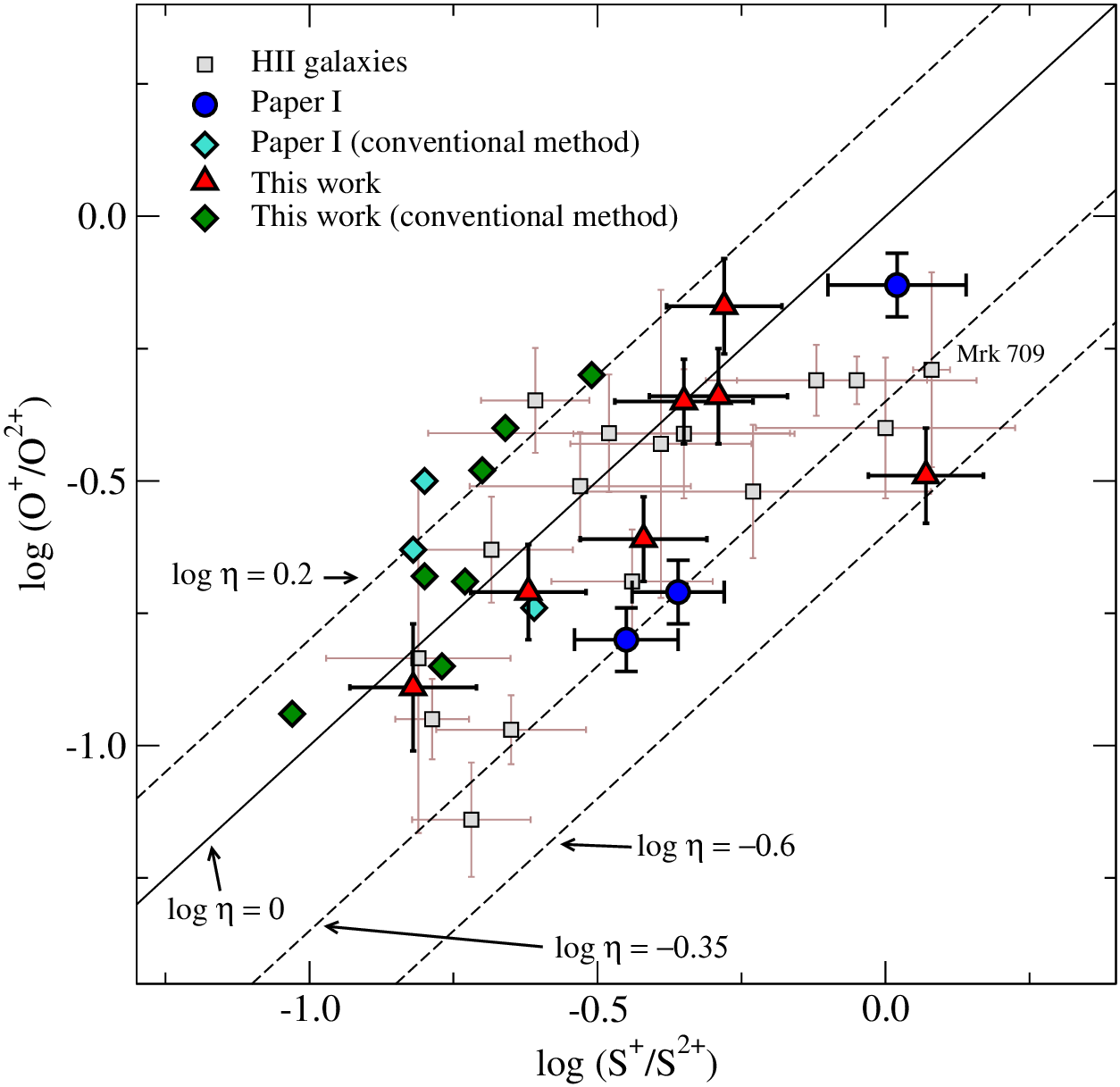}\hspace{0.5cm}
\includegraphics[width=.47\textwidth,angle=0]{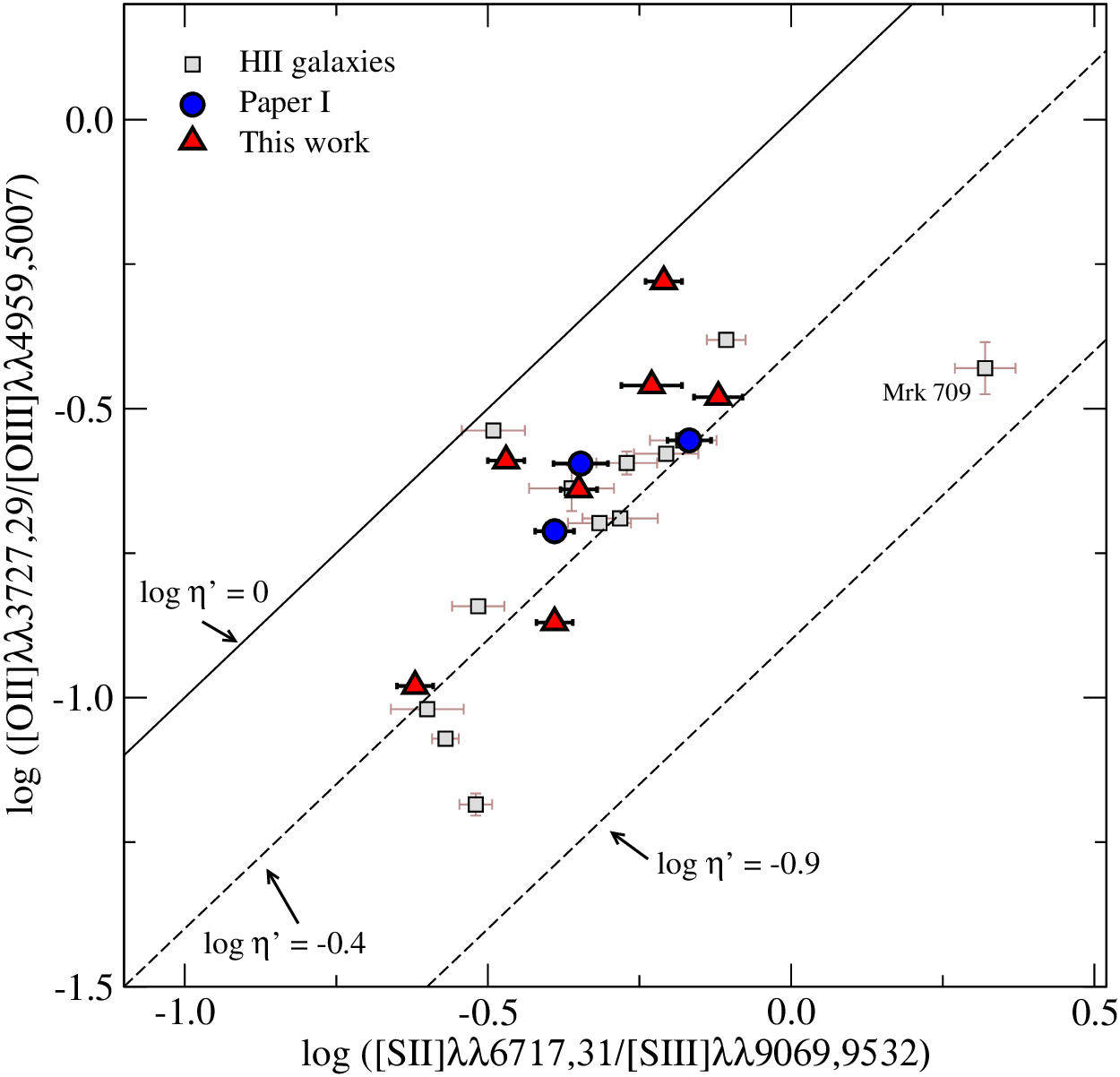}\\
\caption{Panel (a):  log(O$^+$/O$^{2+}$) vs. log(S$^{+}$/S$^{2+}$)  for the
  objects in this paper and Paper I (solid red triangles and blue circles
  respectively) for values calculated using our methodology. Solid diamonds
  represent the values derived using the conventional method for the same
  objects (dark green and turquoise respectively in the electronic version of
  the paper). Open squares are the \HII\ galaxies from P\'erez-Montero \&
  D\'iaz (2005) with derived  oxygen and sulphur ionic ratios. Diagonals in
  this diagram correspond to constant values of $\eta$. Panel (b): log([O{\sc
      ii}]/[O{\sc iii}]) vs. log([S{\sc ii}]/[S{\sc iii}]), symbols as in
  panel (a) and the \HII\ galaxies from P\'erez-Montero \&  D\'iaz (2005) have
  data on the [S{\sc iii}]\,$\lambda\lambda$\,9069,9532\,\AA\ emission
  lines. Diagonals in this diagram correspond to constant values of $\eta$'.} 
\label{eta}
\end{figure*}

\subsection{Ionisation structure}
\label{ionization}

An insight into the ionisation structure of the observed objects can be gained
by means of the O$^+$/O$^{2+}$ vs S$^{+}$/S$^{2+}$ diagram (see Paper I).


In panel (a) of Fig. \ref{eta} we show the location on this diagram of the
observed objects together with those in Paper I (red filled triangles and
blue circles, respectively) and the \HII\ galaxies in Table \ref{temp}. In this
diagram diagonal lines correspond to constant values of the $\eta$ parameter
which can be taken as an indicator of the ionising temperature (V\'ilchez \&
Pagel 1988). In this diagram \HII\ galaxies occupy the region with log\,$\eta$
between -0.35 and 0.2,  which corresponds to high values of the ionising
temperature according to these authors. One of the observed objects, \seis ,
shows a very low value of $\eta$=-0.6. This object however, had the [O{\sc
    ii}] $\lambda\lambda$ 7319,25 \AA\ affected by atmospheric absorption
lines. Unfortunately no previous data of this object exist apart form the SDSS
spectrum. We have retrieved this spectrum and measured the [O{\sc ii}] lines
deriving a t$_e$([O{\sc ii}]) = 1.23$\pm$0.21, the large error being due to
the poor signal to noise in the [O{\sc ii}] $\lambda$ 7319,25 \AA . This lower
temperature would increase the value of O$^+$/O$^{2+}$ moving the data point
corresponding to this objects upwards in the left panel of Figure
\ref{eta}. This would be consistent with the position of the object in the
right panel of the figure which shows log([O{\sc ii}]/[O{\sc iii}]) vs.\
log([S{\sc ii}]/[S{\sc iii}]), the observational equivalent to the $\eta$
plot. The axes in this diagram define log\,$\eta$' as: 
\begin{eqnarray*}
\lefteqn{log\,\eta{\textrm'}\,=\,log\Big[ \frac{[O{\textrm {\sc ii}}]\lambda\lambda\,3727,29\,/\,[O{\textrm {\sc iii}}]\lambda\lambda\,4959,5007}{[S{\textrm {\sc ii}}]\lambda\lambda\,6717,31\,/\,[S{\textrm {\sc iii}}]\lambda\lambda\,9069,9532}\Big]
{} } \nonumber\\ & & {}
\,=\,log\,\eta - \frac{0.14}{t_e} -0.16   {}
\end{eqnarray*} 
\noindent where t$_e$ is the electron temperature in units of 10$^4$. The
value of log\,$\eta$' for \seis\ is -0.36 corresponding to $\eta$ = -0.09 for
t$_e$([O{\sc iii}]) = 1.23.  

Inconsistencies between the values of $\eta$ and $\eta$' are also found if the
ionic ratios are derived using values of electron temperatures obtained
following the prescriptions given by Izotov et al.\ (2006). These values are
represented by solid diamonds in the left panel of Figure \ref{eta} for the
objects in this work and those from paper I. In all cases, higher values of
$\eta$ are obtained which, in conjunction with the measured values of $\eta$', 
would indicate values of electron temperatures much lower than directly
obtained. These higher values of $\eta$ would also imply ionising temperatures
lower than those shown by the measured $\eta$' values.




Metallicity calibrations based on abundances derived according to this
conventional method are probably bound to provide metallicities which are
systematically too high and should therefore be revised.

\section{Summary and conclusions}

We have performed a detailed analysis of newly obtained spectra of seven
\HII\ galaxies selected from the Sloan Digital Sky Survey Data Release 3.
The spectra cover from 3400 to 10400 \AA\ in wavelength at a FWHM resolution 
of about 2000 in the blue and 1500 in the red spectral regions. 
 
The high signal-to-noise ratio of the obtained spectra allows the measurement
of four line electron temperatures: T$_e$([O{\sc iii}]), T$_e$([S{\sc iii}]),
T$_e$([O{\sc ii}]) and T$_e$([S{\sc ii}]), for all the objects of the sample
with the addition of T$_e$([N{\sc ii}]) for one of the objects. These
measurements and a careful and realistic treatment of the observational errors
yield total oxygen abundances with accuracies between 7 and 12\%. The
fractional error is as low as 1\% for the ionic O$ ^{2+} $/H$ ^{+} $ ratio due
to the small errors associated with the measurement of  the strong nebular
lines of [O{\sc iii}] and the derived T$_e$([O{\sc iii}]), but increases to up
to 30\% for the O$^{+}$/H$^{+}$ ratio. The accuracies are lower in the case of
the abundances of sulphur (of the order of 25\% for S$^+$ and 15\% for
S$^{2+}$) due to the presence of larger observational errors both in the
measured line fluxes and the derived electron temperatures. The error for the
total abundance of sulphur  is also larger than in the case of oxygen (between
15\% and 20\%) due to the uncertainties in the ionisation correction factors.

This is in contrast with the unrealistically small errors quoted for line
temperatures other than T$_e$([O{\sc iii}]) in the literature, in part due to the
commonly assumed methodology of deriving them from the measured T$_e$([O{\sc
    iii}]) through a theoretical relation. These relations are found from
photoionization model sequences and no uncertainty is attached to them
although large scatter is found when observed values are plotted; usually the
line temperatures obtained in this way  carry only the observational error
found for the T$_e$([O{\sc iii}]) measurement and does not include the
observed scatter, thus heavily understimating the errors in the derived
temperature.

In fact, no clear relation is found between T$_e$([O{\sc iii}]) and
T$_e$([O{\sc ii}]) for the existing sample of objects confirming our previous
results. A comparison between measured and model derived T$_e$([O{\sc ii}])
shows than, in general, model predictions overestimate this temperature and
hence underestimate the O$^+$/H$^+$ ratio. This, though not very important for
high excitation objects, could be of some concern for lower excitation ones
for which total O/H abundances could be underestimated by up to 0.2 dex. It is 
worth noting that the objects observed with double-arm spectrographs,
therefore implying simultaneous and spatially coincident observations over the
whole spectral range,  show less scatter in the T$_e$([O{\sc
    iii}])\,-\,T$_e$([O{\sc ii}] plane clustering around the N$_e$\,=\,100
cm$^{-3}$ photo-ionisation model sequence. On the other hand, this small
scatter could partially be due to the small range of temperatures shown by
these objects due to possible selection effects. This small temperature range
does not allow either to investigate the metallicity effects found in the
ralations between the various line temperatures in recent photo-ionisation
models by Izotov et al.\ (2006).  

Also, the observed objects, as well as those in Paper I, though showing  Ne/O
and Ar/O relative abundances typical of those found for a large \HII\ galaxy
sample (P\'erez-Montero et al.\ 2007), show higher than typical N/O abundance
ratios that would be even higher if the [O{\sc ii}] temperatures would be
found from photo-ionisation models. We therefore conclude that approach of
deriving the O$^+$ temperature from the O$^{2+}$ one should be discouraged if
an accurate abundance derivation is sought. 

These issues could be addressed by re-observing the objects in Table
\ref{temp} , which cover an ample range in temperatures and metal content,
with double arm spectrographs. This sample should be further extended  to
obtain a self consistent sample of about 50 objects with high S/N and
excellent spectrophotometry covering simultaneously from 3600 to 9900\,\AA\
This simple and easily feasible project would provide important scientific
return in the form of critical tests of photoionisation models. 







The O$^{+} $/O$^{2+} $ and S$^{+} $/S$ ^{2+} $ ratios for all the observed
galaxies, except one,  cluster around a value of the ``softness parameter"
$\eta$ of 1 implying high values of the stellar ionising temperature. For the
discrepant object, showing a much lower value of $\eta$,  the intensity of the
[O{\sc ii}]\,$\lambda\lambda$\,7319,25\,\AA\ lines are affected by atmospheric
absorption lines. When the observational counterpart of the ionic ratios is
used, this object shows a ionisation structure similar to the rest of the
observed ones. This simple exercise shows the potential of checking for
consistency in both the $\eta$ and $\eta$' plots in order to test if a given
assumed ionisation structure is adequate. In fact, these consistency checks
show that the stellar ionising temperatures found for the observed \HII\
galaxies using the ionisation structure predicted by state of the art
ionisation models result too low when compared to those implied by the
corresponding observed emission line ratios.  Therefore, metallicity
calibrations based on abundances derived according to this conventional method
are probably bound to provide metallicities which are systematically too high
and should be revised.

\section*{Acknowledgements}


We wish to express our gratitude to Fabian Rosales for calculating the ionic 
He abundances for our objects using Porter's Helium emissivities.
We are pleased to thank the staff at Calar Alto, and especially Felipe Hoyo,
for their assistance  during the observations. We also thank the Time
Allocation Committee for awarding observing time and an anonymous referee
for her/his careful and constructive revision of the manuscript. 

Funding for the creation and distribution of the SDSS Archive has been
provided by the Alfred P. Sloan Foundation, the Participating Institutions,
the National Aeronautics and Space Administration, the National Science
Foundation, the US Department of Energy, the Japanese Monbukagakusho, and the
Max Planck Society. The SDSS Web site is http://www.sdss.org.

The SDSS is managed by the ARC for the Participating Institutions. The
Participating Institutions are the University of Chicago, Fermilab, the
Institute for Advanced Study, the Japan Participation Group, The Johns Hopkins
University, the Korean Scientist Group, Los Alamos National Laboratory, the
Max Planck Institute for Astronomy (MPIA), the Max Planck Institute for
Astrophysics (MPA), New Mexico State University, the University of Pittsburgh,
the University of Portsmouth, Princeton University, the United States Naval
Observatory, and the University of Washington.

This research has made use of the NASA/IPAC Extragalactic Database (NED) which
is operated by the Jet Propulsion Laboratory, California Institute of
Technology, under contract with the National Aeronautics and Space
Administration and of the SIMBAD database, operated at CDS, Strasbourg,
France.

This work has been partially supported by DGICYT grant AYA-2004-02860-C03. GH
and MC acknowledge support from the Spanish MEC through FPU grants AP2003-1821
and AP2004-0977. AID acknowledges support from  the Spanish MEC through a
sabbatical grant PR2006-0049. Also, partial support from the Comunidad de
Madrid under grant S-0505/ESP/000237 (ASTROCAM) is acknowledged. PM
acknowledges support from CNRS-INSU (France) and its Programme National
Galaxies. Support from the Mexican Research Council (CONACYT) through grant
19847-F is acknowledged by ET and RT. We thank the hospitality of the
Institute of Astronomy of Cambridge where part of this paper was developed. 
GH and MC also thank the hospitality of the INAOE and the Laboratoire
d'Astrophysique de Toulouse-Tarbes. 

When we mentioned to Bernard Pagel the title of this paper he said, with his
characteristic cheeky grin: ``precision abundance? but that's an oxymoron. It
will be nice if you can do it". Dear Bernard, you are sadly missed; we
dedicate this work to your memory.

\bibliographystyle{mn2e}
\bibliography{caha}
\end{document}

%% file: tables/ratios-pap.tex
 \begin{longtable}{lc|ccc|cccc|ccc}
 \caption{Relative reddening corrected line intensities [$F(H\beta)$=$I(H\beta)$=10000]\label{ratiostot 1}} \\
 \hline
  &  & \multicolumn{3}{c|}{SDSS J1455} & \multicolumn{4}{c|}{SDSS J1509} & \multicolumn{3}{c}{SDSS J1528} \\
 \multicolumn{1}{c}{$\lambda$ ({\AA})} & f($\lambda$) & -EW  & $I(\lambda)$ &  $I(\lambda)^\dag$ & -EW  & $I(\lambda)$ &   $I(\lambda)^\dag$ &  $I(\lambda)^\ddag$ & -EW  & $I(\lambda)$ &  $I(\lambda)^\ddag$  \\
  & & (\AA) & & & (\AA) & &  &  & (\AA) & &  \\
 \hline
 \endfirsthead
 
 \multicolumn{ 12}{l}{\small Table~\ref{ratiostot 1} continued}
  \\
 \hline
  &  & \multicolumn{3}{c|}{SDSS J1455} & \multicolumn{4}{c|}{SDSS J1509} & \multicolumn{3}{c}{SDSS J1528} \\
 \multicolumn{1}{c}{$\lambda$ ({\AA})} & f($\lambda$) & -EW  & $I(\lambda)$ &  $I(\lambda)^\dag$ & -EW  & $I(\lambda)$ &  $I(\lambda)^\dag$ &  $I(\lambda)^\ddag$ & -EW  & $I(\lambda)$ &  $I(\lambda)^\ddag$  \\
  & & (\AA) & & & (\AA) & &  &  & (\AA) & &  \\
 \hline
 \endhead
 
 \hline\multicolumn{ 12}{r}{\small\sl continued on next page}
 \endfoot
 
 \hline
 \noalign {\noindent$^a$\,possibly blend with an unknown line; $^b$\,[O{\sc ii}]\,$\lambda\lambda$\,3726\,+\,3729; $^c$\,[Fe{\sc iii}]\,$\lambda\lambda$\,4986\,+\,4987; $^d$\,[O{\sc ii}]\,$\lambda\lambda$\,7318\,+\,7320; $^e$\,[O{\sc ii}]\,$\lambda\lambda$\,7330\,+\,7331. $^\dag$ from \citet{2006A&A...448..955I}; they gave $^\gimel$\,[S{\sc ii}]\,$\lambda\lambda$\,6717\,+\,6731 and $^\beth$\,[O{\sc ii}]\,$\lambda\lambda$\,7319\,+\,7330. $^\ddag$ from \citet{1992A&A...253..349P}.}
 \endlastfoot

 3727 [O{\sc ii}]$^b$        &   0.271   &   91.6 & 11154$\pm$164 & 10920$\pm$400 & 131.7 & 15318$\pm$182 &  20970$\pm$760 & 18620$\pm$1800 &  177.4 & 22882$\pm$294 & 22390$\pm$2160  \\
 3734 H13                    &   0.270   &    1.1 &   130$\pm$31  &   ---         &   3.5 &   278$\pm$56  &    ---         &   ---          &    2.8 &   295$\pm$56  &   ---  \\
 3750 H12                    &   0.266   &    1.8 &   204$\pm$48  &   270$\pm$70  &   2.1 &   203$\pm$48  &    300$\pm$80  &   ---          &    2.9 &   315$\pm$64  &   ---  \\
 3770 H11                    &   0.261   &    1.9 &   222$\pm$27  &   470$\pm$70  &   2.4 &   236$\pm$38  &    460$\pm$80  &   ---          &    5.0 &   510$\pm$53  &   ---  \\
 3798 H10                    &   0.254   &    3.7 &   407$\pm$39  &   520$\pm$80  &   4.7 &   410$\pm$60  &    480$\pm$90  &   ---          &    7.6 &   709$\pm$70  &   ---  \\
 3820 He{\sc i}              &   0.249   &    0.6 &    81$\pm$16  &   ---         &   0.6 &    66$\pm$15  &    ---         &   ---          &   ---  &     ---       &   ---  \\
 3835 H9                     &   0.246   &    5.0 &   574$\pm$68  &   750$\pm$70  &   6.3 &   537$\pm$59  &    710$\pm$90  &   ---          &    7.6 &   820$\pm$58  &   ---  \\
 3868 [Ne{\sc iii}]          &   0.238   &   37.1 &  4792$\pm$105 &  5230$\pm$200 &  32.5 &  3501$\pm$109 &   4120$\pm$180 &  3980$\pm$380  &   39.0 &  4717$\pm$180 & 5370$\pm$520  \\
 3889 He{\sc i}+H8           &   0.233   &   16.1 &  1816$\pm$99  &   ---         &  16.4 &  1622$\pm$81  &    ---         &   ---          &   21.1 &  2151$\pm$74  & ---  \\
 3968 [Ne{\sc iii}]+H7       &   0.216   &   27.6 &  2996$\pm$71  &   ---         &  20.6 &  2262$\pm$113 &    ---         &  2570$\pm$250  &   35.3 &  3231$\pm$143 & 3020$\pm$290  \\
 4026 [N{\sc ii}]+He{\sc i}  &   0.203   &    1.2 &   148$\pm$22  &   ---         &   1.4 &   157$\pm$19  &    ---         &   ---          &    1.9 &   208$\pm$55  & ---  \\
 4068 [S{\sc ii}]            &   0.195   &    1.2 &   158$\pm$11  &   ---         &   1.7 &   199$\pm$19  &    ---         &   ---          &    2.0 &   214$\pm$18  & ---  \\
 4102 H$\delta$              &   0.188   &   24.4 &  2580$\pm$57  &  2590$\pm$120 &  26.8 &  2420$\pm$61  &   2730$\pm$140 &  2820$\pm$270  &   41.6 &  2824$\pm$94  & 2820$\pm$270  \\
 4340 H$\gamma$              &   0.142   &   48.8 &  4562$\pm$70  &  4710$\pm$180 &  54.1 &  4395$\pm$76  &   4860$\pm$200 &  4680$\pm$450  &   62.4 &  4773$\pm$76  & 4790$\pm$460  \\
 4363 [O{\sc iii}]           &   0.138   &    9.8 &  1022$\pm$35  &   990$\pm$60  &   4.2 &   420$\pm$21  &    500$\pm$60  &   550$\pm$80   &    5.4 &   500$\pm$22  & ---  \\
 4471 He{\sc i}              &   0.106   &    4.1 &   380$\pm$23  &   ---         &   5.1 &   454$\pm$32  &    ---         &   ---          &    5.8 &   465$\pm$28  & ---  \\
 4658 [Fe{\sc iii}]          &   0.053   &    0.3 &    28$\pm$4   &   110$\pm$40  &   1.1 &   106$\pm$12  &    110$\pm$50  &   ---          &    1.6 &   125$\pm$17  & ---  \\
 4686 He{\sc ii}             &   0.045   &    0.8 &    75$\pm$9   &   ---         &  ---  &        ---    &    ---         &   ---          &   ---  &       ---     & ---  \\
 4713 [Ar{\sc iv}]+He{\sc i} &   0.038   &    2.7 &   223$\pm$17  &   ---         &  ---  &        ---    &    ---         &   ---          &    1.0 &    79$\pm$24  & ---  \\
 4740 [Ar{\sc iv}]           &   0.031   &    1.2 &   103$\pm$14  &   110$\pm$40  &  ---  &        ---    &    ---         &   ---          &   ---  &        ---    & ---  \\
 4861 H$\beta$               &   0.000   &  132.8 & 10000$\pm$67  & 10000$\pm$340 & 123.4 & 10000$\pm$74  &  10000$\pm$350 & 10000$\pm$960  &  171.4 & 10000$\pm$116 & 10000$\pm$960  \\
 4881 [Fe{\sc iii}]          &  -0.005   &    0.3 &    25$\pm$6   &   ---         &  ---  &       ---     &    ---         &   ---          &   ---  &        ---    & ---  \\
 4921 He{\sc i}              &  -0.014   &    1.3 &    96$\pm$11  &   ---         &   1.1 &    99$\pm$12  &    ---         &   ---          &    1.9 &   114$\pm$17  & ---  \\
 4959 [O{\sc iii}]           &  -0.024   &  254.3 & 20456$\pm$134 & 21040$\pm$680 & 190.7 & 16751$\pm$148 &  16610$\pm$560 & 16600$\pm$1600 &  256.2 & 16557$\pm$173 & 15850$\pm$1530  \\
 4986 [Fe{\sc iii}]$^c$      &  -0.030   &    0.9 &    71$\pm$18  &    90$\pm$40  &   1.3 &   114$\pm$16  &    140$\pm$40  &   ---          &    2.0 &   129$\pm$31  & ---  \\
 5007 [O{\sc iii}]           &  -0.035   &  782.2 & 61355$\pm$336 &   ---         & 571.3 & 49942$\pm$153 &    ---         & 50120$\pm$4840 &  764.2 & 48932$\pm$292 & 47860$\pm$4620  \\
 5015 He{\sc i}              &  -0.037   &    3.1 &   234$\pm$20  &   ---         &   2.9 &   249$\pm$26  &    ---         &   ---          &    5.4 &   324$\pm$32  & ---  \\
 5199 [N{\sc i}]             &  -0.078   &    0.7 &    48$\pm$10  &   ---         &   1.4 &   105$\pm$20  &    ---         &   ---          &   ---  &        ---    & ---  \\
 5270 [Fe{\sc iii}]$^a$      &  -0.094   &    0.4 &    26$\pm$5   &   ---         &  ---  &        ---    &    ---         &   ---          &   ---  &        ---    & ---  \\
 5755 [N{\sc ii}]            &  -0.188   &   ---  &       ---     &   ---         &  ---  &        ---    &    ---         &   ---          &   ---  &        ---    & ---  \\
 5876 He{\sc i}              &  -0.209   &   22.4 &  1140$\pm$35  &  1090$\pm$60  &  21.6 &  1268$\pm$59  &   1120$\pm$70  &  1230$\pm$120  &   28.5 &  1227$\pm$36  & ---  \\
 6300 [O{\sc i}]             &  -0.276   &    5.5 &   257$\pm$11  &   210$\pm$30  &   7.7 &   452$\pm$26  &    400$\pm$40  &   ---          &   11.3 &   438$\pm$18  & 1200$\pm$180  \\
 6312 [S{\sc iii}]           &  -0.278   &    3.6 &   160$\pm$6   &   150$\pm$30  &   2.8 &   164$\pm$9   &    160$\pm$40  &   ---          &    4.3 &   168$\pm$10  & ---  \\
 6364 [O{\sc i}]             &  -0.285   &    2.0 &    95$\pm$19  &   ---         &   2.6 &   146$\pm$19  &    ---         &   ---          &    3.8 &   145$\pm$12  & ---  \\
 6548 [N{\sc ii}]            &  -0.311   &    6.2 &   273$\pm$15  &   ---         &   9.6 &   541$\pm$28  &    ---         &   ---          &   19.4 &   721$\pm$45  & ---  \\
 6563 H$\alpha$              &  -0.313   &  646.8 & 27756$\pm$183 & 28190$\pm$980 & 522.6 & 27969$\pm$141 &  28570$\pm$1020& 28180$\pm$2720 &  823.2 & 28888$\pm$169 & 29510$\pm$2850  \\
 6584 [N{\sc ii}]            &  -0.316   &   18.2 &   792$\pm$19  &   750$\pm$50  &  25.4 &  1387$\pm$37  &   1370$\pm$70  &  1910$\pm$180  &   54.0 &  1959$\pm$43  & 1450$\pm$140  \\
 6678 He{\sc i}              &  -0.329   &    9.7 &   350$\pm$38  &   ---         &   7.2 &   392$\pm$15  &    ---         &   ---          &   10.3 &   350$\pm$10  & ---  \\
 6717 [S{\sc ii}]            &  -0.334   &   21.9 &  1000$\pm$26  &  1720$\pm$70$^\gimel$  &  35.8 &  1966$\pm$48 &  2880$\pm$110$^\gimel$  &  2140$\pm$210  &   57.1 &  1923$\pm$75 & 1320$\pm$200  \\
 6731 [S{\sc ii}]            &  -0.336   &   14.5 &   788$\pm$22  &   ---         &  27.1 &  1488$\pm$40  &    ---         &  1780$\pm$170  &   43.7 &  1422$\pm$58  & 1120$\pm$170  \\
 7065 He{\sc i}              &  -0.377   &    8.3 &   293$\pm$12  &   ---         &   6.0 &   302$\pm$19  &    ---         &   ---          &    9.6 &   297$\pm$17  & ---  \\
 7136 [Ar{\sc iii}]          &  -0.385   &   18.5 &   662$\pm$30  &   590$\pm$40  &  19.1 &   982$\pm$29  &   790$\pm$50   &  1200$\pm$120  &   19.1 &        ---    & ---  \\
 7281 He{\sc i}$^a$          &  -0.402   &    2.1 &    78$\pm$9   &   ---         &  ---  &        ---    &    ---         &   ---          &    2.0 &    60$\pm$5   & ---  \\
 7319 [O{\sc ii}]$^d$        &  -0.406   &    5.3 &   181$\pm$9   & 290$\pm$40$^\beth$ &   4.8 &   203$\pm$9 &   410$\pm$50$^\beth$ & ---         &   10.0 &   296$\pm$17 & ---  \\
 7330 [O{\sc ii}]$^e$        &  -0.407   &    4.1 &   142$\pm$8   &   ---         &   4.0 &   167$\pm$8   &    ---         &   ---          &    8.2 &   238$\pm$12  & ---  \\
 7751 [Ar{\sc iii}]          &  -0.451   &    4.7 &   154$\pm$9   &   ---         &   5.2 &   242$\pm$16  &    ---         &   ---          &    8.1 &   210$\pm$13  & ---  \\
 8446 O{\sc i}               &  -0.513   &    1.7 &    46$\pm$8   &   ---         &   1.6 &    68$\pm$10  &    ---         &   ---          &   ---  &        ---    & ---  \\
 8503 P16                    &  -0.518   &    2.6 &    59$\pm$10  &   ---         &  ---  &        ---    &    ---         &   ---          &   ---  &        ---    & ---  \\
 8546 P15                    &  -0.521   &   ---  &       ---     &   ---         &  ---  &        ---    &    ---         &   ---          &   ---  &        ---    & ---  \\
 8599 P14                    &  -0.525   &   ---  &       ---     &   ---         &   3.5 &   100$\pm$33  &    ---         &   ---          &    4.8 &    72$\pm$13  & ---  \\
 8665 P13                    &  -0.531   &    4.9 &   105$\pm$18  &   ---         &   4.0 &   118$\pm$42  &    ---         &   ---          &    5.7 &   102$\pm$15  & ---  \\
 8751 P12                    &  -0.537   &    7.9 &   147$\pm$17  &   ---         &   5.0 &   154$\pm$47  &    ---         &   ---          &   12.0 &   145$\pm$16  & ---  \\
 8865 P11                    &  -0.546   &    8.3 &   181$\pm$22  &   ---         &  11.4 &   297$\pm$65  &    ---         &   ---          &   24.5 &   199$\pm$43  & ---  \\
 9014 P10                    &  -0.557   &   11.4 &   269$\pm$32  &   ---         &  33.6 &   447$\pm$93  &    ---         &   ---          &   ---  &        ---    & ---  \\
 9069 [S{\sc iii}]           &  -0.561   &   50.8 &  1154$\pm$59  &   ---         &  82.1 &  2546$\pm$120 &    ---         &   ---          &   88.1 &  1688$\pm$127 & ---  \\
 9229 P9                     &  -0.572   &   33.0 &   400$\pm$50  &   ---         &  27.2 &   595$\pm$98  &    ---         &   ---          &   31.9 &   292$\pm$86  & ---  \\
 9532 [S{\sc iii}]           &  -0.592   &  140.1 &  3278$\pm$148 &   ---         & 176.3 &  5181$\pm$256 &    ---         &   ---          &  506.6 &  4049$\pm$289 & ---  \\
 9547 P8                     &  -0.593   &   21.7 &   455$\pm$69  &   ---         &  43.5 &   876$\pm$129 &    ---         &   ---          &   ---  &        ---    & ---  \\
\multicolumn{2}{l|}{I(H$\beta$)(erg\,seg$^{-1}$\,cm$^{-2}$)}    & \multicolumn{2}{c}{1.49\,$\times$\,10$^{-14}$}  &   & \multicolumn{2}{c}{1.35\,$\times$\,10$^{-14}$}  &   &   & \multicolumn{2}{c}{1.73\,$\times$\,10$^{-14}$}  & 
  \\
\multicolumn{2}{l|}{c(H$\beta$)}     & \multicolumn{2}{c}{ 0.13$\pm$0.01 }    &  0.05   & \multicolumn{2}{c}{ 0.07$\pm$0.01 }    & 0.08   &  0.05  & \multicolumn{2}{c}{ 0.04$\pm$0.01 }   & 0.15
  \\

 \end{longtable}

\setcounter{table}{4}

 \begin{longtable}{lc|cccc|ccc|cc}
 \caption{{\it (cont.)} Relative reddening corrected line intensities [$F(H\beta)$=$I(H\beta)$=10000] \label{ratiostot 2}} \\
 \hline
  &  & \multicolumn{4}{c|}{SDSS J1540} & \multicolumn{3}{c|}{SDSS J1616} & \multicolumn{1}{c}{SDSS J1657} \\
 \multicolumn{1}{c}{$\lambda$ ({\AA})} & f($\lambda$) & -EW  & $I(\lambda)$ &  $I(\lambda)^\dag$ &  $I(\lambda)^\S$ & -EW  & $I(\lambda)$ &  $I(\lambda)^\dag$  & -EW  & $I(\lambda)$ \\
  & & (\AA) & & &  &  (\AA) & &  &  (\AA) &  \\
 \hline
 \endfirsthead
 
 \multicolumn{ 11}{l}{\small Table~\ref{ratiostot 2} continued}
  \\
 \hline
  &  & \multicolumn{4}{c|}{SDSS J1540} & \multicolumn{3}{c|}{SDSS J1616} & \multicolumn{2}{c}{SDSS J1657} \\
 \multicolumn{1}{c}{$\lambda$ ({\AA})} & f($\lambda$) & -EW  & $I(\lambda)$ &  $I(\lambda)^\dag$ &  $I(\lambda)^\S$ & -EW  & $I(\lambda)$ &  $I(\lambda)^\dag$  & -EW  & $I(\lambda)$ \\
  & & (\AA) & & &  &  (\AA) & &  &  (\AA) &  \\
 \hline
 \endhead
 
 \hline\multicolumn{ 11}{r}{\small\sl continued on next page}
 \endfoot
 
 \hline
 \noalign {\noindent$^a$\,possibly blend with an unknown line; $^b$\,[O{\sc ii}]\,$\lambda\lambda$\,3726\,+\,3729; $^c$\,[Fe{\sc iii}]\,$\lambda\lambda$\,4986\,+\,4987; $^d$\,[O{\sc ii}]\,$\lambda\lambda$\,7318\,+\,7320; $^e$\,[O{\sc ii}]\,$\lambda\lambda$\,7330\,+\,7331. $^\dag$ from \citet{2006A&A...448..955I}; they gave $^\gimel$\,[S{\sc ii}]\,$\lambda\lambda$\,6717\,+\,6731 and $^\beth$\,[O{\sc ii}]\,$\lambda\lambda$\,7319\,+\,7330. $^\S$ from \citet{2004ApJS..153..429K}; they gave $^\varkappa$\,[O{\sc ii}]\,$\lambda\lambda$\,7319\,+\,7330.}
 \endlastfoot

 3727 [O{\sc ii}]$^b$       &   0.271   &  232.8 &  21793$\pm$256 &   ---         &   ---         &   33.5 &  8491$\pm$196 &   ---         &  120.1 & 18832$\pm$230  \\
 3734 H13                   &   0.270   &    2.6 &    272$\pm$48  &   ---         &   ---         &   ---  &       ---     &   ---         &   ---  &       ---      \\
 3750 H12                   &   0.266   &   ---  &       ---      &   ---         &   ---         &    1.8 &   377$\pm$101 &   ---         &    1.9 &   232$\pm$47   \\
 3770 H11                   &   0.261   &    2.7 &    314$\pm$61  &   ---         &   ---         &    1.5 &   357$\pm$113 &   ---         &    2.3 &   293$\pm$40   \\
 3798 H10                   &   0.254   &    4.8 &    581$\pm$93  &   760$\pm$90  &   810$\pm$280 &    2.1 &   447$\pm$92  &   ---         &    4.1 &   500$\pm$68   \\
 3820 He{\sc i}             &   0.249   &    1.5 &    166$\pm$23  &   ---         &   ---         &   ---  &       ---     &   ---         &   ---  &       ---      \\
 3835 H9                    &   0.246   &    6.2 &    678$\pm$85  &   920$\pm$90  &   940$\pm$260 &    3.1 &   638$\pm$90  &   790$\pm$110 &    7.1 &   780$\pm$93   \\
 3868 [Ne{\sc iii}]         &   0.238   &   15.2 &   2142$\pm$82  &  1860$\pm$90  &   ---         &   17.2 &  4105$\pm$166 &  5450$\pm$230 &   23.1 &  3262$\pm$132  \\
 3889 He{\sc i}+H8          &   0.233   &   11.6 &   1445$\pm$113 &   ---         &   ---         &    7.3 &  1591$\pm$112 &   ---         &   14.0 &  1826$\pm$95   \\
 3968 [Ne{\sc iii}]+H7      &   0.216   &   21.9 &   2309$\pm$144 &   ---         &   ---         &   11.7 &  2563$\pm$125 &   ---         &   22.4 &  2456$\pm$121  \\
 4026 [N{\sc ii}]+He{\sc i} &   0.203   &   20.6 &   2505$\pm$78  &   ---         &   ---         &    0.4 &    94$\pm$15  &   ---         &    1.1 &   155$\pm$16   \\
 4068 [S{\sc ii}]           &   0.195   &    1.7 &    231$\pm$18  &   ---         &   ---         &    0.5 &   108$\pm$10  &   ---         &    1.4 &   198$\pm$15   \\
 4102 H$\delta$             &   0.188   &   23.7 &   2589$\pm$63  &  2570$\pm$120 &  2590$\pm$130 &   13.3 &  2530$\pm$67  & 2800$\pm$150  &   20.9 &  2432$\pm$65   \\
 4340 H$\gamma$             &   0.142   &   44.2 &   4753$\pm$61  &  4670$\pm$170 &  4640$\pm$100 &   26.7 &  4597$\pm$89  & 5000$\pm$200  &   43.1 &  4417$\pm$97   \\
 4363 [O{\sc iii}]          &   0.138   &    4.5 &    291$\pm$17  &   220$\pm$40  &   220$\pm$50  &    4.5 &   851$\pm$26  &  980$\pm$80   &    4.5 &   524$\pm$24   \\
 4471 He{\sc i}             &   0.106   &    4.7 &    463$\pm$33  &   ---         &   ---         &    2.5 &   404$\pm$32  &   ---         &    4.2 &   443$\pm$33   \\
 4658 [Fe{\sc iii}]         &   0.053   &    0.7 &     78$\pm$14  &   ---         &   ---         &   ---  &       ---     &   ---         &    1.0 &   107$\pm$16   \\
 4686 He{\sc ii}            &   0.045   &   ---  &       ---      &   ---         &   ---         &    2.1 &   329$\pm$43  &   310$\pm$60  &    1.2 &   126$\pm$14   \\
 4713 [Ar{\sc iv}]+He{\sc i}&   0.038   &   ---  &       ---      &   ---         &   ---         &   ---  &       ---     &   ---         &   ---  &       ---      \\
 4740 [Ar{\sc iv}]          &   0.031   &   ---  &       ---      &   ---         &   ---         &    0.5 &    69$\pm$22  &   ---         &   ---  &       ---      \\
 4861 H$\beta$              &   0.000   &  122.4 & 10000$\pm$76   & 10000$\pm$340 & 10000$\pm$100 &   83.0 & 10000$\pm$96  & 10000$\pm$350 &  117.8 & 10000$\pm$79   \\
 4881 [Fe{\sc iii}]         &  -0.005   &   ---  &       ---      &   ---         &   ---         &   ---  &       ---     &   ---         &   ---  &       ---      \\
 4921 He{\sc i}             &  -0.014   &    1.2 &   109$\pm$15   &   ---         &   ---         &    0.9 &   116$\pm$13  &   ---         &    0.8 &    75$\pm$14   \\
 4959 [O{\sc iii}]          &  -0.024   &  114.4 & 10480$\pm$75   &  9920$\pm$340 &  9830$\pm$90  &  156.5 & 20492$\pm$147 & 20380$\pm$680 &  152.5 & 14333$\pm$127  \\
 4986 [Fe{\sc iii}]$^c$     &  -0.030   &    0.9 &    84$\pm$12   &   180$\pm$40  &   ---         &   ---  &       ---     &   ---         &    1.4 &   135$\pm$28   \\
 5007 [O{\sc iii}]          &  -0.035   &  348.3 & 30942$\pm$188  &   ---         & 29000$\pm$230 &  480.7 & 61516$\pm$371 &   ---         &  455.1 & 43082$\pm$240  \\
 5015 He{\sc i}             &  -0.037   &    2.7 &   235$\pm$16   &   ---         &   ---         &    1.8 &   225$\pm$26  &   ---         &    2.4 &   222$\pm$23   \\
 5199 [N{\sc i}]            &  -0.078   &    2.0 &   151$\pm$27   &   ---         &   ---         &   ---  &       ---     &   ---         &    2.0 &   157$\pm$26   \\
 5270 [Fe{\sc iii}]$^a$     &  -0.094   &   ---  &       ---      &   ---         &   ---         &   ---  &       ---     &   ---         &   ---  &       ---      \\
 5755 [N{\sc ii}]           &  -0.188   &   ---  &       ---      &   ---         &   ---         &   ---  &       ---     &   ---         &   ---  &       ---      \\
 5876 He{\sc i}             &  -0.209   &   19.2 &  1155$\pm$36   &  1090$\pm$60  &   ---         &   13.7 &  1063$\pm$75  &   970$\pm$60  &   18.9 &  1116$\pm$44   \\
 6300 [O{\sc i}]            &  -0.276   &    6.4 &   335$\pm$14   &   330$\pm$30  &   ---         &    1.7 &   104$\pm$18  &   170$\pm$40  &    8.1 &   438$\pm$16   \\
 6312 [S{\sc iii}]          &  -0.278   &    2.6 &   139$\pm$10   &   110$\pm$30  &   ---         &    3.2 &   186$\pm$11  &   200$\pm$40  &    3.7 &   201$\pm$9    \\
 6364 [O{\sc i}]            &  -0.285   &    2.3 &   122$\pm$17   &   ---         &   ---         &    0.9 &    51$\pm$13  &   ---         &    2.8 &   152$\pm$18   \\
 6548 [N{\sc ii}]           &  -0.311   &   13.9 &   735$\pm$34   &   ---         &   ---         &    2.8 &   160$\pm$13  &   ---         &    9.4 &   464$\pm$23   \\
 6563 H$\alpha$             &  -0.313   &  581.9 & 28626$\pm$162  & 28770$\pm$1010& 28730$\pm$220 &  517.2 & 27892$\pm$145 & 28160$\pm$1000&  571.3 & 27772$\pm$153  \\
 6584 [N{\sc ii}]           &  -0.316   &   42.2 &  2116$\pm$66   &  1960$\pm$90  &   ---         &    8.0 &   434$\pm$23  &   380$\pm$40  &   28.8 &  1428$\pm$47   \\
 6678 He{\sc i}             &  -0.329   &    6.8 &   316$\pm$15   &   ---         &   ---         &    5.9 &   296$\pm$18  &   ---         &    6.7 &   315$\pm$18   \\
 6717 [S{\sc ii}]           &  -0.334   &   52.6 &  2609$\pm$52   &  4330$\pm$130$^\gimel$ & 2450$\pm$30 &   15.3 &   774$\pm$23  &  1400$\pm$70$^\gimel$  &   47.4 &  2207$\pm$57  \\
 6731 [S{\sc ii}]           &  -0.336   &   41.7 &  1919$\pm$50   &   ---         &  1800$\pm$30  &   11.9 &   581$\pm$22  &   ---         &   32.2 &  1598$\pm$43   \\
 7065 He{\sc i}             &  -0.377   &    4.5 &   200$\pm$9    &   ---         &   ---         &    5.1 &   229$\pm$14  &   ---         &    5.6 &   235$\pm$10   \\
 7136 [Ar{\sc iii}]         &  -0.385   &   23.2 &   892$\pm$52   &   880$\pm$50  &   ---         &   17.4 &   736$\pm$44  &   690$\pm$50  &   16.4 &   717$\pm$26   \\
 7281 He{\sc i}$^a$         &  -0.402   &    3.1 &   115$\pm$10   &   ---         &   ---         &   ---  &       ---     &   ---         &    0.9 &    41$\pm$7    \\
 7319 [O{\sc ii}]$^d$       &  -0.406   &    6.3 &   273$\pm$16   &   500$\pm$40$^\beth$  &  460$\pm$40$^\varkappa$ &    3.4 &   138$\pm$11  &  260$\pm$40$^\beth$  &   12.3 &   302$\pm$17  \\
 7330 [O{\sc ii}]$^e$       &  -0.407   &    5.0 &   216$\pm$12   &   ---         &   ---         &    2.2 &    90$\pm$7   &   ---         &    8.8 &   211$\pm$14   \\
 7751 [Ar{\sc iii}]         &  -0.451   &    6.1 &   223$\pm$26   &   ---         &   ---         &    4.5 &   166$\pm$21  &   ---         &    4.6 &   177$\pm$22   \\
 8446 O{\sc i}              &  -0.513   &   ---  &       ---      &   ---         &   ---         &   ---  &       ---     &   ---         &   ---  &       ---      \\
 8503 P16                   &  -0.518   &   ---  &       ---      &   ---         &   ---         &    2.6 &    60$\pm$20  &   ---         &   ---  &       ---      \\
 8546 P15                   &  -0.521   &   ---  &       ---      &   ---         &   ---         &    1.6 &    38$\pm$14  &   ---         &   ---  &       ---      \\
 8599 P14                   &  -0.525   &   ---  &       ---      &   ---         &   ---         &    5.6 &   109$\pm$20  &   ---         &   ---  &       ---      \\
 8665 P13                   &  -0.531   &   ---  &       ---      &   ---         &   ---         &    8.0 &   153$\pm$31  &   ---         &    7.7 &   144$\pm$53   \\
 8751 P12                   &  -0.537   &   ---  &       ---      &   ---         &   ---         &    8.1 &   181$\pm$32  &   ---         &    4.2 &   101$\pm$29   \\
 8865 P11                   &  -0.546   &   ---  &       ---      &   ---         &   ---         &    9.6 &   182$\pm$34  &   ---         &    8.5 &   211$\pm$34   \\
 9014 P10                   &  -0.557   &   12.9 &   237$\pm$68   &   ---         &   ---         &   24.4 &   319$\pm$44  &   ---         &   15.4 &   167$\pm$36   \\
 9069 [S{\sc iii}]          &  -0.561   &   67.2 &  2095$\pm$62   &  2020$\pm$100 &   ---         &   74.3 &  1647$\pm$62  &  1470$\pm$80  &   59.2 &  1400$\pm$99   \\
 9229 P9                    &  -0.572   &   ---  &       ---      &   ---         &   ---         &   14.2 &   272$\pm$51  &   ---         &   16.9 &   263$\pm$47   \\
 9532 [S{\sc iii}]          &  -0.592   &  236.8 &  5331$\pm$331  &   ---         &   ---         &   74.3 &  4011$\pm$153 &   ---         &  157.3 &  3674$\pm$257  \\
 9547 P8                    &  -0.593   &   29.0 &   573$\pm$145  &   ---         &   ---         &   ---  &       ---     &   ---         &   ---  &       ---      \\
I(H$\beta$)(erg\,seg$^{-1}$\,cm$^{-2}$) &   & \multicolumn{2}{c}{0.53\,$\times$\,10$^{-14}$}  &   &   & \multicolumn{2}{c}{1.36\,$\times$\,10$^{-14}$}  &  & \multicolumn{2}{c}{0.63\,$\times$\,10$^{-14}$}
  \\
c(H$\beta$) &     & \multicolumn{2}{c}{0.22$\pm$0.01}    & 0.07   &  0.07  & \multicolumn{2}{c}{0.02$\pm$0.01}     & 0.06  & \multicolumn{2}{c}{0.05$\pm$0.01}
  \\

 \end{longtable}

\setcounter{table}{4}

 \begin{longtable}{lc|cccc}
 \caption{{\it (cont.)} Relative reddening corrected line intensities [$F(H\beta)$=$I(H\beta)$=10000] \label{ratiostot 3}} \\
 \hline
  &  & \multicolumn{4}{c}{SDSS J1729} \\
 \multicolumn{1}{c}{$\lambda$ ({\AA})} & f($\lambda$) & -EW  & $I(\lambda)$ & $I(\lambda)^\dag$ &  $I(\lambda)^\S$ \\
  & & (\AA) &  \\
 \hline
 \endfirsthead
 
 \multicolumn{6}{l}{\small Table~\ref{ratiostot 3} continued}
  \\
 \hline
  &  & \multicolumn{4}{c}{SDSS J1729} \\
 \multicolumn{1}{c}{$\lambda$ ({\AA})} & f($\lambda$) & -EW  & $I(\lambda)$ & $I(\lambda)^\dag$ &  $I(\lambda)^\S$ \\
  & & (\AA) &  \\
 \hline
 \endhead
 
 \hline\multicolumn{6}{r}{\small\sl continued on next page}
 \endfoot
 
 \hline
 \noalign {\noindent$^a$\,possibly blend with an unknown line; $^b$\,[O{\sc ii}]\,$\lambda\lambda$\,3726\,+\,3729; $^c$\,[Fe{\sc iii}]\,$\lambda\lambda$\,4986\,+\,4987; $^d$\,[O{\sc ii}]\,$\lambda\lambda$\,7318\,+\,7320; $^e$\,[O{\sc ii}]\,$\lambda\lambda$\,7330\,+\,7331. $^\dag$ from \citet{2006A&A...448..955I}; they gave $^\gimel$\,[S{\sc ii}]\,$\lambda\lambda$\,6717\,+\,6731 and $^\beth$\,[O{\sc ii}]\,$\lambda\lambda$\,7319\,+\,7330. $^\S$ from \citet{2004ApJS..153..429K}; they gave $^\varkappa$\,[O{\sc ii}]\,$\lambda\lambda$\,7319\,+\,7330.}
 \endlastfoot

 3727 [O{\sc ii}]$^b$       &   0.271   &  135.8 & 17622$\pm$243 &   ---         & --- \\
 3734 H13                   &   0.270   &    1.4 &   175$\pm$54  &   ---         & --- \\
 3750 H12                   &   0.266   &    4.2 &   397$\pm$89  &   370$\pm$70  & --- \\
 3770 H11                   &   0.261   &    7.6 &   713$\pm$88  &   480$\pm$70  & --- \\
 3798 H10                   &   0.254   &    5.2 &   598$\pm$71  &   620$\pm$60  &   370$\pm$200 \\
 3820 He{\sc i}             &   0.249   &   ---  &       ---     &   ---         & --- \\
 3835 H9                    &   0.246   &    7.6 &   721$\pm$64  &   900$\pm$60  &   640$\pm$140 \\
 3868 [Ne{\sc iii}]         &   0.238   &   34.9 &  4787$\pm$172 &  3730$\pm$140 & --- \\
 3889 He{\sc i}+H8          &   0.233   &   16.9 &  2058$\pm$83  &   ---         & --- \\
 3968 [Ne{\sc iii}]+H7      &   0.216   &   24.7 &  2969$\pm$129 &   ---         & --- \\
 4026 [N{\sc ii}]+He{\sc i} &   0.203   &    2.4 &   311$\pm$30  &   ---         & --- \\
 4068 [S{\sc ii}]           &   0.195   &    0.9 &   122$\pm$11  &   ---         & --- \\
 4102 H$\delta$             &   0.188   &   24.9 &  2765$\pm$63  &  2540$\pm$100 &  2430$\pm$100 \\
 4340 H$\gamma$             &   0.142   &   48.7 &  4839$\pm$84  &  4760$\pm$170 &  4750$\pm$80  \\
 4363 [O{\sc iii}]          &   0.138   &    5.9 &   660$\pm$30  &   510$\pm$40  &   510$\pm$40  \\
 4471 He{\sc i}             &   0.106   &    5.2 &   516$\pm$24  &   ---         &   --- \\
 4658 [Fe{\sc iii}]         &   0.053   &    0.8 &    90$\pm$14  &   ---         &   --- \\
 4686 He{\sc ii}            &   0.045   &   ---  &      ---      &   ---         &   --- \\
 4713 [Ar{\sc iv}]+He{\sc i}&   0.038   &    0.7 &    77$\pm$16  &   ---         &   --- \\
 4740 [Ar{\sc iv}]          &   0.031   &    0.4 &    40$\pm$10  &   ---         &   --- \\
 4861 H$\beta$              &   0.000   &  125.6 & 10000$\pm$83  & 10000$\pm$320 & 10000$\pm$80  \\
 4881 [Fe{\sc iii}]         &  -0.005   &   ---  &      ---      &   ---         &   --- \\
 4921 He{\sc i}             &  -0.014   &    1.4 &   122$\pm$12  &   ---         &   --- \\
 4959 [O{\sc iii}]          &  -0.024   &  185.8 & 17097$\pm$146 & 16990$\pm$540 & 17470$\pm$90  \\
 4986 [Fe{\sc iii}]$^c$     &  -0.030   &    0.6 &    56$\pm$18  &    80$\pm$30  &  --- \\
 5007 [O{\sc iii}]          &  -0.035   &  555.4 & 51541$\pm$418 &   ---         & 52320$\pm$300 \\
 5015 He{\sc i}             &  -0.037   &    1.9 &   177$\pm$12  &   ---         &   --- \\
 5199 [N{\sc i}]            &  -0.078   &   ---  &      ---      &   ---         &   --- \\
 5270 [Fe{\sc iii}]$^a$     &  -0.094   &    0.7 &    55$\pm$12  &   ---         &   --- \\
 5755 [N{\sc ii}]           &  -0.188   &    0.8 &    60$\pm$6   &   ---         &   --- \\
 5876 He{\sc i}             &  -0.209   &   17.4 &  1261$\pm$37  &  1130$\pm$50  &   --- \\
 6300 [O{\sc i}]            &  -0.276   &    4.0 &   253$\pm$16  &   260$\pm$30  &   --- \\
 6312 [S{\sc iii}]          &  -0.278   &    2.7 &   176$\pm$11  &   170$\pm$30  &   --- \\
 6364 [O{\sc i}]            &  -0.285   &    1.2 &    78$\pm$11  &   ---         &   --- \\
 6548 [N{\sc ii}]           &  -0.311   &   12.5 &   771$\pm$21  &   ---         &   --- \\
 6563 H$\alpha$             &  -0.313   &  476.9 & 28700$\pm$163 & 28510$\pm$960 & 28530$\pm$150 \\
 6584 [N{\sc ii}]           &  -0.316   &   35.8 &  2189$\pm$52  &  2080$\pm$80  &   --- \\
 6678 He{\sc i}             &  -0.329   &    6.0 &   355$\pm$12  &   ---         &   --- \\
 6717 [S{\sc ii}]           &  -0.334   &   21.7 &  1286$\pm$32  &  2160$\pm$80$^\gimel$  & 1220$\pm$20 \\
 6731 [S{\sc ii}]           &  -0.336   &   17.3 &  1009$\pm$30  &   ---         & 950$\pm$20 \\
 7065 He{\sc i}             &  -0.377   &    5.7 &   265$\pm$18  &   ---         &   --- \\
 7136 [Ar{\sc iii}]         &  -0.385   &   18.0 &   864$\pm$41  &   950$\pm$50  &   --- \\
 7281 He{\sc i}$^a$         &  -0.402   &    1.5 &    75$\pm$11  &   ---         &   --- \\
 7319 [O{\sc ii}]$^d$       &  -0.406   &    4.6 &   233$\pm$8   &   430$\pm$40$^\beth$  & 390$\pm$20$^\varkappa$ \\
 7330 [O{\sc ii}]$^e$       &  -0.407   &    3.9 &   194$\pm$9   &   ---         &   --- \\
 7751 [Ar{\sc iii}]         &  -0.451   &    4.5 &   210$\pm$10  &   ---         &   --- \\
 8446 O{\sc i}              &  -0.513   &   ---  &      ---      &   ---         &   --- \\
 8503 P16                   &  -0.518   &    3.0 &   102$\pm$32  &   ---         &   --- \\
 8546 P15                   &  -0.521   &   ---  &      ---      &   ---         &   --- \\
 8599 P14                   &  -0.525   &    2.9 &    96$\pm$17  &   ---         &   --- \\
 8665 P13                   &  -0.531   &    4.7 &   140$\pm$20  &   ---         &   --- \\
 8751 P12                   &  -0.537   &    7.0 &   205$\pm$35  &   ---         &   --- \\
 8865 P11                   &  -0.546   &   ---  &      ---      &   ---         &   --- \\
 9014 P10                   &  -0.557   &    8.1 &   236$\pm$54  &   ---         &   --- \\
 9069 [S{\sc iii}]          &  -0.561   &   58.5 &  2092$\pm$95  &   ---         &   --- \\
 9229 P9                    &  -0.572   &   19.9 &   400$\pm$64  &   ---         &   --- \\
 9532 [S{\sc iii}]          &  -0.592   &  181.5 &  4718$\pm$250 &   ---         &   --- \\
 9547 P8                    &  -0.593   &   20.3 &   480$\pm$69  &   ---         &   --- \\
     I(H$\beta$)(erg\,seg$^{-1}$\,cm$^{-2}$) &   & \multicolumn{2}{c}{2.40\,$\times$\,10$^{-14}$} \\
     c(H$\beta$) &     & \multicolumn{2}{c}{0.03$\pm$0.01}     & 0.02   & 0.04  \\

 \end{longtable}

%% file: tables/temden-pap.tex
 \begin{table*}
 {\small
 \caption{Electron densities and temperatures for the     observed galaxies.}
 \label{temden}
 \begin{center}
 \begin{tabular}{lcccccc}
 \hline
  & n([S{\sc ii}]) & t$_e$([O{\sc iii}]) &  t$_e$([O{\sc ii}]) & t$_e$([S{\sc iii}]) & t$_e$([S{\sc ii}]) &  t$_e$([N{\sc ii}])
  \\
 \hline

SDSS J1455                &  94 $\pm$ 40 & 1.40 $\pm$ 0.02 & 1.33 $\pm$ 0.07 & 1.37 $\pm$ 0.05 & 1.31 $\pm$ 0.11 &       ---       \\
SDSS J1509                &  85 $\pm$ 45 & 1.09 $\pm$ 0.01 & 1.18 $\pm$ 0.05 & 1.02 $\pm$ 0.04 & 0.89 $\pm$ 0.07 &       ---       \\
SDSS J1528                &  60:         & 1.16 $\pm$ 0.01 & 1.17 $\pm$ 0.05 & 1.21 $\pm$ 0.06 & 0.99 $\pm$ 0.07 &       ---       \\
SDSS J1540                &  47 $\pm$ 38 & 1.13 $\pm$ 0.02 & 1.15 $\pm$ 0.06 & 0.97 $\pm$ 0.04 & 0.85 $\pm$ 0.05 &       ---       \\
SDSS J1616                &  54:         & 1.30 $\pm$ 0.01 & 1.29 $\pm$ 0.09 & 1.29 $\pm$ 0.06 & 1.21 $\pm$ 0.12 &       ---       \\
SDSS J1657                &  30:         & 1.23 $\pm$ 0.02 & 1.33 $\pm$ 0.07 & 1.45 $\pm$ 0.08 & 0.88 $\pm$ 0.05 &       ---       \\
SDSS J1729                & 109 $\pm$ 47 & 1.26 $\pm$ 0.02 & 1.16 $\pm$ 0.04 & 1.13 $\pm$ 0.05 & 0.82 $\pm$ 0.06 & 1.40 $\pm$ 0.09 \\

 \hline
 \multicolumn{7}{l}{densities in $cm^{-3}$                 and temperatures in 10$^4$\,K}
 \end{tabular}
 \end{center}}
 \end{table*}

%% file: tables/abundHe-pap.tex
$He^+/H^+$        & 4471    & 0.079$\pm$0.004 & 0.092$\pm$0.006 & 0.095$\pm$0.005 & 0.094$\pm$0.006 & 0.084$\pm$0.006 & 0.091$\pm$0.007 & 0.106$\pm$0.004
  \\
                  & 5876    & 0.089$\pm$0.002 & 0.094$\pm$0.004 & 0.093$\pm$0.002 & 0.087$\pm$0.002 & 0.082$\pm$0.006 & 0.086$\pm$0.003 & 0.096$\pm$0.003
  \\
                  & 6678    & 0.098$\pm$0.010 & 0.103$\pm$0.004 & 0.094$\pm$0.002 & 0.084$\pm$0.004 & 0.081$\pm$0.005 & 0.086$\pm$0.005 & 0.097$\pm$0.003
  \\
                  & 7065    & 0.104$\pm$0.005 & 0.122$\pm$0.008 & 0.120$\pm$0.007 & 0.082$\pm$0.004 & 0.087$\pm$0.006 & 0.093$\pm$0.005 & 0.099$\pm$0.008
  \\
               & adopted~B99    & 0.090$\pm$0.010 & 0.100$\pm$0.012 & 0.095$\pm$0.013 & 0.086$\pm$0.005 & 0.083$\pm$0.002 & 0.088$\pm$0.003 & 0.098$\pm$0.004
  \\
               & P05        & 0.096$\pm$0.014 &0.102 $\pm$0.011 & 0.096$\pm$0.010& 0.089$\pm$0.010 & 0.084$\pm$0.012 & 0.089$\pm$0.009& 0.100$\pm$0.010
  \\
$He^{2+}/H^+$     & 4686    & 0.0007$\pm$0.0001 &         ---       &         ---       &         ---       & 0.0029$\pm$0.0004 & 0.0011$\pm$0.0001 &         ---      
  \\
\bf{(He/H)}       &         & 0.090$\pm$0.010 &        ---      &        ---      &        ---      & 0.086$\pm$0.002 & 0.089$\pm$0.003 &        ---     
  \\

%% file: tables/ionic-abund-pap.tex
 12+$\log(O^+/H^+)$        &  7.16$\pm$0.09 &  7.48$\pm$0.08 &  7.67$\pm$0.09 &  7.67$\pm$0.09 &  7.07$\pm$0.12 &  7.37$\pm$0.09 &  7.57$\pm$0.07
  \\
 12+$\log(O^{2+}/H^+)$     &  7.87$\pm$0.02 &  8.10$\pm$0.02 &  8.00$\pm$0.02 &  7.84$\pm$0.02 &  7.96$\pm$0.02 &  7.87$\pm$0.02 &  7.92$\pm$0.02
  \\
 12+$\log(S^+/H^+)$        &  5.36$\pm$0.08 &  6.02$\pm$0.10 &  5.89$\pm$0.10 &  6.19$\pm$0.08 &  5.30$\pm$0.10 &  6.07$\pm$0.08 &  5.95$\pm$0.10
  \\
 12+$\log(S^{2+}/H^+)$     &  5.98$\pm$0.05 &  6.44$\pm$0.05 &  6.18$\pm$0.07 &  6.47$\pm$0.06 &  6.13$\pm$0.05 &  6.00$\pm$0.06 &  6.31$\pm$0.06
  \\
 12+$\log(N^+/H^+)$        &  5.90$\pm$0.06 &  6.28$\pm$0.06 &  6.43$\pm$0.06 &  6.47$\pm$0.07 &  5.67$\pm$0.09 &  6.15$\pm$0.06 &  6.30$\pm$0.07
  \\
 12+$\log(Ne^{2+}/H^+)$    &  7.20$\pm$0.03 &  7.44$\pm$0.03 &  7.47$\pm$0.04 &  7.17$\pm$0.04 &  7.24$\pm$0.03 &  7.22$\pm$0.04 &  7.35$\pm$0.04
  \\
 12+$\log(Ar^{2+}/H^+)$    &  5.50$\pm$0.05 &  5.94$\pm$0.05 &  5.73$\pm$0.09 &  5.95$\pm$0.06 &  5.59$\pm$0.06 &  5.49$\pm$0.06 &  5.78$\pm$0.06
  \\
 12+$\log(Ar^{3+}/H^+)$    &  4.58$\pm$0.07 &       ---      &       ---      &       ---      &  4.49$\pm$0.13 &       ---      &  4.28$\pm$0.12
  \\
 12+$\log(Fe^{2+}/H^+)$    &  4.81$\pm$0.08 &  5.71$\pm$0.06 &  5.69$\pm$0.07 &  5.52$\pm$0.09 &       ---      &  5.54$\pm$0.08 &  5.44$\pm$0.08
  \\

%% file: tables/total-abund-pap.tex
 \bf{12+log(O/H)}          &  7.94$\pm$0.03 &  8.19$\pm$0.03 &  8.17$\pm$0.04 &  8.07$\pm$0.05 &  8.01$\pm$0.03 &  7.99$\pm$0.04 &  8.08$\pm$0.04
  \\
 ICF($S^++S^{2+}$)         &  1.51$\pm$0.08 &  1.42$\pm$0.06 &  1.22$\pm$0.04 &  1.14$\pm$0.03 &  1.71$\pm$0.16 &  1.32$\pm$0.05 &  1.23$\pm$0.03
  \\
 \bf{12+log(S/H)}          &  6.25$\pm$0.06 &  6.73$\pm$0.07 &  6.45$\pm$0.08 &  6.71$\pm$0.07 &  6.42$\pm$0.07 &  6.46$\pm$0.07 &  6.55$\pm$0.07
  \\
 \bf{log(S/O)}             & -1.69$\pm$0.06 & -1.46$\pm$0.07 & -1.72$\pm$0.09 & -1.36$\pm$0.08 & -1.59$\pm$0.07 & -1.53$\pm$0.08 & -1.52$\pm$0.08
  \\
 \bf{log(N/O)}             & -1.26$\pm$0.10 & -1.21$\pm$0.10 & -1.24$\pm$0.11 & -1.20$\pm$0.11 & -1.40$\pm$0.15 & -1.23$\pm$0.11 & -1.27$\pm$0.10
  \\
 ICF($Ne^{2+}$)            &  1.08$\pm$0.01 &  1.08$\pm$0.01 &  1.10$\pm$0.01 &  1.12$\pm$0.01 &  1.07$\pm$0.01 &  1.09$\pm$0.01 &  1.10$\pm$0.01
  \\
 \bf{12+log(Ne/H)}         &  7.23$\pm$0.03 &  7.48$\pm$0.03 &  7.51$\pm$0.04 &  7.22$\pm$0.04 &  7.27$\pm$0.03 &  7.25$\pm$0.04 &  7.39$\pm$0.04
  \\
 \bf{log(Ne/O)}            & -0.71$\pm$0.04 & -0.72$\pm$0.05 & -0.66$\pm$0.06 & -0.84$\pm$0.07 & -0.74$\pm$0.04 & -0.74$\pm$0.06 & -0.69$\pm$0.05
  \\
 ICF($Ar^{2+}$)            &       ---      &  1.18$\pm$0.03 &  1.11$\pm$0.01 &  1.11$\pm$0.01 &       ---      &  1.13$\pm$0.02 &       ---     
  \\
 ICF($Ar^{2+}$+$Ar^{3+}$)  &  1.03$\pm$0.01 &       ---      &       ---      &       ---      &  1.02$\pm$0.01 &       ---      &  1.06$\pm$0.01
  \\
 \bf{12+log(Ar/H)}         &  5.56$\pm$0.06 &  6.01$\pm$0.05 &  5.78$\pm$0.07 &  6.00$\pm$0.06 &  5.64$\pm$0.08 &  5.54$\pm$0.06 &  5.82$\pm$0.06
  \\
 \bf{log(Ar/O)}            & -2.39$\pm$0.06 & -2.19$\pm$0.06 & -2.39$\pm$0.08 & -2.07$\pm$0.08 & -2.38$\pm$0.08 & -2.45$\pm$0.07 & -2.25$\pm$0.07
  \\
 ICF($Fe^{2+}$)            &  6.08$\pm$1.10 &  5.28$\pm$0.88 &  3.77$\pm$0.66 &  3.21$\pm$0.57 &       ---      &  4.49$\pm$0.79 &  3.82$\pm$0.56
  \\
 \bf{12+log(Fe/H)}         &  5.59$\pm$0.10 &  6.43$\pm$0.09 &  6.27$\pm$0.10 &  6.03$\pm$0.12 &       ---      &  6.19$\pm$0.11 &  6.03$\pm$0.10
  \\